\documentclass[journal,article,accept,pdftex,moreauthors]{Definitions/mdpi} 
\usepackage{pdfpages}
\usepackage{amsmath}
\firstpage{1} 
\pubvolume{1}
\issuenum{1}
\articlenumber{0}
\pubyear{2022}
\copyrightyear{2022}
\datereceived{} 
\dateaccepted{} 
\datepublished{} 
\hreflink{https://doi.org/} 
\pdfoutput=1

\Title{Deep learning of quasar lightcurves in the LSST era}

\TitleCitation{Deep learning of quasar lightcurves in the LSST era}

\Author{Andjelka B. Kova{\v c}evi{\'c} $^{1,2}$\orcidA{}*,  Dragana Ili{\'c} $^{1,3}$\orcidB{}, Luka {\v C} Popovi{\'c} $^{4,1,2}$\orcidC{},  Nikola Andri{\'c} Mitrovi{\'c} $^{5}$, Mladen Nikoli{\'c} $^{1}$, Marina S. Pavlovi{\'c}$^{6}$, Iva {\v C}vorovi{\'c}-Hajdinjak$^{1}$,  Miljan Kne{\v z}evi{\'c}$^{1}$, and {Dj}ordje V. Savi{\'c}$^{4,7}$}

\AuthorNames{Andjelka B. Kova{\v c}evi{\'c}, Dragana Ili{\'c}, Luka {\v C} Popovi{\'c}, Nikola Andri{\'c} Mitorvi{\'c}, Mladen Nikoli{\'c}, Marina S. Pavlovi{\'c}, Iva {\v C}vorovi{\'c}-Hajdinjak, Miljan Kne{\v z}evi{\'c} and Djordje V. Savi{\'c}}

\AuthorCitation{Kova{\v c}evi{\'c}, A. B., Ili{\'c}, D.; Popovi{\'c}, L. {\v C}.; Andri{\'c} Mitrovi{\'c}, N.;   Pavlovi{\'c}, M.S.;, {\v C}vorovi{\'c}-Hajdinjak, I. Nikoli{\'c}, M.;  Kne{\v z}evi{\'c}, M., Savi{\'c}, Dj. V.}

\address{$^{1}$ \quad Department of Astronomy, Faculty of Mathematics, University of
Belgrade, Studentski trg 16, 11000 Belgrade, Serbia; \\
$^{2}$ \quad PIFI Research Fellow, Key Laboratory for Particle Astrophysics, Institute of High Energy Physics, Chinese Academy of Sciences,19B Yuquan Road, 100049 Beijing, China\\
$^{3}$ \quad Humboldt Research Fellow, Hamburger Sternwarte, Universitat Hamburg, Gojenbergsweg 112, 21029 Hamburg, Germany; \\
$^{4}$ \quad Astronomical Observatory, Volgina 7, 11000 Belgrade, Serbia; \\
$^{5}$ \quad University of Padova, Department of Mathematics "Tullio Levi Civita", Via Trieste, 63 - 35121 Padova, Italy;\\
$^{6}$ \quad Mathematical institute of the Serbian Academy of Sciences and Arts,  Kneza Mihaila 36, 11 000 Belgrade, Serbia;\\
$^{7}$ \quad Institut d’Astrophysique et de Géophysique, Université de Liège, Allée du 6 Août 19c, 4000 Liège, Belgium;}

\corres{Correspondence: andjelka.kovacevic@matf.bg.ac.rs}

\abstract{{Deep learning techniques are required for the analysis of synoptic (multi-band and multi-epoch) light curves in massive data of quasars, as expected from the Vera C. Rubin Observatory Legacy Survey of Space and Time (LSST).
In this follow-up study, we introduced an upgraded version of a conditional neural process (CNP) embedded in a multistep approach for analysis  of large data of quasars in the LSST Active Galactic Nuclei Scientific Collaboration data challenge database.
We present a case study of a stratified set of \texttt{u}-band light curves for 283 quasars with very low variability $\sim 0.03$.
In this sample, CNP average mean square error is found to be $\sim 5\% $($\sim 0.5$ mag). Interestingly, beside similar level of variability there are indications that individual light curves show flare like features.   According to preliminary structure function analysis, these occurrences may be associated to microlensing events with larger time scales $5-10$ years.}
}


\DeclareUnicodeCharacter{2212}{-}
\begin{document}



\section{Introduction}

\begin{figure}[H]
\includegraphics[width=0.8\textwidth]{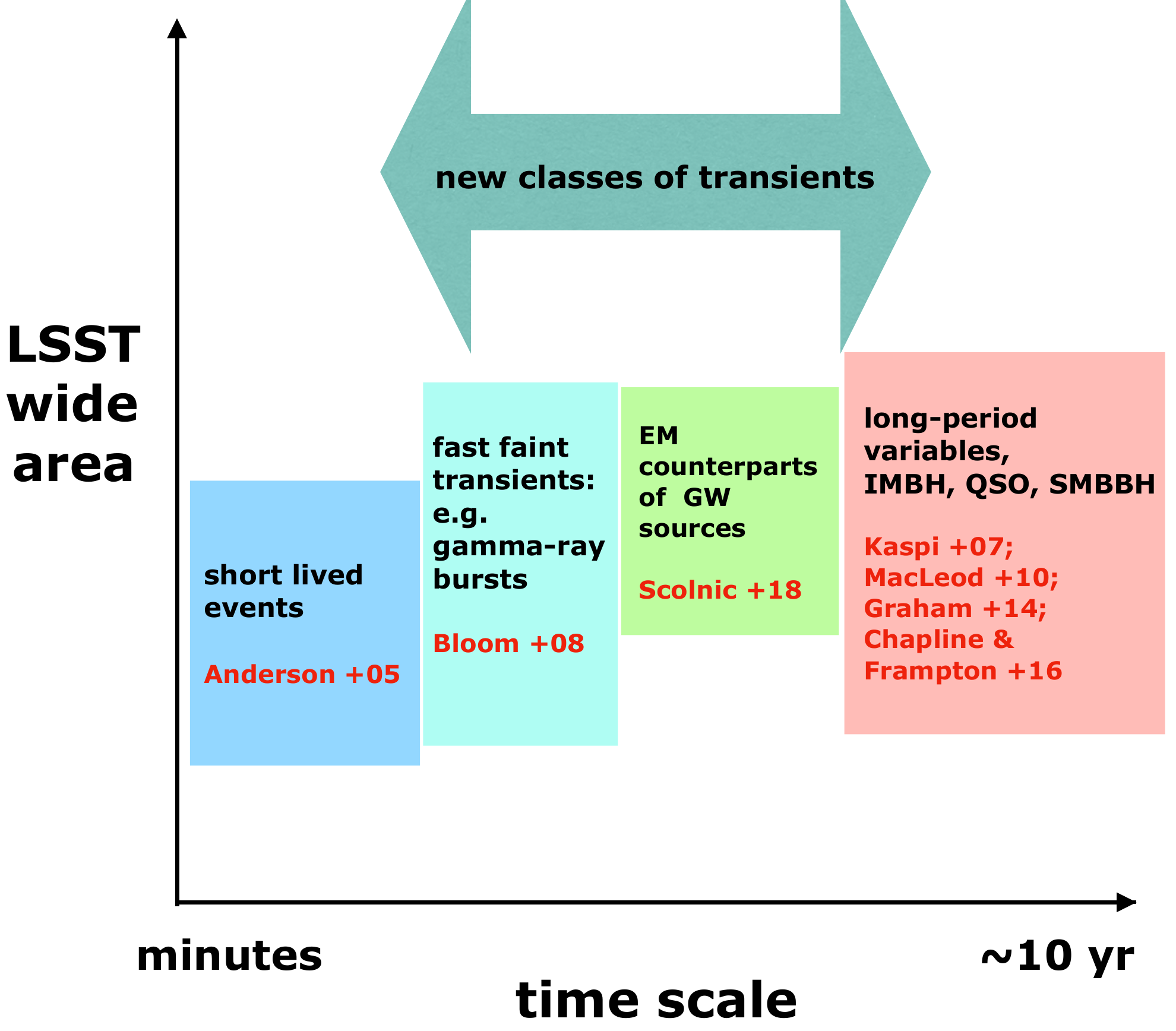}
\caption{Schematic view of the LSST third scientific pillar-
Exploring the transient optical sky as defined in \citet{2019ApJ...873..111I}. The broad range of probed time scales and sky area may allow for the search for general variability properties across different types of objects, similar to study by \citet{10.1126/science.abg9933} which is suggesting a common process for all accretion disks. The references \citet{2005AJ....130.2230A,
2008AN....329..284B,
2018ApJ...852L...3S,
2007ApJ...659..997K,
2010ApJ...721.1014M,
2014MNRAS.439..703G,
2016JCAP...11..042C} given in boxes are further explained in the text. \label{lsst}}
\end{figure}

The launch of the Legacy Survey of Space and Time (LSST), which will be conducted by the Vera C. Rubin Observatory, is currently scheduled to take place in the first half of 2024.
The cadences of the LSST, in concert with its large observational coverage, will capture a wide variety of intriguing time domain events, some of which are periodic signals of interest \citep{2019ApJ...873..111I}.
LSST should probe time series with cadences ranging from one minute to ten years across not only a vast portion of the sky, but also  across five photometric bands (see Fig. \ref{lsst}). 

Such synoptic (multiband and multi-epoch) cadences combined with the large
coverage will enable us to detect very short-lived events such as eclipses in ultracompact double-degenerate binary systems \citep{2005AJ....130.2230A}, fast faint
transients-such as optical phenomena associated with gamma-ray  bursts \citep{2008AN....329..284B}, and  electromagnetic counterparts to gravitational wave sources \citep{2018ApJ...852L...3S, 10.1093/astrogeo/atab077}. In contrast, the LSST decadal data catalogues will make it possible to investigate long-period variables, intermediate-mass black holes (IMBH), and quasars (QSO) \citep{2007ApJ...659..997K, 2010ApJ...721.1014M, 2014MNRAS.439..703G, 2016JCAP...11..042C, 10.1126/science.abg9933}.

Quasars are an important population to study in order to have a better grasp of the physics behind the accretion of the matter when it is subjected to extremely harsh conditions.  Moreover, studies are showing that they may be used as cosmological probes  \citep[e.g.][]{2015ApJ...815...33R,2021IAUS..356...66M}. Up to this point, several hundred thousand quasars have been spectroscopically confirmed, and numerous efforts have been done to identify the properties of their temporal flux variability \citep{Tachibana_2020}. Proposed  physical mechanisms underlying the optical/UV
variability  range from the superposition of supernovae \citep[e.g.,][]{1998ApJ...504..671K}, microlensing \citep{2007A&A...462..581H, 2004A&A...420..881Z}, thermal fluctuations
from magnetic field turbulence \citep{2009ApJ...698..895K} up to instabilities in the accretion disk \citep{1998ApJ...504..671K}.
{The amplitude of quasar observed optical variability is typically a few tenths of a magnitude \citep[e.g.][found the SDSS quasars variability is $\sim 0.03$ mag]{2007AJ....134.2236S} with a characteristic time-scale of several months, but it can also show larger variations over longer time-scales \citep[see][]{MacLeod_2012, 2017A&A...597A.128K}, with statistical description via a damped random walk (DRW) model \citep[e.g.,][]{Kelly_2009,Kelly_2014, 2017A&A...597A.128K}. The long lasting flare like events (extreme tails of the variability distribution) are less clearly defined and represent a distinct kind of modeling problem \citep[see e.g.,][and references therein]{10.1093/mnras/stx1456}.
On top of this, the light curves have different topologies which are superimposed on different type of cadences,  which imposes many difficulties in their modeling and extracting knowledge from them.  \footnote{Gaining knowledge from large astronomical databases is a complex procedure, including various deep learning algorithms and procedures so we will use 'deep learning' in that wide context.}}

Specifically, the LSST will provide a breakthrough in quasar observations  in survey area and depth \citep{10.1093/mnras/stab1856}  as well as variability information of light curves sampled at a relatively high cadence (the order of days),  over the course of a decade of the operations.
 Because of these new qualities, the LSST will be able to search even for supermassive black hole (SMBH) binaries having shorter  periods ($<5$ years), which are significantly more uncommon.
For instance, \citet{10.1093/mnras/stab1856}, suggested that it might be possible to identify ultra-short-period SMBH binaries (periods $<3$ days) in the LSST quasar catalogue. These binaries are thought to be so compact that they will 'chirp', or evolve in frequency, into the gravitational wave band, where the Laser Interferometer Space Antenna \citep[LISA][]{2017arXiv170200786A} will be able to detect them in the middle of the 2030s. 
Expectations of such exciting discoveries are probable \citep{10.1093/mnras/stab1856},
as massive binary SMBHs are predicted to spend $O(10^{5})$ years in orbits with periods of the order of a year if orbital decay is caused by either gravitational wave emission or negative torques exerted on the viscous time-scale by the surrounding gas disc \citep{Haiman_2009}.

Even while the LSST will have  cadences that are unprecedented in contrast to those of its predecessors, these cadences will primarily take the shape of more or less regular samplings that are separated by various short or lengthy periods (seasons) in which there are no observations.
\unskip

When attempting to evaluate the variable features of quasar light curves, these frequent gaps represent one of the most challenging obstacles to overcome \citep[along with the obviously irregular cadences, see][]{Kelly_2014}. In quasar time domain analysis, there are two main ways to deal with sampling that is not even in stochastic light curves \citep{Kelly_2014}: the first approach is to use Monte Carlo simulations to forward model in frequency domain \citep[see e.g.,][]{2013MNRAS.433..907E}, whereas the second approach  is to fit the light curve in the time domain mostly using Gaussian Processes \citep[GP, see][]{2013ApJ...779..187K}.

Both approaches are viable, although they can be computationally expensive \citep[see detailes][and references therein]{Kelly_2014}. If the cadences are similar to those found in the LSST with seasonal gaps, the first method is a costly one to compute because it necessitates either creating a highly dense light curve at the optimal sampling rate or segmenting the light curve and computing the periodograms of each segment separately. The likelihood function of GP, on the other hand, is computationally expensive (scales $\sim O(n^{3})$) for the second approach because it requires inverting the $n \times n$ covariance matrix of the light curve, where $n$ is the number of data points.
A first-order continuous-time autoregressive process (CAR(1)) or an Ornstein-Uhlenbeck process, which could represent quasar light curves, is a special class of GP for which the computational complexity only scales linearly with the length of the light curve \citep{Kelly_2009}.
Application of CAR(1) to modeling typical  active galactic nuclei (AGNs) optical light curves is questioned \citep[see][]{2017A&A...597A.128K},  as some studies have found evidence for deviations from the CAR(1) process for optical light curves of AGN \citep[see e.g.,][]{2011ApJ...743L..12M, 2014MNRAS.439..703G, Smith_2018}.
Due to the nature of this issue, it was necessary to develop more complex Gaussian random process models, such as the continuous auto-regressive moving-average  models \citep[CARMA,][]{Kelly_2014}. 
Moreover, it has been demonstrated by \citet{yu22} that a second-order stochastic process, a damped harmonic oscillator (DHO), is a more accurate way to characterize the variability of AGNs.
Based on previous examples, the evolution of algorithms used to model AGN light curves typically emphasizes an increase in the total number of needed parameters.
All of these models, however, are based on information gathered before the LSST era, which has a tendency to favor more luminous and nearby AGNs.
Therefore, in order to make use of the tens of millions of the LSST AGN multiband light curves effectively,  it is highly desirable to employ flexible data driven machine
learning algorithms.

At the moment, kernel methods (like GPs) and deep neural networks are seen as two of the most remarkable machine learning techniques \citep{9815027}. The relationship between these two approaches has been the subject of a great deal of research in recent years \citep{9815027,2022arXiv221110305D}.
In this light, here we present the AGN light curve modeling unit, which was created as preprocessing module of the SER-SAG\footnote{SER-SAG is Serbian team that contributes  to AGN investigation and participates in the LSST  AGN  Scientific Collaboration} team's LSST in-kind contribution of time-domain periodicity mining pipeline and combines the best of two machine learning worlds.
The neural latent variable model \citep[Neural Process-NP][]{88fae17e4b14416d84edf0b23a013f2a} is at the heart of this modeling unit. NPs, like GPs, create distributions over functions, can adapt quickly to new observations, and can assess the uncertainty in their predictions \citep{88fae17e4b14416d84edf0b23a013f2a}. NPs, like neural networks, are computationally efficient during training and assessment, but they also learn to adapt their priors to data as well \citep{88fae17e4b14416d84edf0b23a013f2a}.
The NP module has been trained  on the quasars light curves found in a dedicated database that arose  from a challenge focused on the future use of the LSST quasar data \citep[\texttt{LSST\_AGN\_DC}][]{yu2022}. During the course of the operation, we also came to the realization that it is possible to distinguish the variable properties of quasars. 

{In our previous work \citep[see][hereafter Paper I]{2022AN....34310103A} we adapted conditional neural process (CNP) for modeling general variability of quasar light curves on smaller sample of tens of objects. In this work, we complement the study of Paper I by upgraded version of CNP on much larger database of ($\sim 4\times 10^{5}$) quasars (\texttt{LSST\_AGN\_DC}) which demanded 'deep learning'  through multistep process and provide a case study example of how complex procedure of 'deep learning' of quasar variability may  lead to surprising results, i.e. detection of larger collection of quasars with flare like events. These flare-like incidents occur over a longer time period, and preliminary structure function analysis suggests that they may be related to microlensing events.}

It is anticipated that the LSST would lead to an increase of at least one order of magnitude in the number of  lensed quasars that are known \citep{2010MNRAS.405.2579O}. Thus optimizing the analysis or selecting sub-samples of those systems is becoming important \citep[e.g.,][]{10.1093/mnras/staa1208}. 

The structure of the paper is as follows. In Section \ref{materials} we describe the  data used for the machine learning experiments. In Section \ref{methods} we provide a concise explanation of the machine learning methods that were applied, while in Section \ref{resdisc} we report and discuss in further detail the series of experiments that were carried out. In the final Section \ref{conclusion}, we summarize our findings.

\section{Materials}\label{materials}

\subsection{Description of quasars data in \texttt{LSST\_AGN\_DC} }

The dataset of AGN light curves used for the demonstration of our neural process modeling unit is selected from the LSST AGN data challenge 2021 dataset \citep[\texttt{LSST\_AGN\_DC}, see details in][]{yu2022}.
The \texttt{LSST\_AGN\_DC} mimics the future LSST data release catalogs as much as possible \citep[see also][]{Savic}.
{Calculations  were run on NVIDIA T4, 2560 CUDA cores, Compute capability 7.5, Memory 16GB  GDDR6, Max memory bandwidth  300 GB$/$sec.}

The total number of objects in the \texttt{LSST\_AGN\_DC} is $\sim 440 000$, with stars, galaxies and 39173 quasars drawn from two main survey fields, an expanded Stripe 82 area and the XMM-LSS region. The total number of epochs for all objects is $\sim 5\times 10^{6}$ \citep{yu2022}. {Quantity of  features (parameters)  of objects  is 381,  sorted as \citep[see details in][]{yu2022, Savic} astrometry (celestial equatorial coordinates, proper motion, parallax); photometry (point and extended source photometry in  AB magnitudes and fluxes (in nano-Jansky); color (derived from flux ratio between different photometric bands); morphology (continuous number in range $[0,1]$,  extended sources has morphology closer to 1, while  point-like sources closer to 0); light curve features (extracted from SDSS); spectroscopic and photometric redshift; and class labels (Star/Galaxy/quasar). The time domain data of light curves include observation epochs, photometric magnitudes and  errors in \texttt{u,g,r,i,z} bands as well as   periodic and non periodic features \citep[see  description of features in][]{2011ApJ...733...10R}.}

 \subsection{Sample selection}

\begin{figure}[H]
\includegraphics[width=0.7\textwidth]{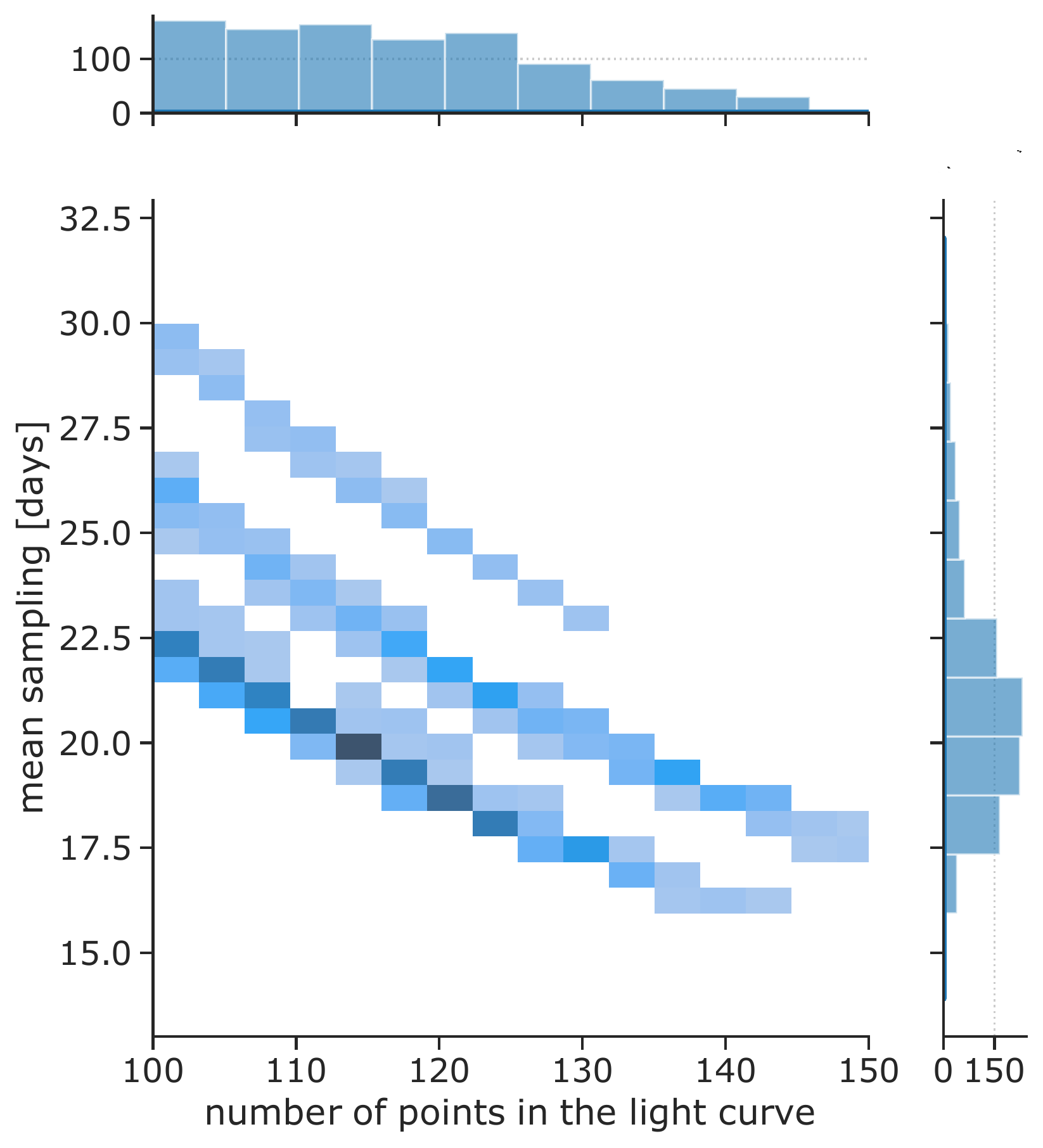}
\caption{Statistical description of  1006 quasar light curves in the \texttt{u} band having $\geq$ 100 points. The primary joint distribution plot of the number of points in the light curves and mean sampling makes the stratification about mean sampling a very obvious. The non-gaussianity of the mean sample and the number of points in the light curves are brought to light by the marginal distributions that are presented on the side plots. The marginals are presented as counts in this instance. Upper marginal: y axis counts a number of light curves having a number of points given in  x-axis  in the main plot. Right marginal: y axis counts a number of light curves having mean sampling  in  y-axis  in the main plot. \label{fig2}}
\end{figure}

We chose 1006 quasars, {that are spectroscopically
confirmed and which have}  $\geq$ 100 epochs in \texttt{u}-band light curves (see Fig. \ref{fig2}) from this large initial database. We employ this criterion because earlier research has demonstrated that this number of points is acceptable for both modeling light curves and extracting periodicity \citep[see e.g.,][]{2021arXiv210512420K}. We selected to study  \texttt{u}-band light curves, since they are less deformed by photometric filters \citep{2022ApJS..262...49K}. In terms of mean sampling, chosen quasar subsample  exhibits a  stratification into three non-intersecting branches of mean sampling
 (Fig. \ref{fig2}).

Both non-gaussianity of marginal distributions and stratification in the phase space of mean sampling and the number of points in the light curves indicate that we are encountering a highly diverse sample.

{
\citet{10.1093/mnras/stv1230} provided the first case study  of the relevance of using modeling procedures individually on stratified light curves, as opposed to "one-size-fits-all" approach, which allows to account for the prevailing physical processes heterogeneity. The authors categorized the Kepler light curves of 20 objects based on visual similarities and found that the light curves falls into five broad strata: stochastic-looking, somewhat stochastic-looking+weak oscillatory features, oscillatory features dominant, flare features dominant, and not-variable.
Certain light curves appear to change from one state of variability to another.}

\begin{figure}[H]
\includegraphics[width=\textwidth]{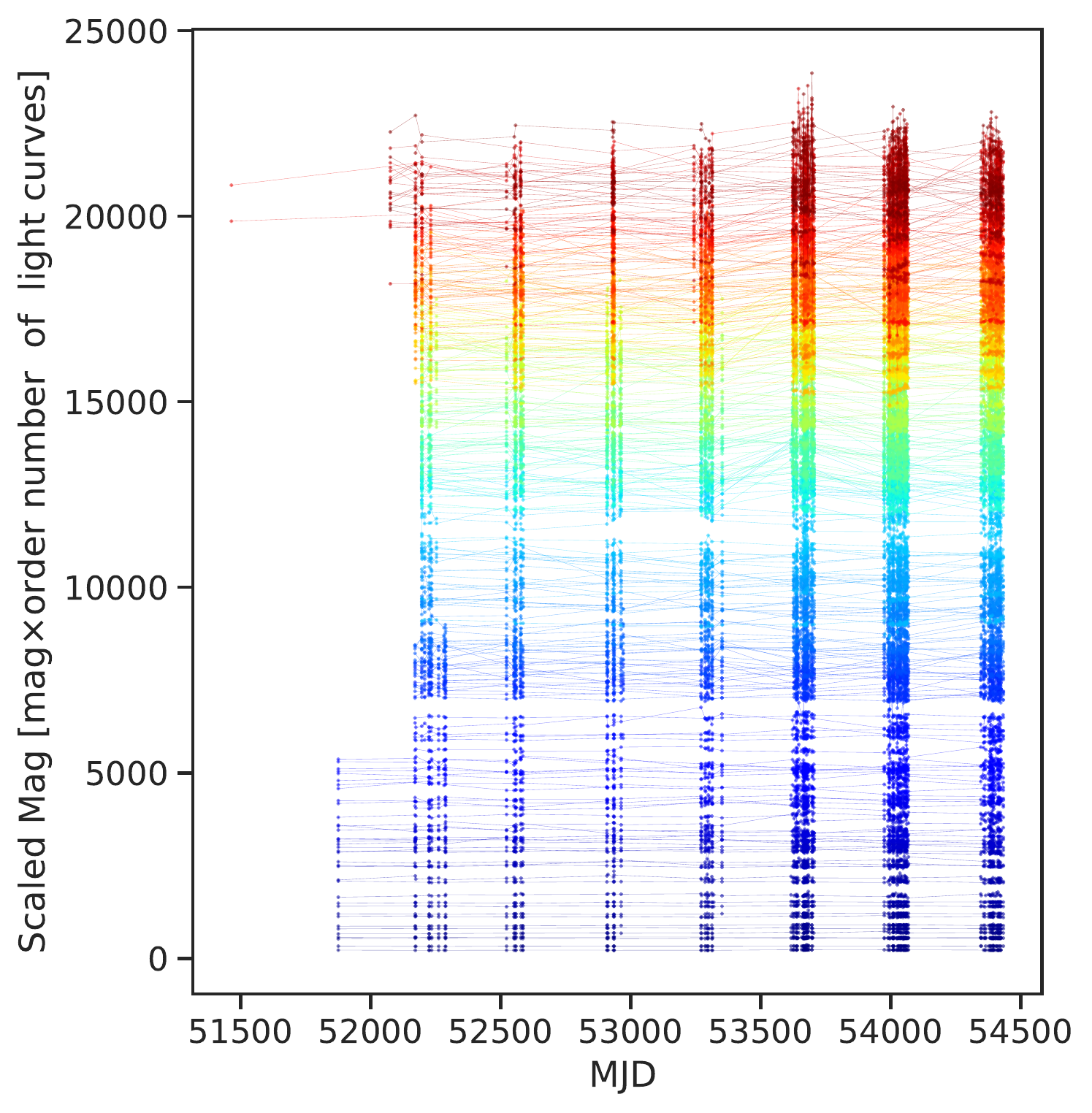}
\caption{Cluster of  283 \texttt{u}-band light curves with low variability obtained via SOM algorithm.  Observed magnitudes are  given as points, while solid lines serve as an eye guide. The magnitudes of each light curve have been multiplied by their ordinal number for clarity. Warmer colors correspond to higher ordinal numbers.   \label{fig3}}
\end{figure}

{ Motivated by  \citet{10.1093/mnras/stv1230} example and  large number of  our preselected objects (1006) which can not be visually stratified,}  we  employed the Self-Organizing
Maps (SOM) algorithm \citep{vettigliminisom} to stratify (cluster) light curves  with similar topological patterns. 
From 36 clusters obtained from SOM, we chose an interesting strata containing 310 light curves with apparent low variability (see Figure \ref{fig3}).

To check thoroughly the characteristics of 283 sources, we calculated  the fractional root mean square (rms) variability amplitude $F_{var}$ and the optical
luminosities ($L_{op}$).  

The uncertainties of the individual magnitude measurements
will contribute an additional variance which is captured in $F_{var}$ \citep[see][]{1990ApJ...359...86E, 10.1046/j.1365-2966.2003.07042.x} as:
\begin{align}
S^{2}&=\frac{1}{N-1}\sum^{N}_{i=1} (mag_{i}-<mag>)^{2}\\
F_{var}&=\sqrt{\frac{S^{2}-<\sigma>^{2}}{<mag>^{2}}}
\end{align}
\noindent where $N$ is the number of points in the light curves, $mag_{i}, i=1,...,N$ are observed magnitudes, $<mag>=\frac{1}{N}\sum^{N}_{i=1} mag_{i}$, $<\sigma>^{2}=\frac{1}{N}\sum^{N}_{i=1} {\sigma}^{2}_{i}$ and ${\sigma}_{i}, i=1,...,N$ are measurement errors\footnote{$F_{var}$ behaves as normalized variance so it is more robust to outliers, flares.}

The black hole masses ($M$) were randomly assigned using the probability distribution based on absolute magnitude $M_{u}$ by \citep{Macleod2010}:
\begin{equation}
P(\log_{10}M|M_{u})=\frac{1}{\sqrt{2\pi\sigma_{M}^{2}}}
exp^{-\frac{(\log_{10}M-\log_{10}<M>)^{2}}{2\sigma_M}}
\end{equation}
\noindent where $\log_{10}<M>=2.0-0.27M_{u}$, $\sigma_{M}=0.58+0.011M_{u}$, and $M_{u}$ is an absolute magnitude and  is calculated using the known \texttt{u}-band magnitude and K-correction, $K(z)=-2.5(1+\delta)\log(1+z)$, with the canonical spectral index $ \delta=-0.5$  as in \citet{Solomon_2022}. The assigned masses of  black holes serve as proxies that complement other inferred quasar properties.

\unskip

\begin{figure}[H]
\includegraphics[width=\textwidth]{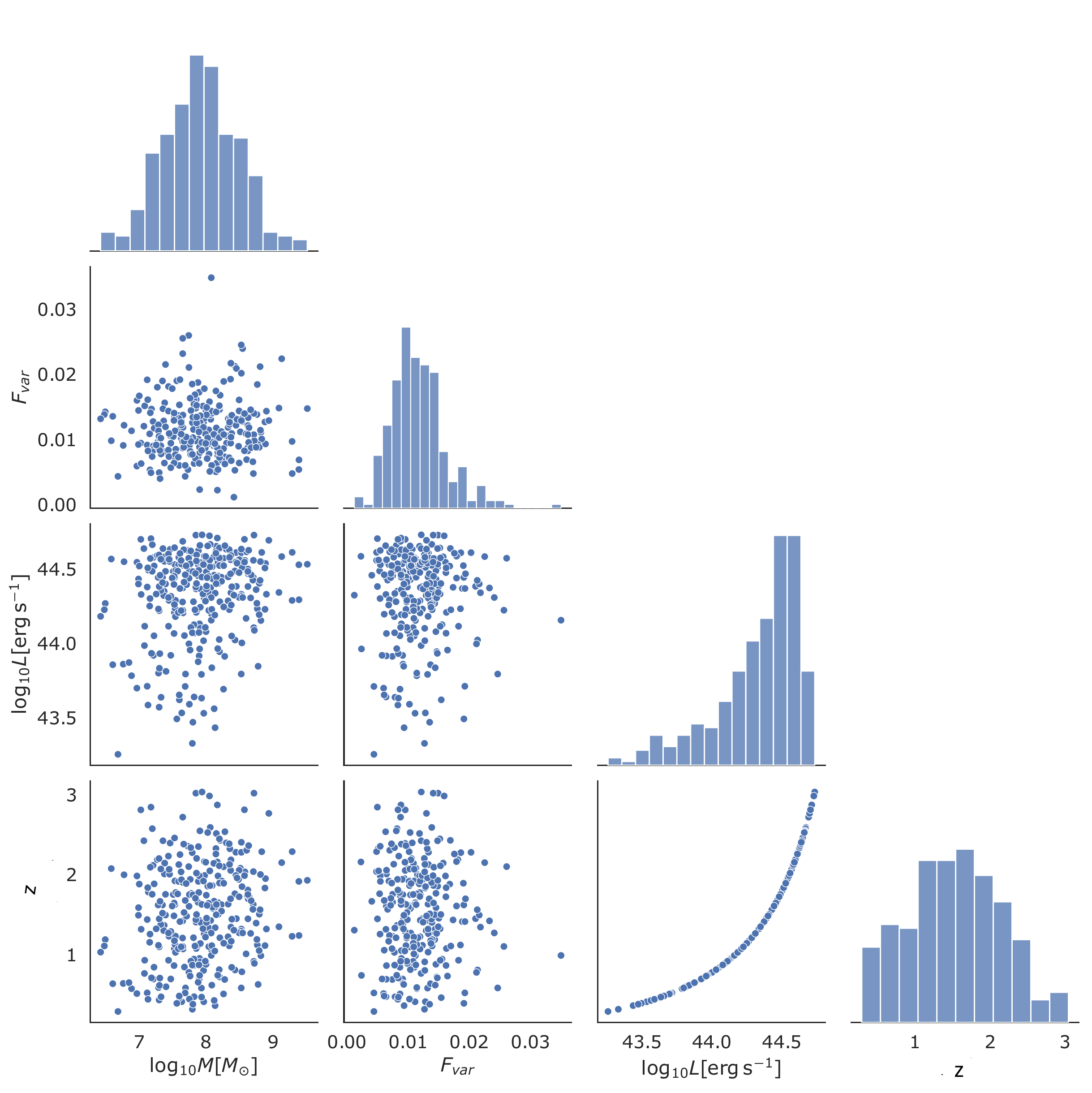}
\caption{ 
Corner plot of the probability distribution of quasars parameters in selected Cluster 36. Because the parameter distribution is multi-dimensional (only four are seen here),  the information is unfolded into a succession of 1D and 2D representations. The histograms of the marginalized probability for each parameter are given on the diagonal. The scatter plots for all pairwise parameter combinations are given on the off-diagonal.
  \label{fig31}}
\end{figure}   

We approximate the optical luminosity ($L_{op}$)  of the sampled objects via \citep[see][]{Tachibana_2020}
\begin{equation}
L_{op}[\mathrm{erg}\,\mathrm{s^{-1}}]=4\pi d^{2}_{L} F_{0, \lambda}\lambda_{eff} 10^{-2.5(<mag>-A)}
\end{equation}

\noindent where $d_{L}$ is the estimated luminosity distance  for a flat universe \citep[using \texttt{astropy} module,][]{Condon_2018} with standard cosmological parameters $H_{0}=67.4 \mathrm{km}\,\mathrm{s}^{-1}\mathrm{Mpc}^{-1}$ and $\Omega_{m}=0.315$ \citep{Planck}, $c$ is the light speed, $z$ is the redshift of the object, $F_{0,\lambda}=3.75079\times 10^{-9} \mathrm{erg},\mathrm{cm}^{2}\,\mathrm{s}^{-1}${\AA}$^{-1}$ is  the zero point flux density, $\lambda_{eff}=3608.04${\AA}  is
the effective wavelength of the SDSS filter system \citep[see][]{Rodrigo1, Rodrigo2}, $<mag>$ is a mean \texttt{u}-band magnitude, and A is the Galactic absorption at the effective wavelength along the line of sight. However, for our purposes we did not take into account $A$.

We present a corner plot of the four parameters distributions ($M, F_{var}, L_{op}, z$)  in Figure \ref{fig31}.
Selected objects are indeed characterized by small variability $F_{var}\sim[0.,0.03]$ and larger redshift $1\leq z \leq 3$.
Moreover, the two-dimensional plots reveal some scatter in the parameter distributions.
In contrast, the 2D distribution of luminosity and redshift of objects is strongly nonlinear, as expected \citep[see also][]{Tachibana_2020}.  
Finally, {after excluding objects with an exact number of 100 points in the light curves, } our sample contained 283 light curves with $>100$ data points that were used for NP modeling.

\section{Methods}\label{methods}
In this section we provide motivation and description of computational model.
\subsection{Motivation}
{
The stochastic variability seen in quasar flux time series is thought to be caused by emission from an accretion disc with local 'spots' that contribute more or less flux than the disc's mean flux level. These spots appear at random and dissipate over a specific physical time scale \citep{2011ApJ...727L..24D}. Because the spots do not dissipate instantly within the disc, some long-term correlations may exist, which can be described by power spectrum density (PSD $\sim f^{-2}$, where $f$ is frequency) consistent with the Autoregressive (1) model (AR(1)), or damped random walk, or simplest form of Gaussian process characterized by  the relaxation time, and the variability
on timescales much shorter than relaxation time \citep[see][]{Kelly_2009}. Some ground-based studies \citep{2014MNRAS.439..703G, 2013ApJ...765..106Z} show that AGN light curves could have PSD slopes steeper than AR(1) PSD on very short time scales, indicating that the damped random walk process oversimplifies  optical quasar variability \citep[see][]{2014MNRAS.439..703G,2017ApJ...834..111C}. Also, \citet{10.1093/mnras/stv1230} developed the damped power-law (DPL) model by generalizing the PSD of AR(1) as $\sim 1/f^{\gamma}$. If $\gamma <2$, the process exhibits weaker autocorrelation on short time scales than the AR(1), resulting in a less smooth time series. When $\gamma>2$, the process exhibits stronger autocorrelation on short time scales than the AR(1), resulting in a smoother time series. 
However, the light curves show a wide variety of different types of behaviour, even superposition of at least two features \citep[e.g., stochastic+flare][]{10.1093/mnras/stv1230}.
\citet{2012ApJ...760...51R} modeled blazar variability flare-like features with an AR(1) process using observed 101 blazar light curves from the Lincoln Near-Earth Asteroid Research (LINEAR) near-Earth asteroid survey. Also \citet{10.1093/mnras/stv1230} tried to model flare like features in Kepler light curves with DPL and AR(1) and point out that both models are  unable to model flare like features.
Beside variety of the strength of correlation in the AGN light curves, different topologies of light curves, the next issue which should be taken into account is the cadence gaps (long time ranges without observation) of variable size. As we want to predict data in these gaps, it is important that we do not introduce any additional relation which could be reflected in PSD \citep[see][]{Smith_2018}.
The various types of Neural processes can  be considered, such as attentive processes which can introduce attentive mechanisms for correlation among data points. However due to large cadence gaps in quasar light curves, in Paper I, we presented successful  application of the Conditional neural process (CNP,  general Neural process which does not introduce correlation),   combining the neural network and general Gaussian processes capabilities, to model tens of stochastic light curves with large gaps and without flares. 
Here we show upgraded version of the code, which is applied on the 
quasar light curves obtained from the largest database mimicking LSST survey (\texttt{LSST\_AGN\_{DC}}, containing about 40000 quasars).}

\subsection{Conditional neural process}
 {Here, we briefly summarize the conditional neural process description; for a detailed mathematical description,  the reader is referred to the given literature.  }
It is commonly accepted in the field of machine learning that models must be "trained" with a large number of examples before they can make meaningful predictions about data they have never seen before. However, there are several instances where we do not have enough data to meet this demand: acquiring a substantial volume of data may prove prohibitively expensive, if not impossible. For example, it is not possible to obtain homogeneous cadences of observations with any ground-based telescope, including the LSST. Nonetheless, there are compelling grounds to believe that this is not a side effect of learning. Humans are known to be particularly good at generalizing after only seeing a small number of examples, according to what we know ("few-shot estimate").
In current meta-learning terminology, NPs and GPs are examples of approaches for "few-shot function estimates "\citep{88fae17e4b14416d84edf0b23a013f2a}.
NPs, as opposed to GPs, are metalearners \footnote{We underline the distinction between NPs and both classic neural networks and GPs previously applied to quasars' light curves. Classical neural network fit a single model across points based on learning from a large data collection, whereas GP fits a distribution of curves to a single set of observations (i.e. one light curve). NP combines both approaches, taking use of neural network ability to train on a large collection and GP's ability to fit the distribution of curves because it is a metalearner.} \citep[see][]{foong}.

{We assume the latent continuous time light curve $(\mathbf{x}, \mathbf{y})_{\mathcal{L}}=\left\{(x^{l}, y^{l})\right\}^{L}_{l=1}$ with time instances $\mathbf{x}$ and fluxes $\mathbf{y}$, is a realization of a stochastic  process, so that observed points are sampled from it at irregular instances $(\mathbf{x}, \mathbf{y})_{\mathcal{O}}=\left\{(x^{o}, y^{o})\right\}^{O}_{o=1})$ \citep[see][]{tak2017bayesian},and  which can be learned through Neural Process,   as it is  a} way to meta-learn a map from datasets to predictive stochastic processes using neural networks \citep{foong}.

If we are given target inputs of time instances $(\mathbf{x}_{\mathcal{T}}=\left\{x^{t}\right\}^{T}_{t=1})$, and corresponding unknown fluxes $\mathbf{y}_{\mathcal{T}}=\left\{y^{t}\right\}^{T}{_{t=1}}$,   we need  a distribution over predictions $(p(\mathbf{y}_{\mathcal{T}}| \mathbf{x}_{\mathcal{T}}; \mathcal{C}^{\sim}))$\footnote{In neural processes, the predicted distribution over functions is typically a Gaussian distribution, parameterized by a mean and variance.}, where ${\mathcal{C}^{\sim}}=\left\{(\mathbf{x}_\mathcal{C},\mathbf{y}_\mathcal{C})=(x^{c},y^{c}), c=1,...C,\right\}$ are context points from the light curve (used for training the model).
The $p(\mathbf{y}_{\mathcal{T}}| \mathbf{x}_{\mathcal{T}}; \mathcal{C}^{\sim})$  distribution  is called the stochastic process.

If we chose predictors at random from this distribution, each one would be a plausible way to fit the data, and the distribution of the samples would show how uncertain our predictions are.
So, the NP can be seen as using neural networks to meta-learn a map from datasets to predictive stochastic processes.

Generally, NPs are constructed to firstly map the entire context set to a representation $R$, ${Enc}_{\theta}: \, \mathcal{C}^{\sim}\rightarrow R$\footnote{{We will use a subscript $\theta$
 to denote all the parameters of the neural network such as number of layers, learning rate, size of batches, etc.}}, using an encoder $\mathrm{Enc}_{\theta}(\mathcal{C}^{\sim}) = \rho \left ( \sum_{c=1}^C  \phi(x^{(c)}, y^{(c)}) \right)$, where $\rho, \phi$ are defined by neural network \citep{foong}.
The sum operation in the encoder is a key as it ensures  that the resulting $R$ “resides” in the same space regardless of the number of context points $\mathcal{C}$.
The predictive distribution at any set of target inputs $\mathbf{x}_{\mathcal{T}}$ is factorised, conditioned on lower dimensional representation $R$. 
Having this in mind, one can write that $p_{\theta}(\mathbf{y}_{\mathcal{T}} | \mathbf{x}_{\mathcal{T}}; \mathcal{C}^{\sim}) = \prod_{t=1}^T p_{\theta}(y^{(t)} | x^{(t)}, R)$.
In the next step, NP calls the decoder, $\mathrm{Dec}_{\theta}$,  which is the map parametrizing the predictive distribution using the target input $x^{(t)}$ and the encoding of the context set $(R)$.
Typically the predictive distribution is multivariate Gaussian, meaning that the decoder predicts a mean $\mu^{(t)}$ model and a variance $(\sigma^{2(t)})$ \citep{foong}.

\begin{figure}[H]
\includegraphics[width=0.7\textwidth]{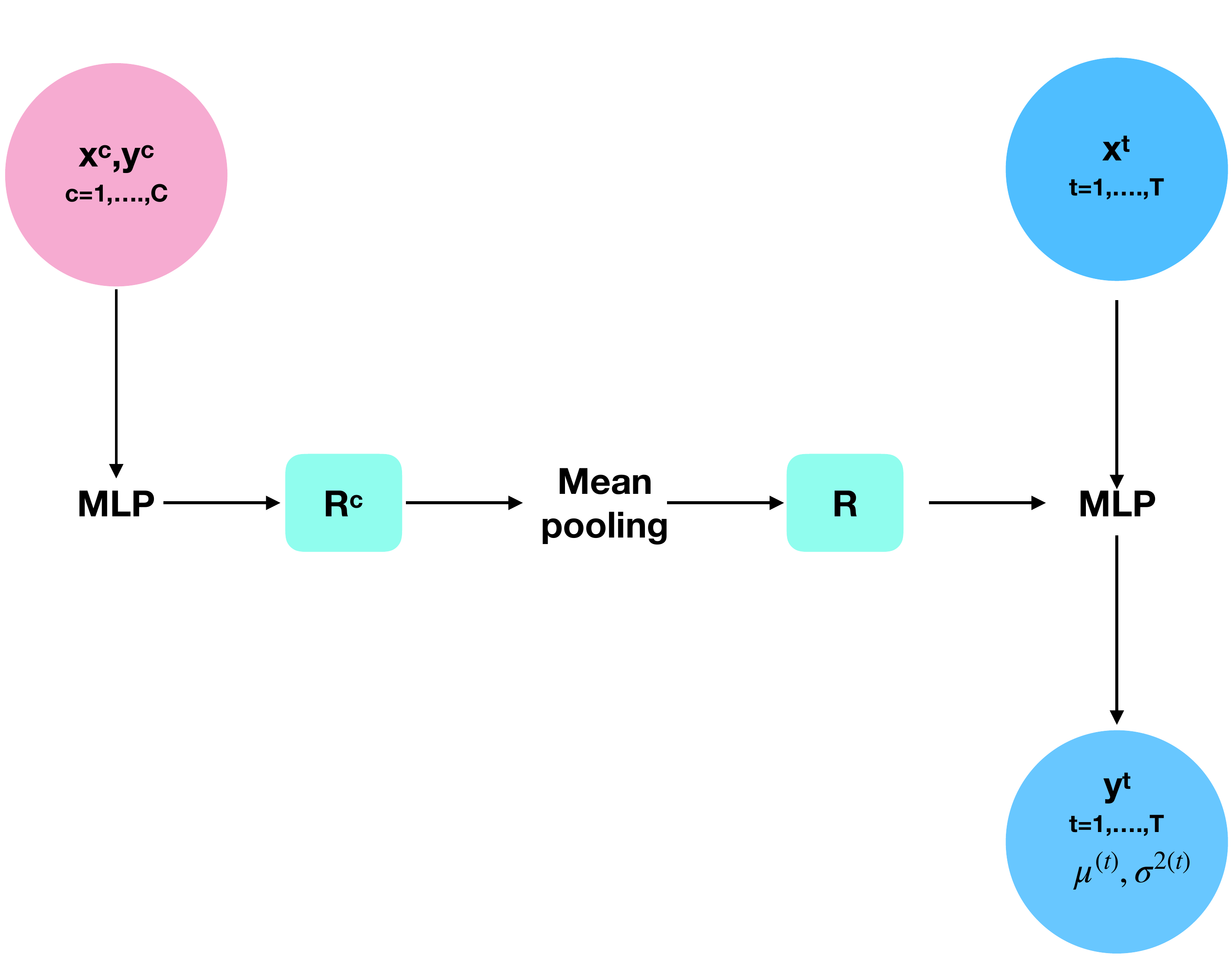}
\caption{Forward pass (left to right) computational graph of the conditional neural process \citep[see][]{foong}. Context (observed points) are given in a red circle. MLP is a multilayer perceptron. $R^{c}$ is local encoding of context points, whereas $R$ is general encoding. For given target time instances $x^{t}$ in a blue circle, the decoder which is an MLP produces the predictive distribution of the target output ($\mu^{(t)}$, $\sigma^{{2}(t)}$) and  predicted values $y^{t}$. \label{fig4}}
\end{figure}

Specifically, the scheme of the particular member of NP called Conditional Neural Process \citep[CNP][]{88fae17e4b14416d84edf0b23a013f2a} is given in Fig. \ref{fig4}.  Each  pair in the context set  is locally encoded by a multilayer perceptron:

\begin{align}
    R^{c} = \mathrm{Enc}_{\theta}(\mathcal{C}^{\sim}) =  \mathrm{MLP} \left( [x^{(c)}; y^{(c)}] \right)
\end{align}
Here comes the main difference between the CNP and other NP: the local encodings $R^{c}$ are then aggregated by a mean pooling to a global representation $R$:

\begin{align}
  R = \mathrm{Enc}_{\theta}(\mathcal{C}^{\sim}) = \frac{1}{C} \sum_{c=1}^{C} \mathrm{MLP} \left( [x^{(c)}; y^{(c)}] \right)
\end{align}

Finally, the global representation $R$ is fed along with the target input $x^{t}$ into a decoder MLP  to yield the mean and variance of the predictive distribution of the target output:

\begin{align}
(\mu^{(t)}, \sigma^{2(t)}) = \mathrm{Dec}_{\theta}(R,x^{(t)}) =  \mathrm{MLP} \left( [R,x^{(t)}] \right)
\end{align}

{Encoder consists of  1-hidden layer (dimension $2\times128$) that encodes the features of time instances, followed by a 3 hidden layers (dimension $128\times128$ that locally encodes each feature-value pair (time instances, magnitudes) with final activation ReLU layer. Decoder has a 4 hidden layer MLP of the dimension ($128\times128$) that predicts the distribution of the target value, with last layer consisting of \texttt{softmax} activation function. We note that the the number of layers, batch sizes, learning rates and optimizers are  chosen as a balance  between the hyperparameters found in literature where multilayer approach is more feasible \citep[see] [also]{Tachibana_2020} and by our experimentation.}
The aim of the training CNP is to minimize the negative conditional log probability (or loss) \citep[more detialed explanation is given in][]{88fae17e4b14416d84edf0b23a013f2a,2022AN....34310103A}: 
\begin{equation*}
\mathcal{L} = -\frac{1}{N}\sum_{\mathbf{y}_{\mathcal{T}}\in \mathcal{T}} 
\log(p(\mathbf{y}_{\mathcal{T}}| \mathbf{x}_{\mathcal{T}}; \mathcal{C})) 
\label{}
\end{equation*}
\noindent where $p(\mathbf{y}_{\mathcal{T}}| \mathbf{x}_{\mathcal{T}}; \mathcal{C})$ defines  {a posterior distribution for target values}  \footnote{An assumption  on  $\mathcal{P}$ is that all finite sets of function evaluations of
$f$ are jointly Gaussian distributed. This class of random
functions are known as Gaussian Processes (GPs).} over functions which could be fitted through observed data points, and $N$ is a cardinality of a randomly chosen subset of observations used for conditioning.
{ An increase in the log-probability  indicates that the predicted distribution better describes the data sample statistically. \footnote{{ We note that the our loss function  works quite similarly to the Cross-Entropy. In the \texttt{PyTorch} ecosystem, Cross-Entropy Loss is obtained by combining  a log-softmax layer and  loss. }} }

The computational cost of prediction estimates for $\mathcal{T}$ target points conditioned on $\mathcal{C}$ context points with is $\mathcal{O}(\mathcal{T}+\mathcal{C})$, which is more efficient than for GP \citep{{88fae17e4b14416d84edf0b23a013f2a},foong}.

Our initial adaptation of the CNP for the purposes of the LSST quasars light curve modeling was described in \citet {2022AN....34310103A,LINCC}.
Using the above concept we fully developed NP-module for LSST quasar light  curve modeling, which is upgraded to \texttt{pytorch} and refactorised into  6 subunits:
\begin{enumerate}
\item	model architecture;
\item	definition of  dataset class   and  collate function;
\item	 metrics (loss and  mean squared error-MSE);
\item  training  and calculation of   training and  validation metrics (loss and MSE);
\item saving model in predefined repository;
\item   upload of  trained  model  so that  prediction can be done anytime.
\end{enumerate}
The following features (from the above list) are new in contrast to the earlier version of the CNP module  \citep[see][]{2022AN....34310103A}: (2), the MSE  given in (3), as well as (5), and (6).


\section{Results and discussion}\label{resdisc}

In this section, we present and discuss   {results of our procedures for 'gaining the knowledge from large data' }comprising of  the training of CNP on strata (Section \ref{traincnp}), CNP modeling of variability of light curves in strata (Section \ref{modeledlc}) and modified structure function analysis of observed and modeled light curves  (Section \ref{variability}).
\subsection{Training of CNP}\label{traincnp}

{Following prescription for  splitting data set into training, testing and validating subsamples  \citep[see e.g.,][]{Tachibana_2020}, }
we divided {randomly}  strata of \texttt{u}-band light curves (seen in Figure \ref{fig4}), into a training dataset with 80\% of the total number of objects, a test dataset  with 10\% of the total number of objects, and a validation dataset with 10\% of the total number of objects. {The training, target and validation time instances, originally given as modified julian date (MJDs), are transformed to a $[-2,2]$ range alongside of corresponding   magnitudes  and measured errors }\footnote{{Data have been transformed using min-max scaler adapted to the range $\left[a,b\right]$
$$\text{scaled\_value} = \frac{(\text{original\_value} - \text{original\_min\_value}) \times (\text{b} - \text{a})}{{\text{original\_max\_value} - \text{original\_min\_value}}} + \text{a}$$
where $a=-2, b=2$ and $\text{original\_value}$ stands for the input data (time instances, magnitudes, magnitudes errors), $\text{original\_max\_value}$ is the maximum of the  $\text{original\_value}$, and $\text{original\_min\_value}$ is the minimum of the  $\text{original\_value}$.
This linear transformation (or more precisely affine) preserves original distribution of data, does not reduce the importance of outliers and preserves covariance structure of the data. We used the range of $\left[-2, 2\right]$
for  enabling direct comparisons with \citep{88fae17e4b14416d84edf0b23a013f2a} original testing data, ensuring consistency in our analysis.}
}. 
{The training, validating and targeting light curves are  given as tensors of size $128\times N$ where 128 is a batch size and $N$ is corresponding number of  epoch in the light curves \footnote{{The N is the maximum number of points in the light curves in the given batch; missing values are zeropadded for shorter light curves. We emphasize that our sample of light curves is well balanced as a result of SOM clustering \citep[see][for an counter example]{Sanchez-Saez_2021}, so that the number of points per light curve covers a fairly limited range [103,127] points, requiring negligible padding. } 
}. About one hundred times, we independently carried out the method of dividing the data and applying our algorithm to the training, validating, and testing data.}

\begin{figure}[H]
\includegraphics[width=0.45\textwidth]{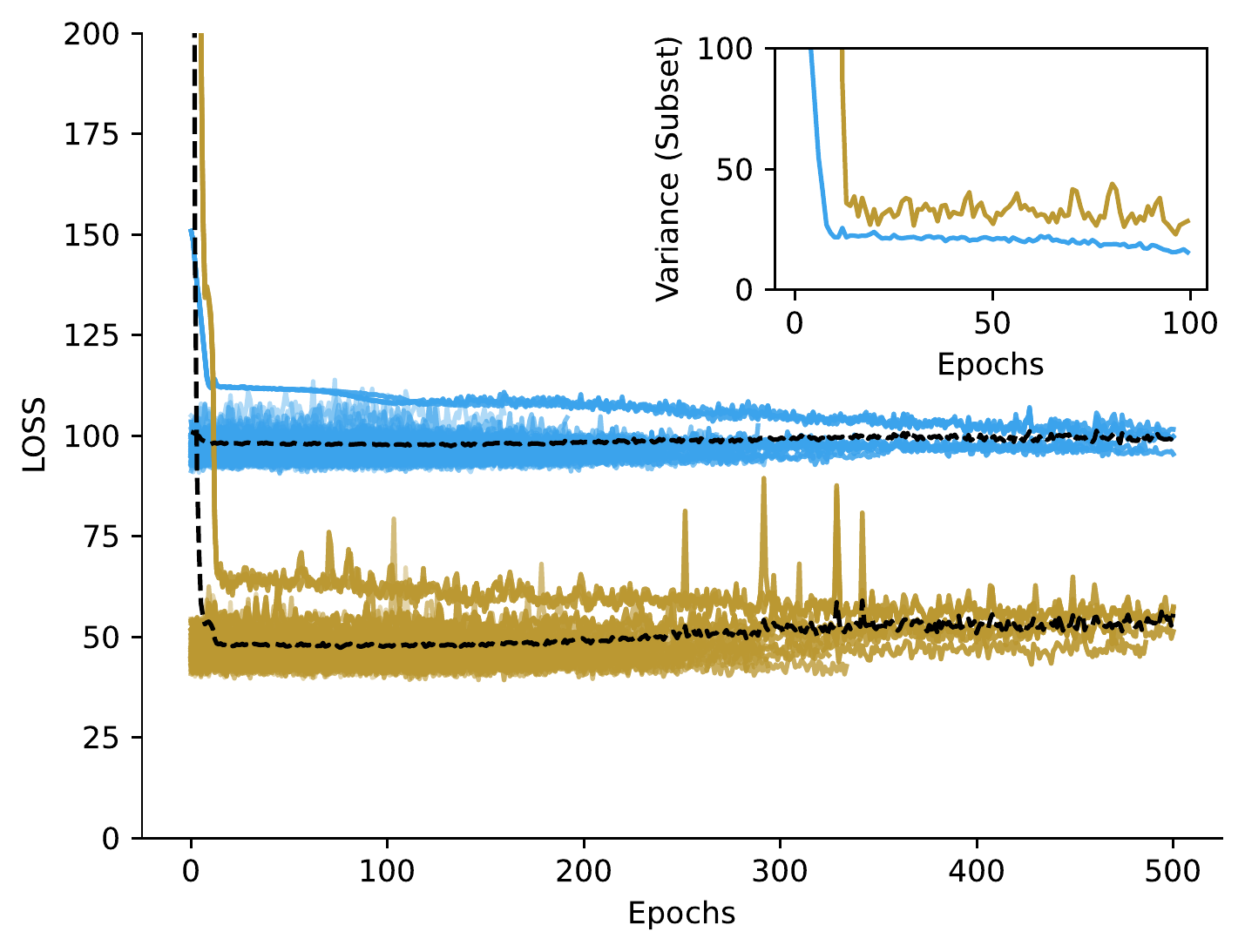}
\includegraphics[width=0.45\textwidth]{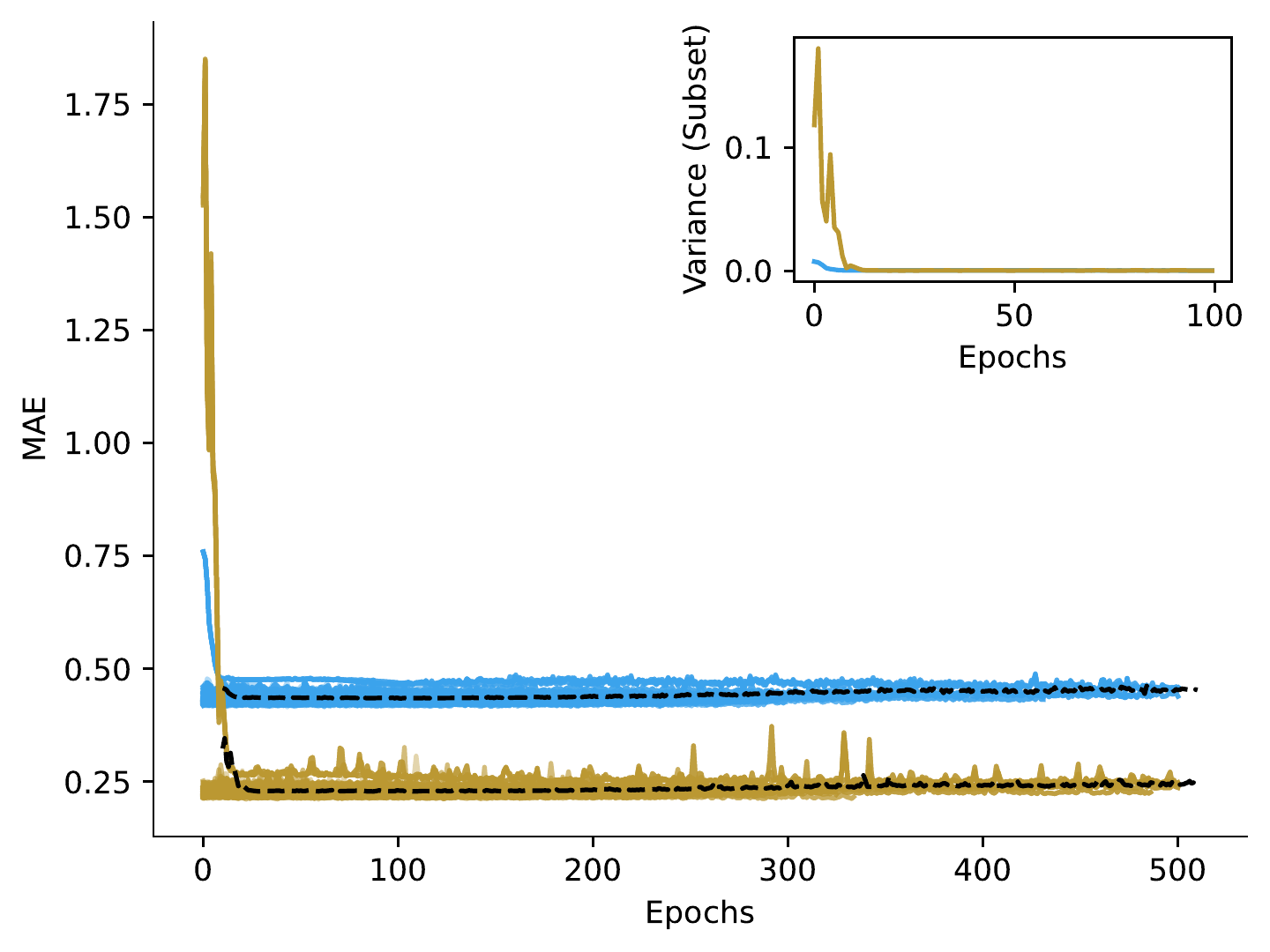}
\includegraphics[width=0.45\textwidth]{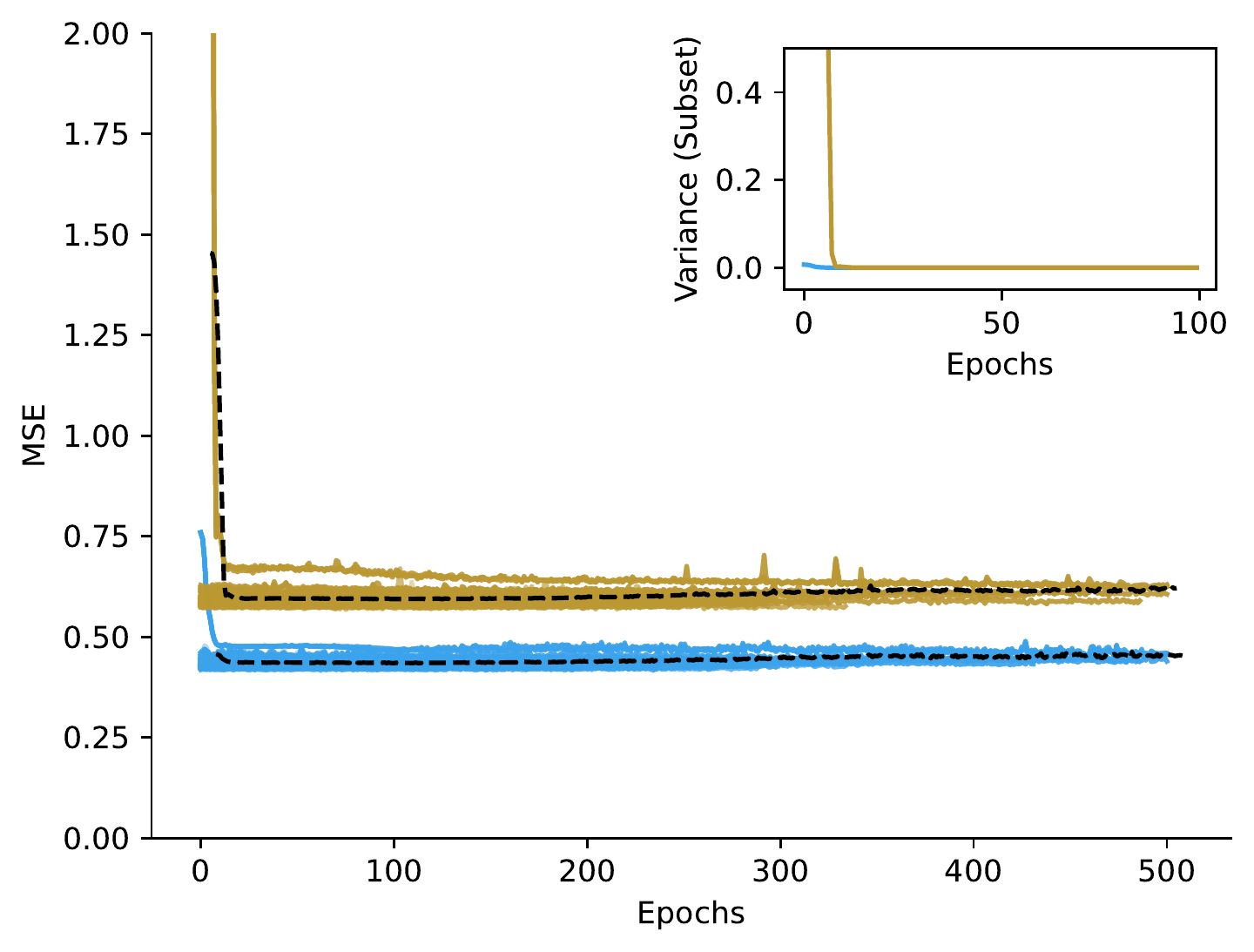}
\caption{Loss (the upper left panel), {mean absolute error (MAE, the upper right panel)} and MSE (the bottom right panel)  for the training  (orange) and for the validation (blue) data set during 100 runs of training. {The dashed black lines represent the mean values of metrics at each epoch over 100 runs.} Training loss is estimated in each iteration within each epoch, but validation loss is obtained at the end of each epoch. {Inset plots show corresponding variances for training (orange) and validation (blue) data sets ( a subset is shown for clarity).} {The metrics are expected to be high in the early epochs (since the network is initialized at random and the network's behavior differs from the desired one in the early epochs). The behavior of metrics at larger extent of epochs justifies inclusion of early stopping criterion in the CNP.} \label{fig5}}
\end{figure}

During the training process, the training set was augmented by the addition of extra curves that were generated from the original by adding and subtracting measured uncertainty from observed points.
{The method of adding noise to neural network inputs during training has been known for a long time. Many theoretical studies have been demonstrated that it allowed
increasing generalization capabilities of the network       \citep[e.g.,][]{Holm,Matsuoka}. \citet{Bishop} has  shown that use of
this method was equivalent to Tikhonov regularization. \citet{Wang} showed that
if this method was applied, the network also trained faster. Most often it is considered as one of the
methods to avoid ANN overtraining \citep[see][]{Zur}. Currently, this method is also used when training deep neural networks \citep{goodfellow2016deep}.
Noise can be introduced into a training neural network in four different phases: input data, model parameters, loss function, and sample labels \citep[see][]{math11020330, Smithing}. For injecting the noise it is necessary that  probability distribution from which noise is drawn corresponds to the real world situation of observed data \citep[see for application in astronimical light curves][]{2018NatAs...2..151N}. As photometric errors in observed light curves are mostly following Gaussian distribution \citep[see e.g.,][]{MacLeod_2012}, the added Gaussian noise ($N(0,\sigma_{i})$) to the input training light curves magnitudes corresponds to estimated measurement error($\sigma_i$) at each time step  \citep[see for application in astronimical light curves][]{2018NatAs...2..151N}.}
The {performance metric}  values for the training
dataset and the validation data set for 100 runs  are shown in Figure \ref{fig5}. {The lower training loss but higher training  MSE compared to the validation loss and MSE can be attributed to the sensitivity of MSE to  particularly the sharp peaks in the light curves. Switching to mean absolute error (MAE) led to a more typical behavior, with the training MAE lower than the validation MAE.}

As already mentioned, as a balance between hyperparameters found in literature and our experimentation, we used Adam optimization algorithm \citep[see][and references therein]{2022AN....34310103A} implemented in Python package \texttt{torch.optim} with a learning rate of
$0.0001$ \citep[see][]{2018NatAs...2..151N, Tachibana_2020} and a batch size of 32.
{Adam is probably the most frequently used optimization algorithm for
training deep learning models thanks to its adaptive step size, which
in practice most often leads to decreased oscillations of the
gradients and faster convergence. It combines the best aspects  of the \texttt{AdaGrad} and \texttt{RMSProp} algorithms to provide an optimization  that can handle sparse gradients on noisy problems as we encounter in quasar light curves \citep[see][]{2018NatAs...2..151N,Tachibana_2020}.}
{Once again we emphasize that general model has been introduced in Paper I, and here we provide its upgraded version, with application on quasar light curves strata \citep[see][]{10.1093/mnras/stv1797}, such  granular application can potentially help understanding of physical properties of categories of objects  \citep[see][]{10.1093/mnras/stv1797}, as we also demonstrated here. Also CNP is well generalizable on various data sets (strata) as it inherits GP ability to  determine predictive  distribution of data. }

The left panel in
Figure \ref{fig5} shows that both the validation and train loss decrease rapidly until epoch 2000, after which both losses stabilize. 
{The loss and MSE are expected to be high in the early epochs (since the network is initialized at random and the network's behavior differs from the desired one in the early epochs) and thus inconsequential. The larger extent of epochs are depicted for illustration purposes that justifies inclusion of early stopping criterion so that CNP is trained at epochs $< 2000$.}

{Overfitting might be indicated by a decreasing training loss and an increasing or plateauing validation loss.} However, our loss curves do not show this behavior. To prevent underfitting, {we employed data augmentation with noise,  and early stopping}.

\subsection{CNP modeling of quasar variability} \label{modeledlc}

A detailed catalogue of CNP models of light curves in our  sample of 283 low variable and high redshift  quasars  ($F_{var}\sim 0.03$, $1\leq z \leq 3$) is given in the Appendix (see Figures \ref{fig111}-\ref{fig142}). Each plot shows the modeling   performance of the CNP.
The most notable quality is that the CNP catches the overall trends and major flare like events. 

Both the autoencoder  neural network constructed by \citet[compare to Figure 7 in][]{Tachibana_2020}, and CNP  model the
quasar temporal behavior purely based on the characteristics of the data without any prior assumptions. However, the main difference is that CNP also inherits the flexibility of stochastic process modeling such as GP. To assess  the modeling accuracy of the CNP each plot presents MSE and loss values along the $95\%$ confidence interval of the model.
As we included observational errors in the CNP training process, the confidence bands for the regions of light curves dominated by points with larger errors are wider. Furthermore, we discovered that practically all light curves contain flare-like occurrences and even outliers, which also raise the obtained confidence band. 
{We note that MSE is of comparable value to the MSE found in other studies \citep[see e.g.,][]{2018NatAs...2..151N}. MSE is calculated on original (nontransformed data) and its value of $\sim 0.5$ mag corresponds to  $\sim5\%$. Given that MSE represents variance, it is also more resistant to outliers (flares).  Because loss is measured as the log of probability density, it is particularly susceptible to large gaps and outliers in our light curves. We emphasize that  deep learning studies of astronomical time series report MSE \citep[see][]{2018NatAs...2..151N, Tachibana_2020} frequently, so we will provide both MSE and loss. }

\begin{figure}
\includegraphics[width=\textwidth]{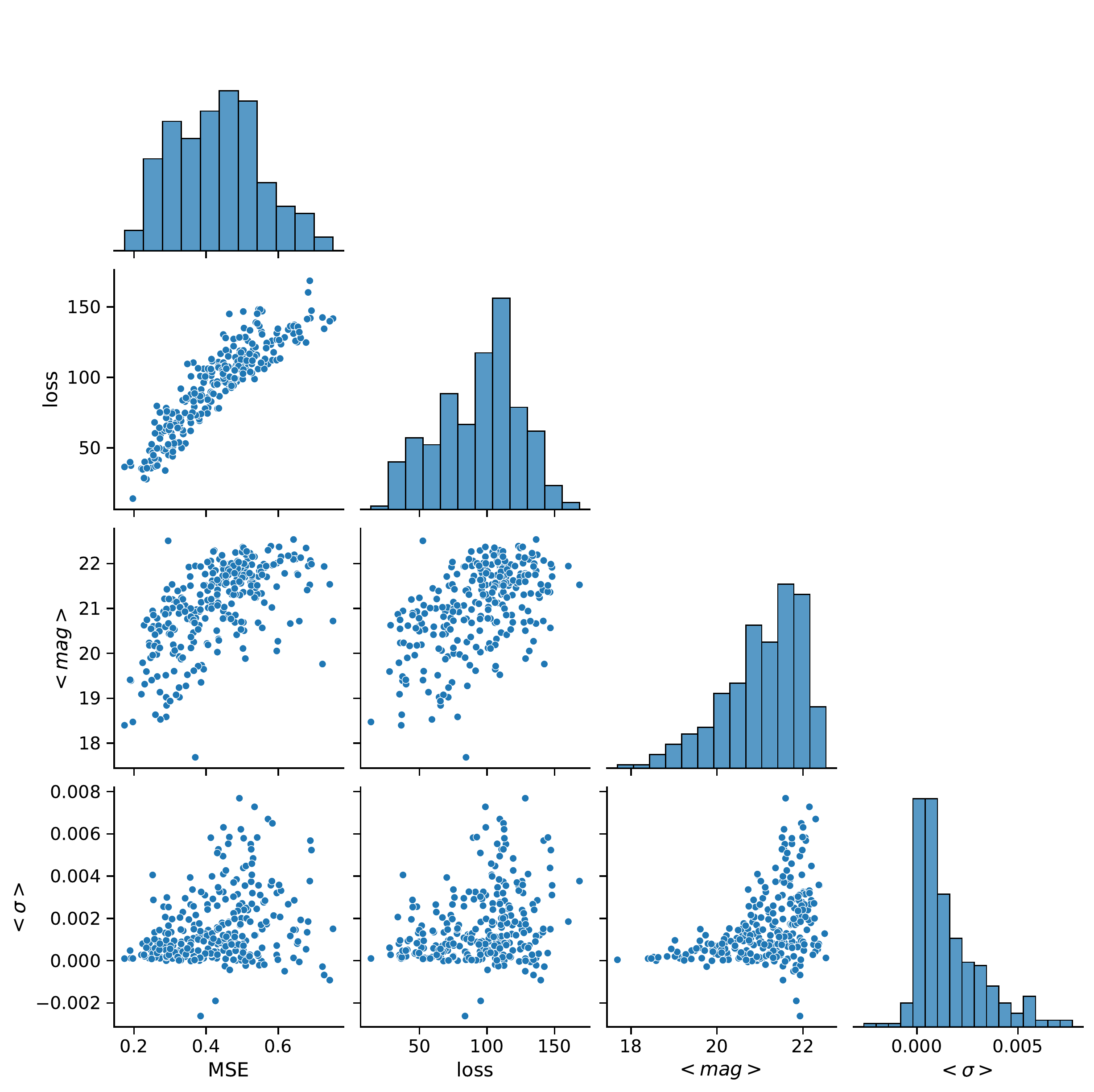}
\caption{Corner plot of the probability distribution of modeling parameters (MSE, loss) for corresponding observed light curves parameters (mean magnitude $<mag>$, mean photometric error $<\sigma>$). Histograms of the marginalized probability for each parameter are given on the diagonal. 
  \label{corner2}}
\end{figure}

The corner plot of mean square error, loss of each model fitting and corresponding mean magnitude and mean photometric error of observed light curves are given in Figure \ref{corner2}. 
According to the individual plots, a higher MSE ($>0.5$) is coupled with mean magnitudes in the range [20,22] and mean photometric errors greater than 0.002. The tail of the marginal mean error distribution likewise contains these mean magnitudes.

In our previous analysis  CNP  has been applied to the light curves with significant changes of gradients, having inhomogeneous cadences and infrequent flare like features \citep{2022AN....34310103A}. However, as demonstrated in \citep{10.1093/mnras/stv1797} and  here, {deeper analysis of flare-like patterns is needed}.  We are testing additional CNP alternatives, but we are treading carefully to avoid unintended introduction of relations that do not exist in data \citep[see][]{Smith_2018}.

\subsection{Modified structure function analysis of 
observed and modeled light curves}\label{variability}

The results from the neural network clustering suggest that
there is no significant variability in the quasar light
curves of the chosen sample. In order to test and further investigate this finding, we used the structure
functions \citep[SF, see e.g.][and references therein]{1998ApJ...504..671K, 10.1093/mnras/stv1797,2022A&A...664A.117D}.
The SF could be defined as \citep{10.1046/j.1365-8711.2002.04939.x}:
\begin{equation}
S(\tau)=\frac{1}{N(\tau)}\sum_{i<j} (m(t_{j})-m(t_{i}))^{2}
\end{equation}
\noindent where  $m(t_{i})$ is the magnitude measured at the epoch $t_i$, and summation runs
across the $N(\tau)$ epochs for which is satisfied  $t_{j}-t_{i}=\tau$.
 In addition to $S$ defined
above, we will also
 use the two modified structure functions $S_{+}$ and $S_{-}$ introduced by \citet{1998ApJ...504..671K}. 
  For $S_+$, the integration only includes pairings of magnitudes for which the flux increases $m(t_{j})-m(t_{i})>0$, whereas for $S_{-}$, the integration only includes pairs of magnitudes for which the flux becomes dimmer $m(t_{j})-m(t_{i})<0$. 
 
Both modified structure functions  measure the underlying
asymmetry of the emission process as manifested in the light curves \citep{1998ApJ...504..671K}.
A comparison of modified structure functions  could, in fact, disclose distinct mechanisms that are responsible for the variability of light curves, as was indicated in
 \citet{10.1046/j.1365-8711.2002.04939.x}.
 To be more specific, the disc instability model will show itself in the form of an asymmetry in the light curves, such that the relation $S_{-}>S_{+}$ will be observed at shorter time scales $\tau$. On the other hand, transient events like supernovae will show up as  the case that $S_{+}>S_{-}$ as the time scale gets shorter. In the scenario of microlensing, which is a fundamentally symmetrical process in the setting of quasar variability, the two functions will be indistinguishable, denoted by the equation, i.e.,  $S_{-}=S_{+}$ \citep{10.1046/j.1365-8711.2002.04939.x}. Even though it looks paradoxical for quasars,  \citet{10.1046/j.1365-8711.2002.04939.x} explained this with a model  where  the fluctuations in the light curves originating from the accretion disc become smaller with increasing luminosity, while the effects of microlensing become more pronounced at higher redshift, which for quasars typically means higher luminosity. 
 
\begin{figure}[H]
\includegraphics[width=10.5 cm]{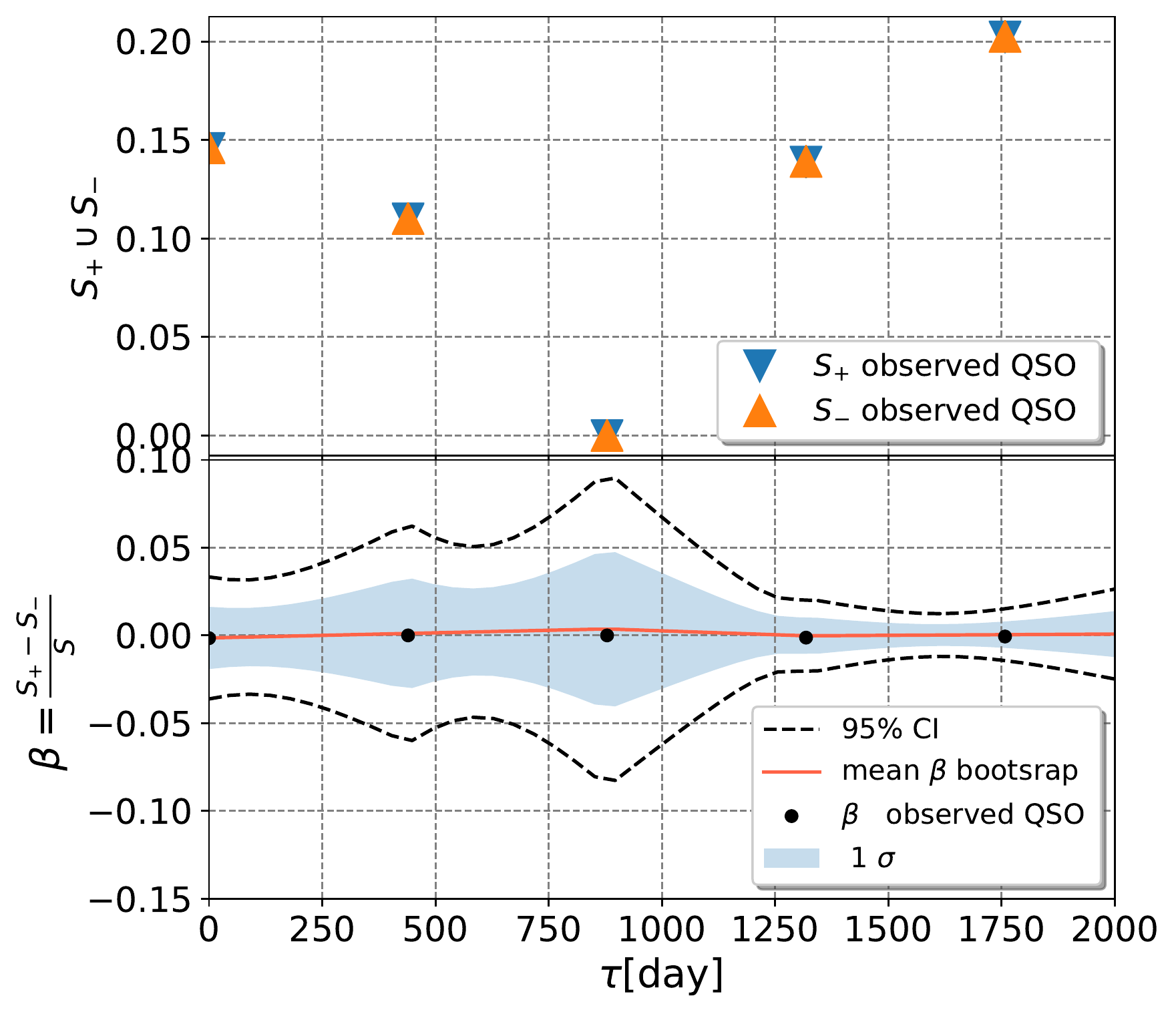}
\caption{ Time symmetry in the light curves of selected quasar sample. \textit{Top}: $S_{-}$ and $S_{+}$ for quasar sample are indicated by upward and downward pointing triangles, respectively. \textit{Bottom}: The normalized relative difference $\beta=\frac{S_{+}-S_{-}}{S}$ for observed quasar sample (black dots), mean $\beta$ (red line), $1\sigma$(blue band) and $95 \%$ (dashed lines)confidence intervals as inferred from bootstrapping quasar sample \label{figbeta}}
\end{figure}

Figure  \ref{figbeta} displays $S_{+}$ and $S_{-}$, and  their relative difference normalized by  standard  SF ($S$, bottom panel) for observed quasar sample. 
 Because of the regularity of large seasonal gaps in the observations, we were able to partition the data into five distinct time bins.
 In the top panel, both modified functions $S_{+}$ and $S_{-}$ are overlapping.  The normalized relative difference between them is also consistent with time symmetry.

The striking feature of both modified structure functions is their zero value at the time lag of around 800 days.
Going back to the overview plot of observed light curves in
Figure \ref{fig3}, we could see that this time lag comprises of combinations of vertical columns of data points after $MJD=52000$ such as the first and third vertical, the second and  forth, the third and the sixth, etc., which are more similar in variability than other combinations which could be made. Also we note that in the range $52000 \leq MJD\leq 53500$ is evidently smaller number of data points than for $MJD> 52500$ which affects the bootstraping method to have larger uncertainty.

\begin{figure}[H]
\includegraphics[width=10.5 cm]{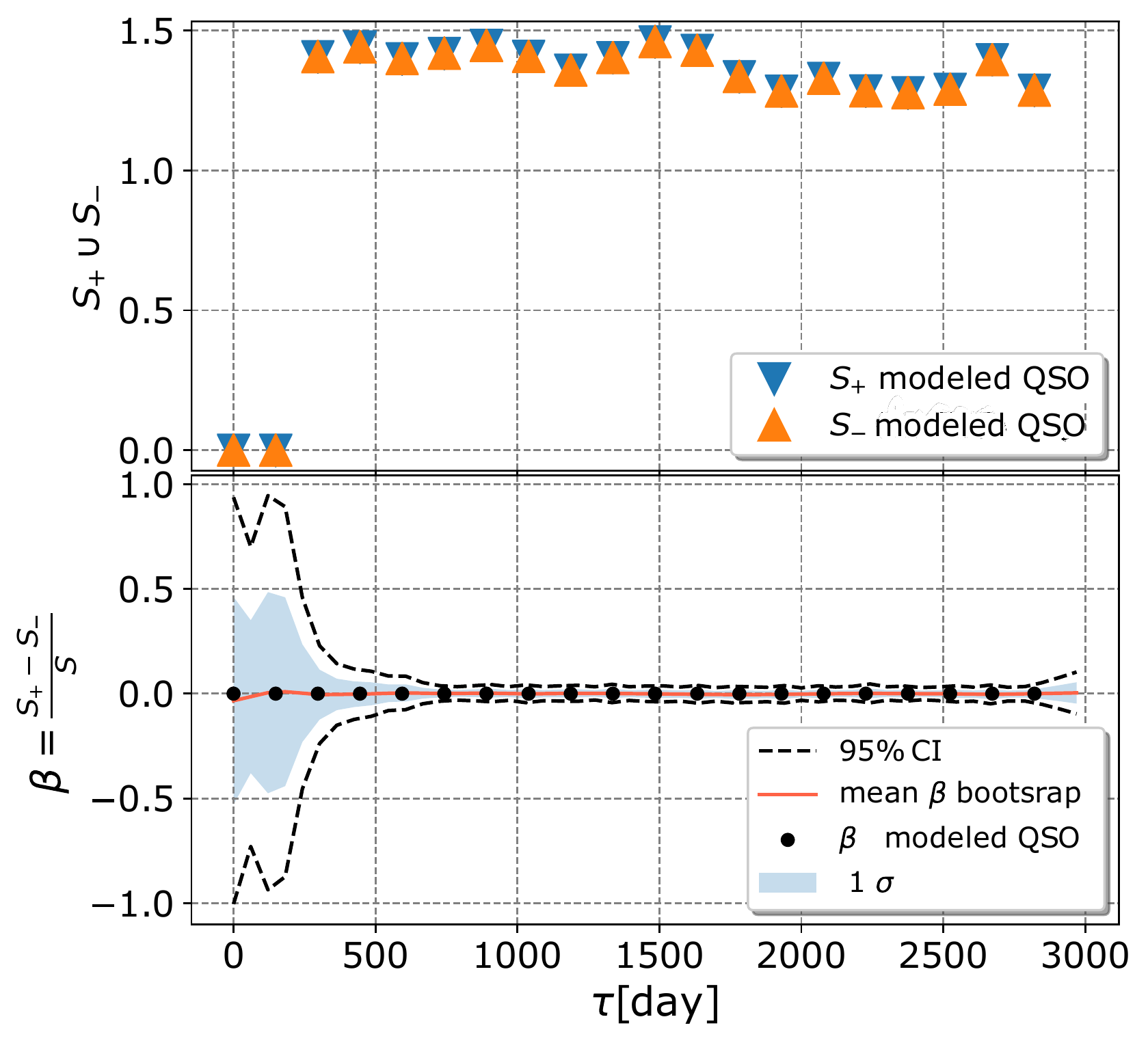}
\caption{ The same as Figure \ref{figbeta} but for CNP modeled light curves.\label{figbeta2}}
\end{figure}

We can further compare the CNP modeled light curves of the selected quasar sample by testing for time
asymmetries (see Figure \ref{figbeta2}). It is important to note that modeled light curves made it possible to construct structure functions using a finer bin grid that contained 20 bins.
As we have seen, $S_{+}$ and $S_{-}$ of modeled light curves are practically identical, which lends further support to the idea that microlensing is at play here. We can see that the CNP modeling does not modify the fundamental variability characteristics of the  light curves that have been observed.

For the structure functions of modeled light curves, the variability
in the shortest timescale ($<200$ days) appears very different  from that at a longer scale. 
This is an
inevitable consequence of the fact that CNP is more uncertain about time lags corresponding to cadence gaps at epochs $<52700$ (see Figure \ref{fig3} and Figures \ref{fig111}-\ref{fig142}). We could see also that the structure functions confidence interval is rather large for these time lags.

We bring to  attention the fact that our quasar sample was selected by our neural network algorithm in the absence of any other criteria (e.g., quasar parameter relationships),
and that our clustering method based on the SOM could segregate objects with specific variability characteristics.

 It is worth mentioning that microlensing can become obvious in multiply lensed quasar. This happens when some variations can only be observed in one image and appear to predominate over fluctuations that can be seen in all images over lengthy time scales. The light curves that are produced by microlensing can be simulated in a variety of different ways. Especially relevant to our collection of light curves are some interesting simulations by \citet{1993MNRAS.261..647L}.
In microlensing simulations, the source size has a significant impact on the look of the light curves, which become smoother and more rounded as the source size increases \citep{1993MNRAS.261..647L,10.1046/j.1365-8711.2002.04939.x}. Light curves of our quasar sample given in Figures \ref{fig111}-\ref{fig142} share striking similarity with   Figures  2(a) and 3(b) from \citet{1993MNRAS.261..647L} regarding dominating non-smoothed flare like patterns.   

\unskip

According to the findings of \citet{Tachibana_2020}, modified structure functions and $\beta (\tau)$ for their sample of observed quasars exhibit asymmetry. They also noted that their simulated light curves as damped random walk (DRW) processes do not exhibit any substantial deviation from symmetric processes when compared to the bounds of confidence intervals. On the other hand, when we move deeper into the confidence interval, the values of the modified structure functions and $\beta (\tau)$ begin to fluctuate. Additionally, their simulated DRW light curves do not reflect the many flare-like events along the time baseline of light curves that are observed in our sample.

Based on the face value of these results, we  find that the photometric variability of this sample of quasars  could be explained with the microlensing model \citep{10.1046/j.1365-8711.2002.04939.x}. A justification that might be called   is a model in which the fluctuations in the emission coming from the accretion disc becomes smaller as the luminosity of the object increases, but the effects of microlensing become more pronounced at higher redshifts, which for quasars typically coincide with higher luminosities \citep[see][]{10.1046/j.1365-8711.2002.04939.x}.
{We also note that low variability corresponds to high bolometric luminosity.
The bolometric quasar luminosity  is closely tied to the accretion rate of the SMBH. A bolometric quasar luminosity function (QLF)   was constructed by \citet{Hopkins_2007} and updated  for the bolometric QLF at $z = 0-7$ by \citet{10.1093/mnras/staa1381}.}

\begin{figure}[H]
\includegraphics[width=13.5 cm]{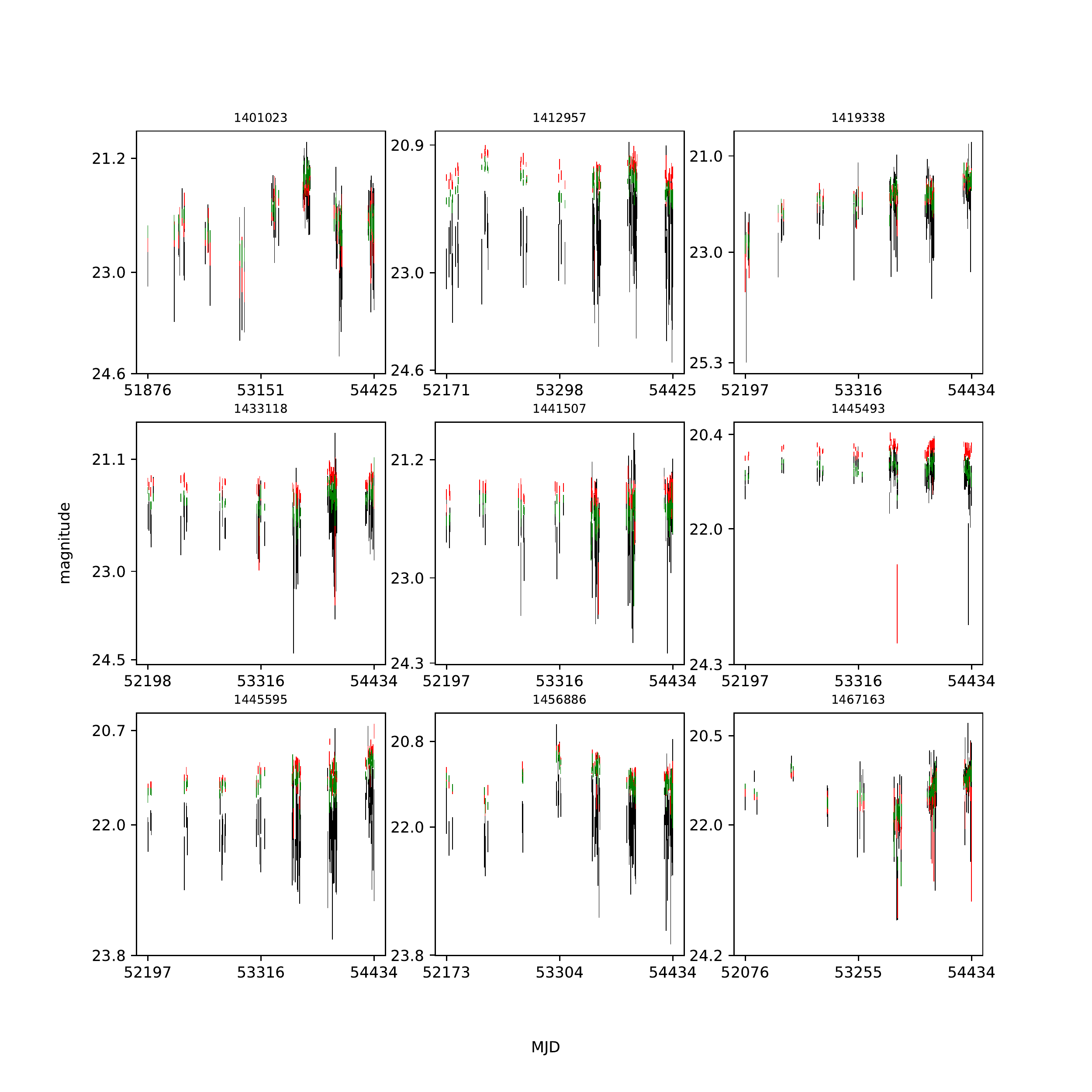}
\caption{Comparison  of presence of flare like patterns in \texttt{u} (black), \texttt{g}(green) and \texttt{r} (red) bands. Object identification number  is given in the corresponding plot title. \label{figbands}}
\end{figure}

{Importantly, the flare like patterns remained present in \texttt{g} and \texttt{r} curves (see an example in Figure \ref{figbands}).  This is a hint that the flare like events  are achromatic, since the gravitational bending of light is independent of the frequency of
the radiation\citep{2020MNRAS.499L..87D}.}

  Nonetheless, it may be too soon to conclude that microlensing is only plausible explanation. This is due to the fact that photometric monitoring for a period of four years is not yet sufficient to construct meaningful statistics based on the cadences that are available. {Extreme optical extragalactic transients caused by explosion of
supernovae, tidal disruption events around dormant
black holes, rare blazars, stellar mass black holes mergers in quasars disks, and intrinsic accretion outbursts in
quasars may also be possible mechanisms behind observed flares \citep[see][]{2023ApJ...942...99G}.}

The discovery of such a distinct  cluster of quasars in the LSST AGN data challenge database that makes  $\sim 0.8\%$  could have interesting implications. 

Assuming that the LSST will harvest  $n \sim \mathcal{O}(2\times10^{7}-10^{8})$ quasars  \citep[see][and reference therein]{10.1093/mnras/stab1856}, a very simple extrapolation  indicates that future LSST data releases may contain a population of microlensed quasars that is not negligible  ($N\sim \mathcal{O}(10^{5}-10^{6})$).  Interestingly, the microlensing duration should be shorter in the X-ray  (several months) than in the UV/optical emission range \citep[several years, see][]{2008MNRAS.386..397J}. Moreover, cosmologicaly distributed lens objects may contribute significantly to the X-ray variability of high-redshifted QSOs \citep[$z>2$,][]{2004A&A...420..881Z}. 
 Also, we could estimate $N$  assuming  the microlensing rate $\Gamma$, i.e. the number of events expected to be detected per quasar and per year. $\Gamma$ was  evaluated  for particular quasar J1249+3449 as $\Gamma\sim 2\times 10^{-4} (m/0.1M_{\odot})^{-0.5}$ events per quasar per year, where $m$ is the lens mass \citep[see][]{2020MNRAS.499L..87D}.
 The value of parameter $m$ could be in the range of $(0.01-0.5)M_{\odot}$, if the host galaxy of quasar has stars with velocities $200-400 \mathrm{km\, s}^{−1}$, and a source lens distance in the range $1-10$ kpc.
 Even if we caution that
$\Gamma$ was derived for a particular AGN \citep{2020MNRAS.499L..87D}, we could use it as a proxy to estimate  within a time interval $\Delta t$, the total number $N$ of expected microlensing events as simply $N\sim \epsilon \Gamma_{total}\Delta t$ \citep[see][]{10.1111/j.1365-2966.2010.17511.x}. For simplicity we assume that  for maximal survey efficiency $\epsilon\sim 1 $, if $n$ sources are monitored over a time interval $\Delta t$ the total rate is $\Gamma_{total}\sim n \times \Gamma$. For $\Delta t$ we could expect  to be within the range of weeks up to decades \citep{wambsganss_2001}. For example, \citet{2007A&A...462..581H} found that   if a lower limit of the time scale of   $\sim 24$  years was supposed to be caused by microlensing,  it would correspond to a minimum mass for microlensing bodies of  $0.4 M_{\odot}$ \citep{2007A&A...462..581H}. 
Taking into account that  $S_{+}$ and $S_{-}$ are more certain in our study for time scales $>800$days, and since the LSST operation time is $\sim 10$ yr, we estimated that the number of microlensing quasars in the LSST data releases may be $N\sim \mathcal{O}(10^{4}-10^{5})$. Because of these assumptions, our estimates are most likely an upper bound on the number of microlensing quasars produced by LSST.

Nonetheless, in order to extract the physical properties of variability, the microlensing quasar population should be handled independently during the analysing process of the LSST quasar data, particularly the detection of binary candidates. Furthermore, it is probable that this population will have an influence on how quasars are classified in the LSST data pipelines.
 
\section{Conclusion}\label{conclusion}

{
In this follow-up study, we presented an improved version of a conditional neural process (CNP, Paper I) that was embedded in a multistep approach for learning (stratification of light curves via neural network, deep learning  of each strata, statistical analysis of observed and modeled light curves in each strata) from large amounts quasars's contained in the LSST Active Galactic Nuclei Scientific Collaboration data challenge database. The main observations are:
\begin{itemize}
    \item Individual light curves of 1006 quasars having more than 100 epochs in  LSST Active Galactic Nuclei Scientific Collaboration data challenge database  exhibit a variety of behavior, which can be generally stratified via neural network into 36 clusters.  
    \item A case study of one of  stratified sets of \texttt{u}-band light curves for 283 quasars with very low variability $F_{var}\sim0.03$ is presented here. CNP model has an average mean square error of  $\sim 5\% $(0.5 mag) on this strata.
 Interestingly, all of the light curves in this strata show features resembling the flares. An initial modified structure-function analysis  suggests that these features may be linked to microlensing events that occur over longer time scales of five to ten years.
\item As many of light curves in LSST AGN data challenge data base could be  modeled with  CNP- still there are enough objects having interesting features in the light curves (as our case study suggests) to urge a more extensive investigation.
\end{itemize}}

With the help of this scientific case, we were also able to demonstrate the importance of CNP (along with other deep learning methods) for data-driven modeling in contexts where considerable samples of objects may have variability patterns that differ from the DRW. In the future, investigations of this nature should receive a greater amount of attention.

\vspace{6pt} 



\authorcontributions{Conceptualization, A.B.K.,D.I. and L. {\v C}. P. and M.N.; LSST SER-SAG team members,  L. {\v C}. P., D.I.,A.B.K.,M.N.;
supervising and conceptualization of machine learning methodology, M.N., software design, M. N., A.B.K., N.A-M., I. {\v C}-H. and M.P.; supervising of mathematical analysis of  light curves,  M.K.;  writing—original draft preparation, A.B.K, D.I., L.{\v C}.P, M.P , M.N., N.A-M., I. \v C-H., M.K., and  Dj.V.S, writing—review and editing, A.B.K, D.I., L.{\v C}.P, M.P , M.N., N.A-M., I. \v C-H., M.K., and  Dj.V.S. All authors have read and agreed to the published version of the manuscript.}

\funding{A.B.K., D.I. and L.{\v C}.P. acknowledge funding provided by University of Belgrade-Faculty of Mathematics  (the contract {451-03-47/2023-01/200104}), through the grants by the Ministry of Science, and Technological Development and Innovation   of the Republic of Serbia.
A.B.K. and L.{\v C}.P. thank the support by  Chinese Academy of Sciences President's International Fellowship Initiative (PIFI) for visiting scientist. 
{L.{\v C}.P.} and Dj. V.S. acknowledges funding provided by Astronomical Observatory (the contract $451-03-68/2022-14/200002$), through the grants by the Ministry of Education, Science, and Technological Development of the Republic of Serbia.}

\institutionalreview{Not applicable.}

\informedconsent{Not applicable.}

\dataavailability{The data are available from the corresponding author upon reasonable request and with the permission of the LSST AGN Scientific Collaboration.} 

\acknowledgments{We sincerely thank Gordon T. Richards, and  Weixiang Yu for their essential efforts in the construction of LSST AGN data challenge  within the Rubin-LSST Science Collaborations. This work was conducted as a joint action of  the Rubin-LSST Active Galactic Nuclei (AGN) and  Transients and Variable Stars  (TVS) Science Collaborations.  The authors  express their gratitude to the Vera C. Rubin LSST AGN and TVS Science Collaborations for fostering cooperation and the interchange of ideas and knowledge during their numerous meetings. }

\conflictsofinterest{The authors declare no conflict of interest. The funders had no role in the design of the study; in the collection, analyses, or interpretation of data; in the writing of the manuscript; or in the decision to publish the~results.} 



\abbreviations{Abbreviations}{
The following abbreviations are used in this manuscript:\\

\noindent 
\begin{tabular}{@{}ll}
AGN &active galactic nuclei \\
LSST & Legacy Survey of Space and Time\\
\texttt{LSST\_AGN\_{DC}} & LSST  AGN data challenge database\\
SMBH & Super massive black hole\\
\end{tabular}
}

\appendixtitles{no} 
\appendixstart
\appendix
\section[\appendixname~\thesection]{Catalogue of CNP models of \texttt{u}-band light curves}
\begin{figure}[H]
\begin{adjustwidth}{-\extralength}{0cm}
\centering
\includegraphics[page=1,width=1.3\textwidth]{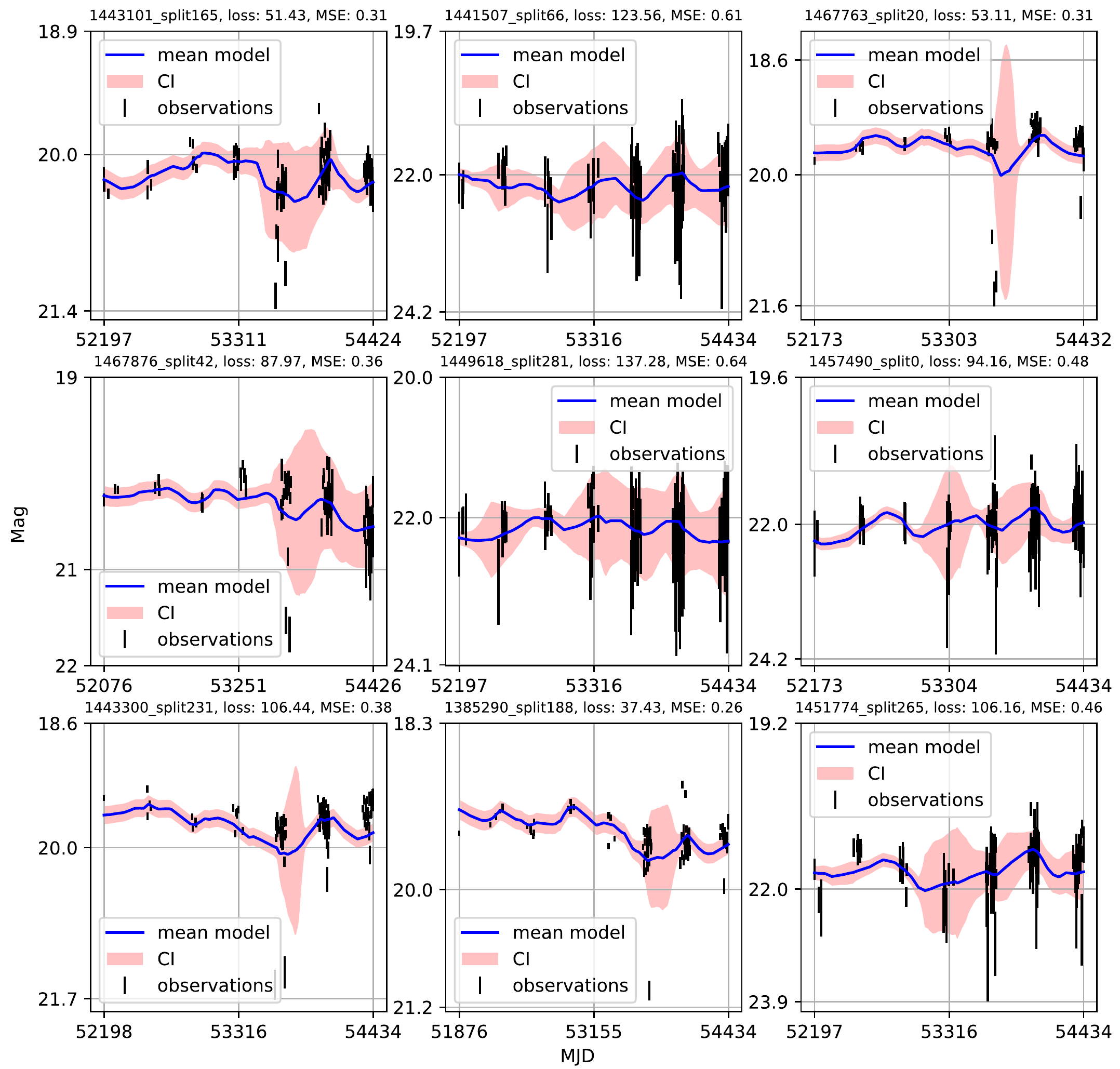}
\end{adjustwidth}
\caption{CNP modeling of light curves. The subtitle of each plot shows the object ID from the LSST AGN DC database with a tag indicating the iteration when it is selected for training, testing, or validation during the training process, the logarithmic value of probability loss, and the MSE value. Black error bars indicate observations with measurement uncertainty. Solid blue lines are model fits to the data. The red band represents the $1\, \sigma$ confidence interval. \label{fig111}}
\end{figure}

\begin{figure}[H]
\begin{adjustwidth}{-\extralength}{0cm}
\centering
\includegraphics[page=2,width=1.3\textwidth]{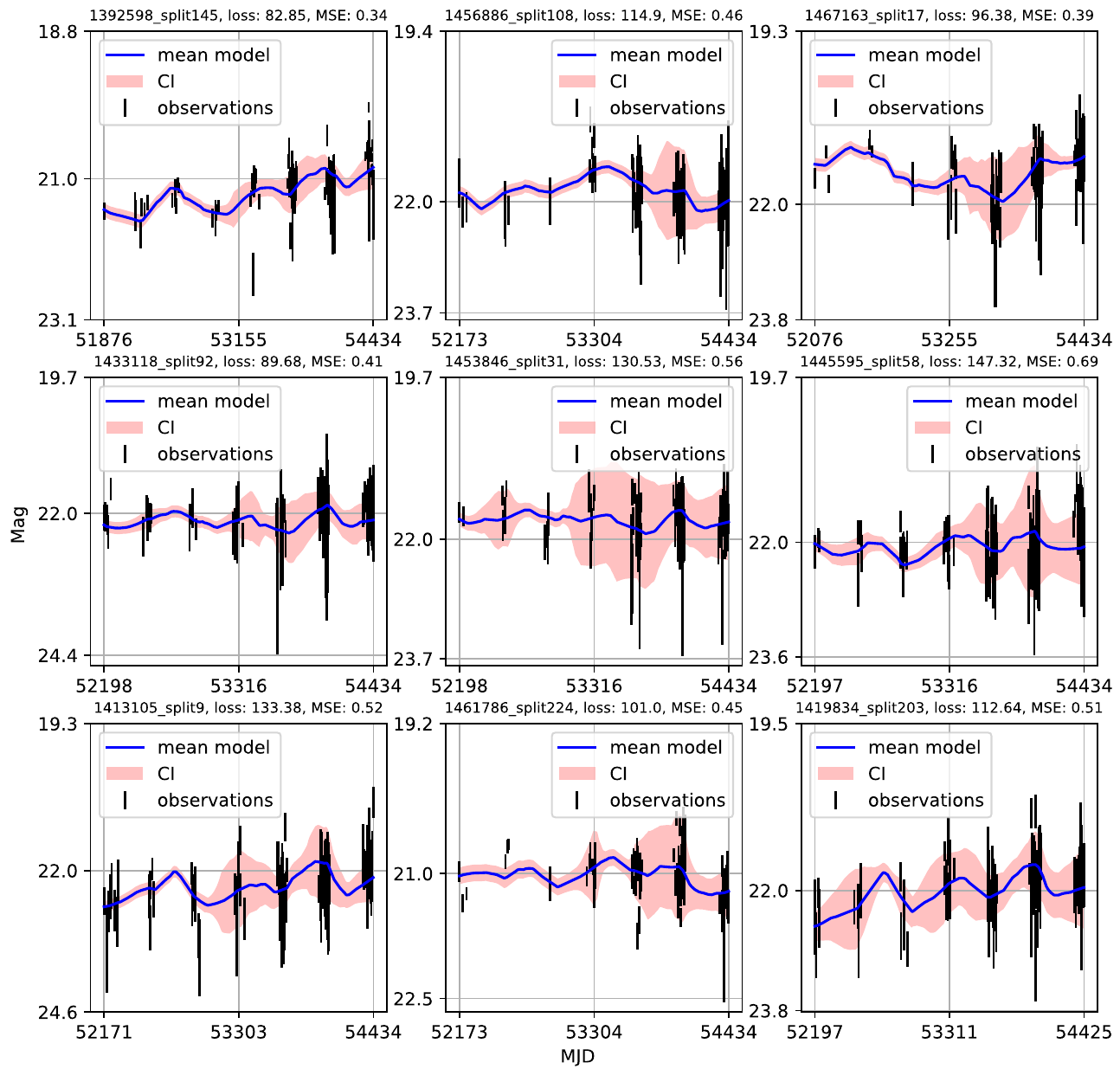}
\end{adjustwidth}
\caption{The same as Figure \ref{fig111}.\label{fig112}}
\end{figure}

\begin{figure}[H]
\begin{adjustwidth}{-\extralength}{0cm}
\centering
\includegraphics[page=3,width=1.3\textwidth]{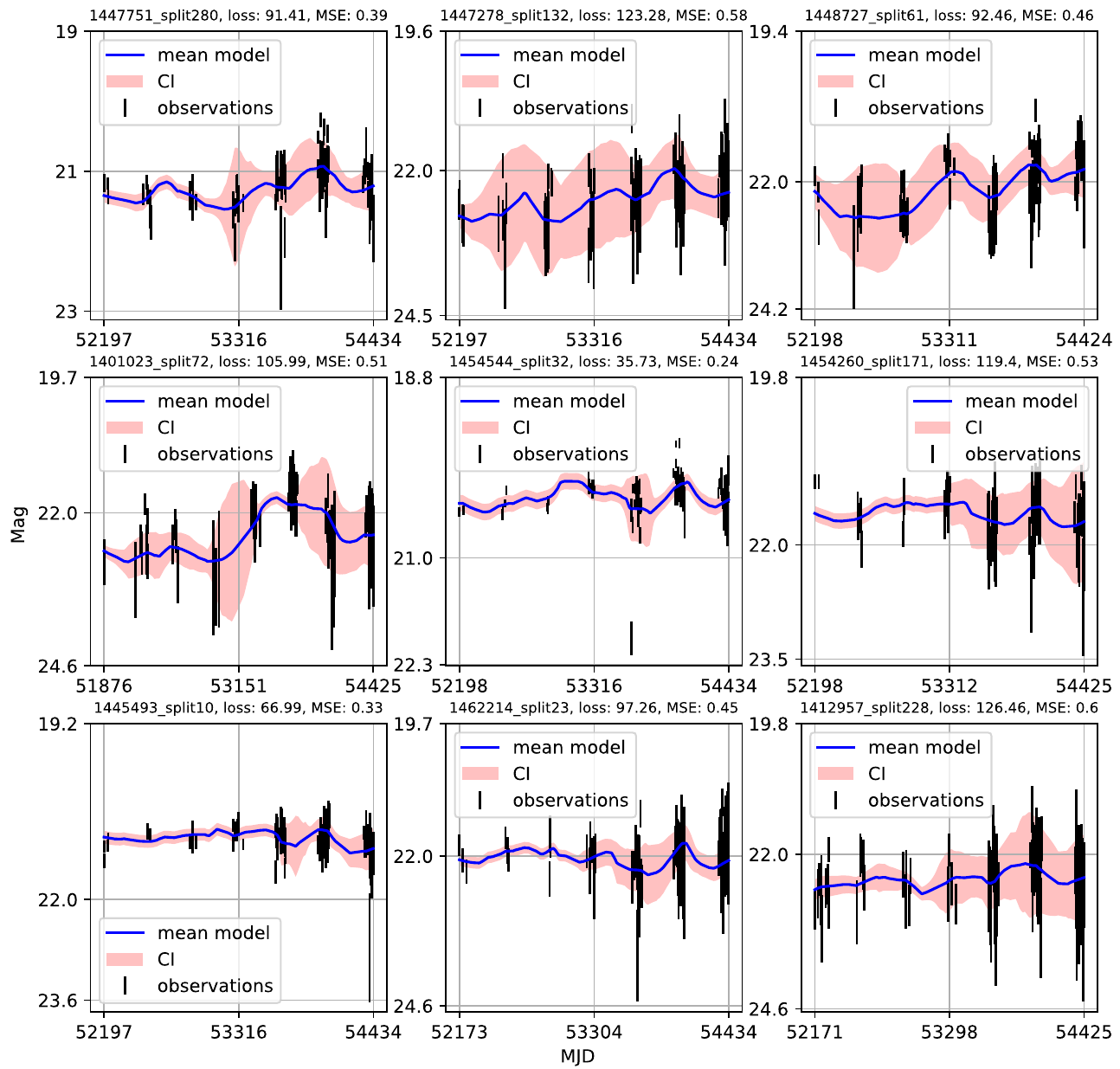}
\end{adjustwidth}
\caption{The same as Figure \ref{fig111}.\label{fig113}}
\end{figure}

\begin{figure}[H]
\begin{adjustwidth}{-\extralength}{0cm}
\centering
\includegraphics[page=4,width=1.3\textwidth]{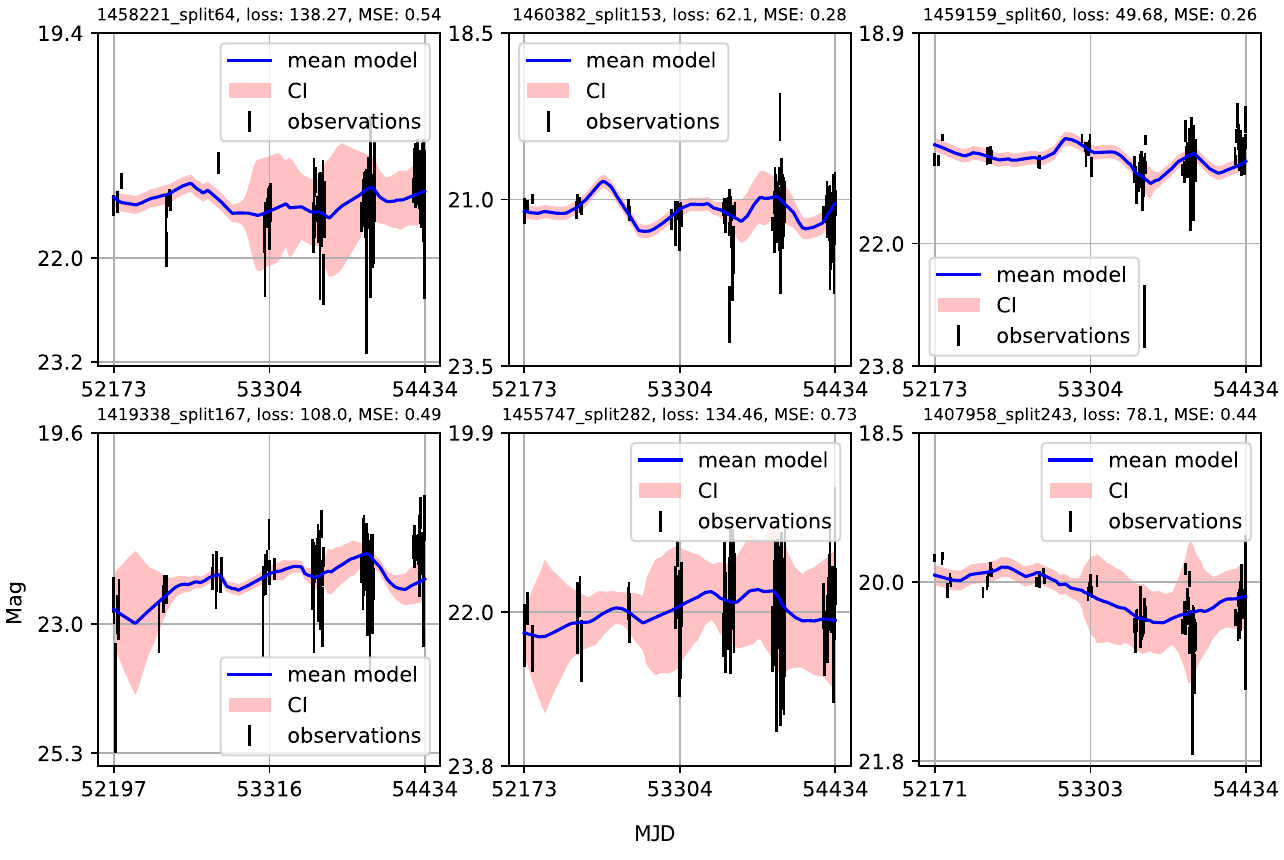}
\end{adjustwidth}
\caption{The same as Figure \ref{fig111}.\label{fig114}}
\end{figure}

\begin{figure}[H]
\begin{adjustwidth}{-\extralength}{0cm}
\centering
\includegraphics[page=1,width=1.3\textwidth]{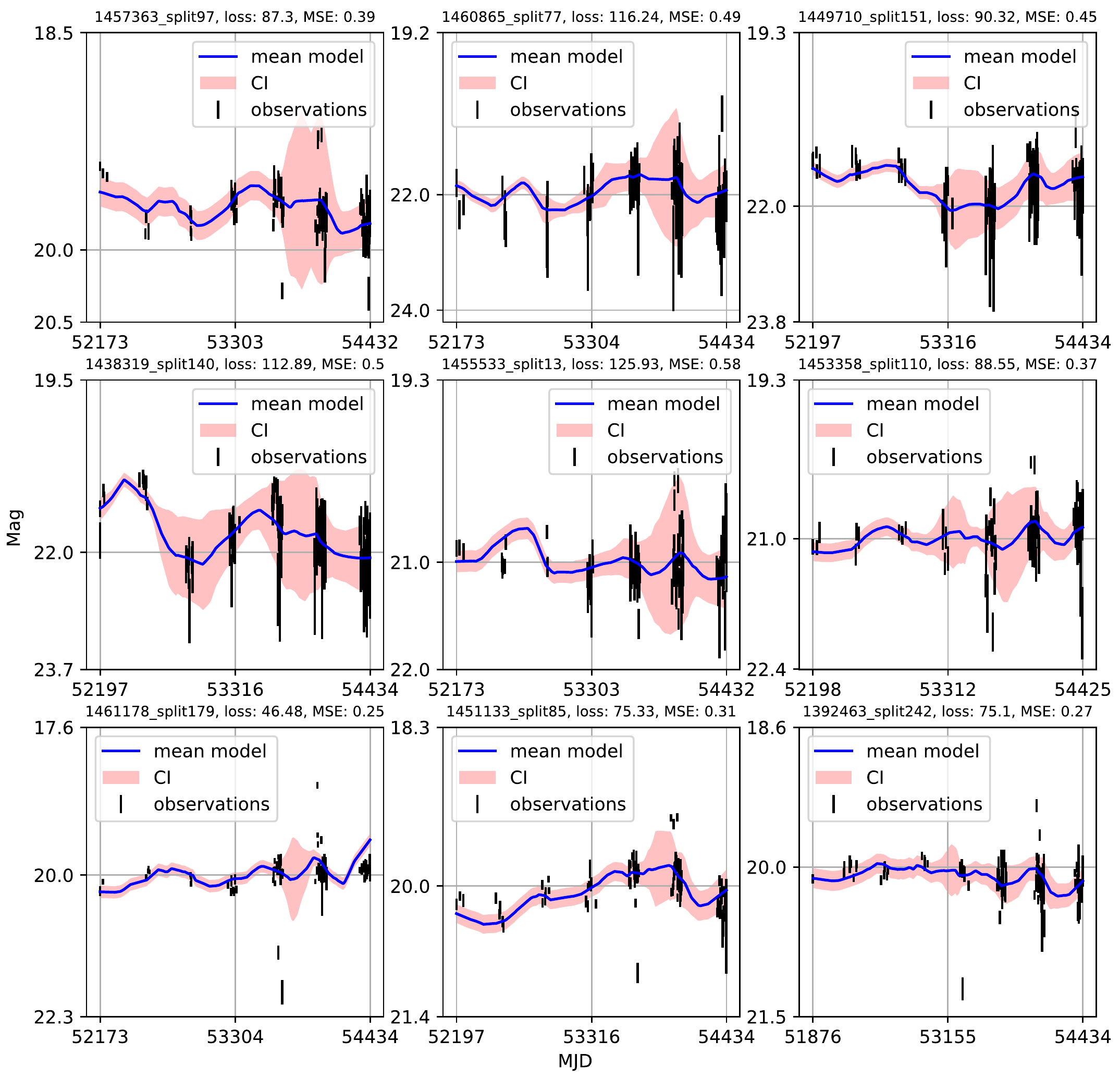}
\end{adjustwidth}
\caption{The same as Figure \ref{fig111}.\label{fig115}}
\end{figure}

\begin{figure}[H]
\begin{adjustwidth}{-\extralength}{0cm}
\centering
\includegraphics[page=2,width=1.3\textwidth]{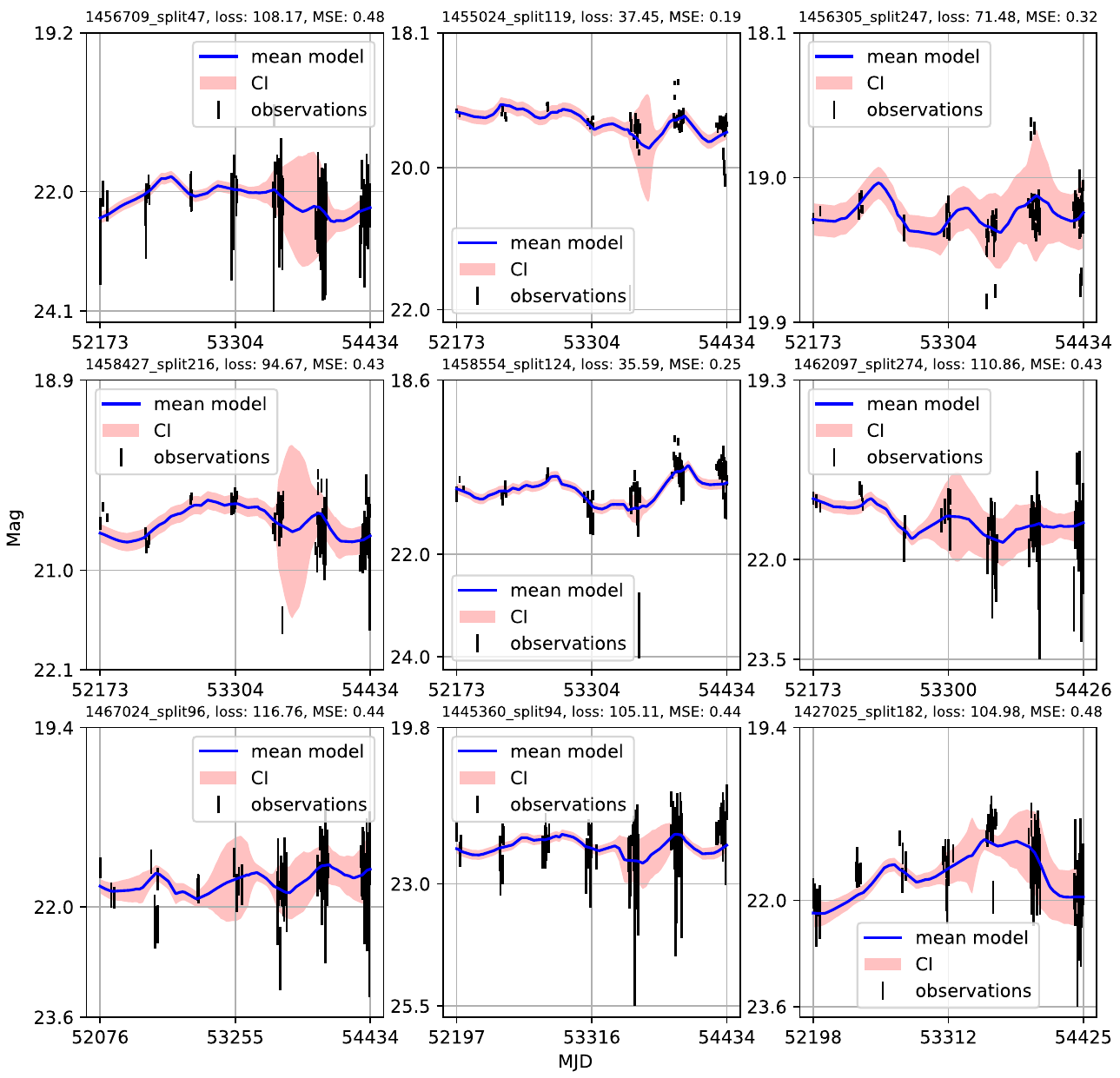}
\end{adjustwidth}
\caption{The same as Figure \ref{fig111}.\label{fig116}}
\end{figure}

\begin{figure}[H]
\begin{adjustwidth}{-\extralength}{0cm}
\centering
\includegraphics[page=3,width=1.3\textwidth]{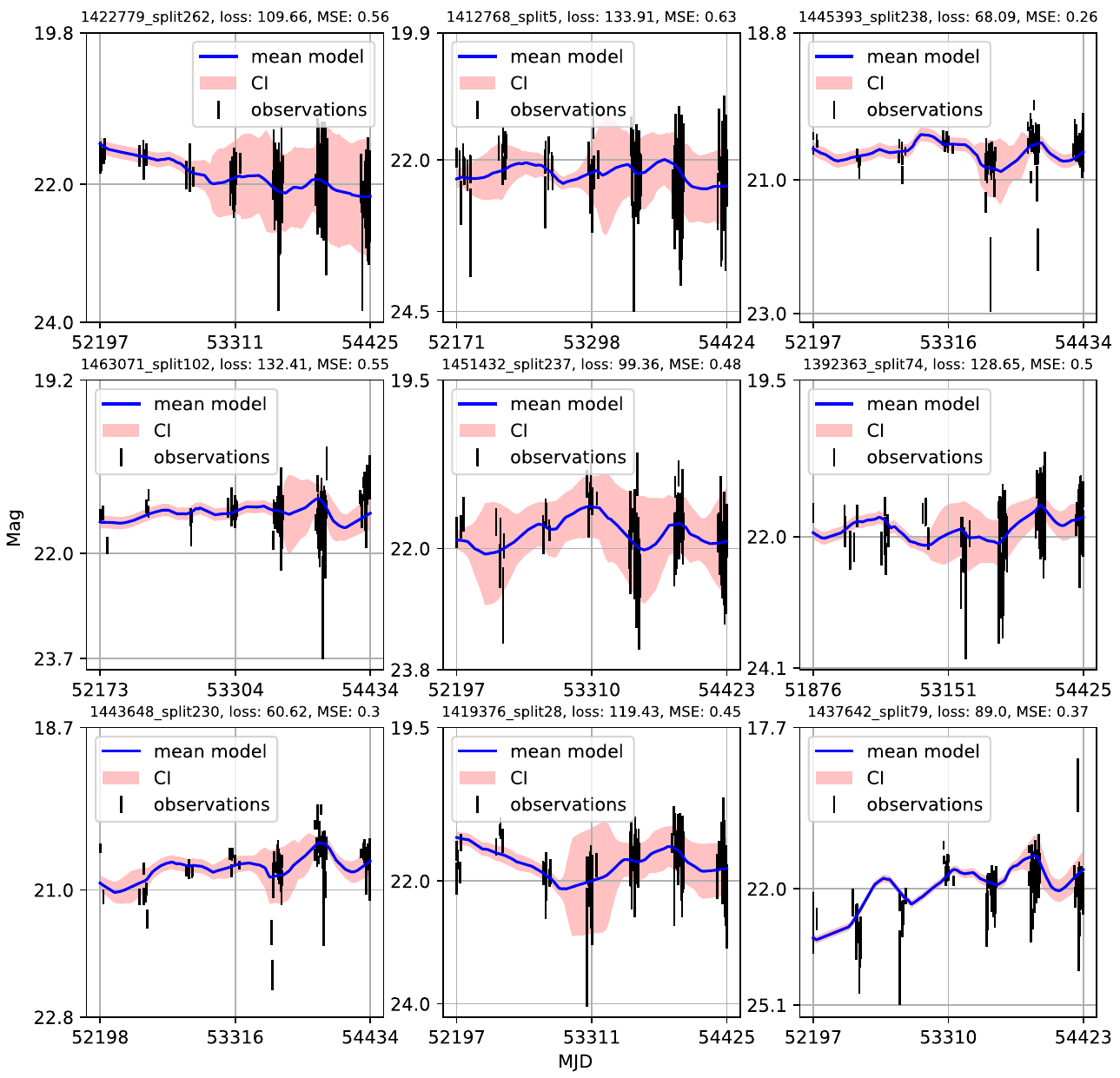}
\end{adjustwidth}
\caption{The same as Figure \ref{fig111}.\label{fig117}}
\end{figure}

\begin{figure}[H]
\begin{adjustwidth}{-\extralength}{0cm}
\centering
\includegraphics[width=1.3\textwidth]{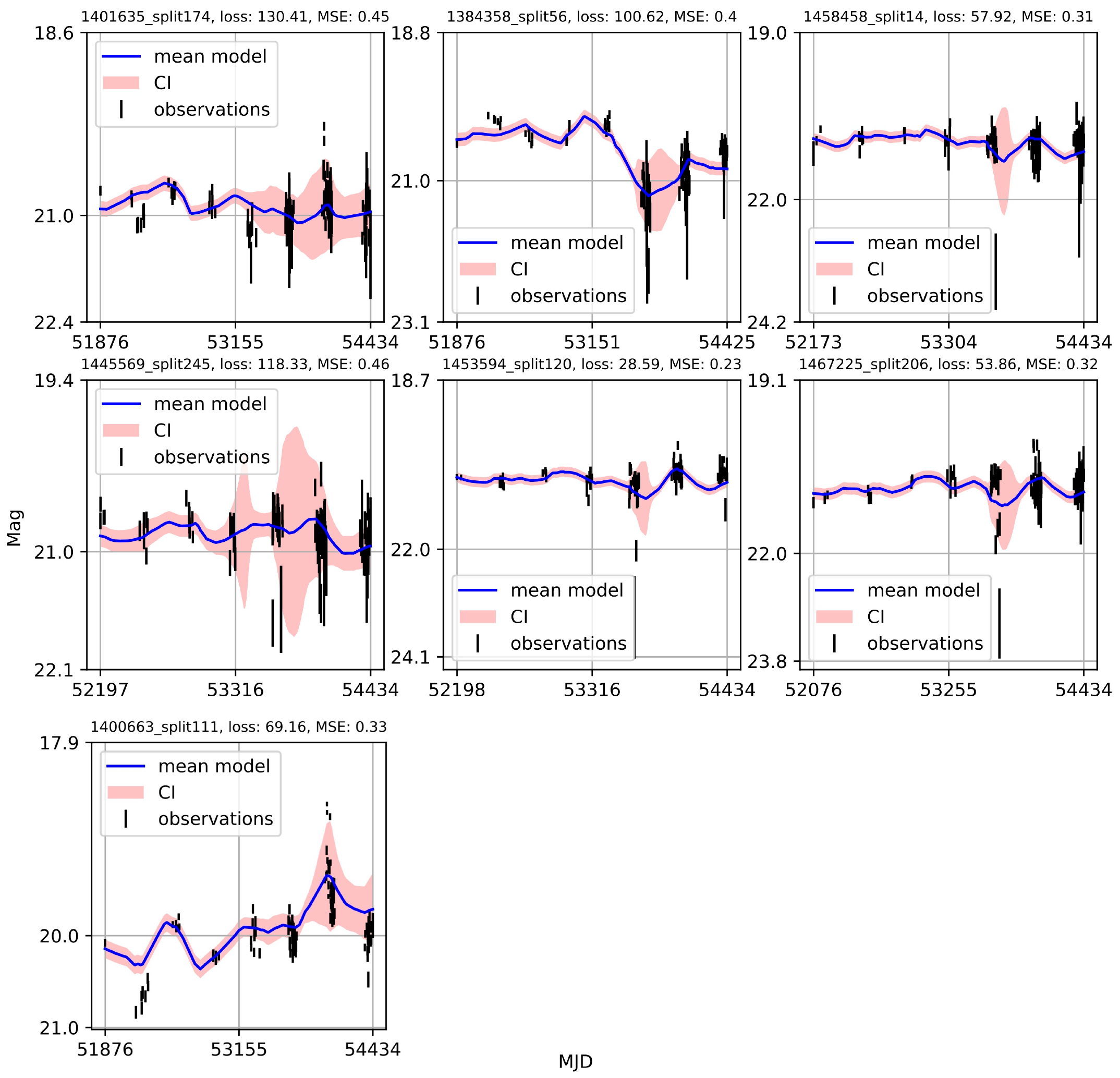}
\end{adjustwidth}
\caption{The same as Figure \ref{fig111}.\label{fig118}}
\end{figure}

\begin{figure}[H]
\begin{adjustwidth}{-\extralength}{0cm}
\centering
\includegraphics[page=1,width=1.3\textwidth]{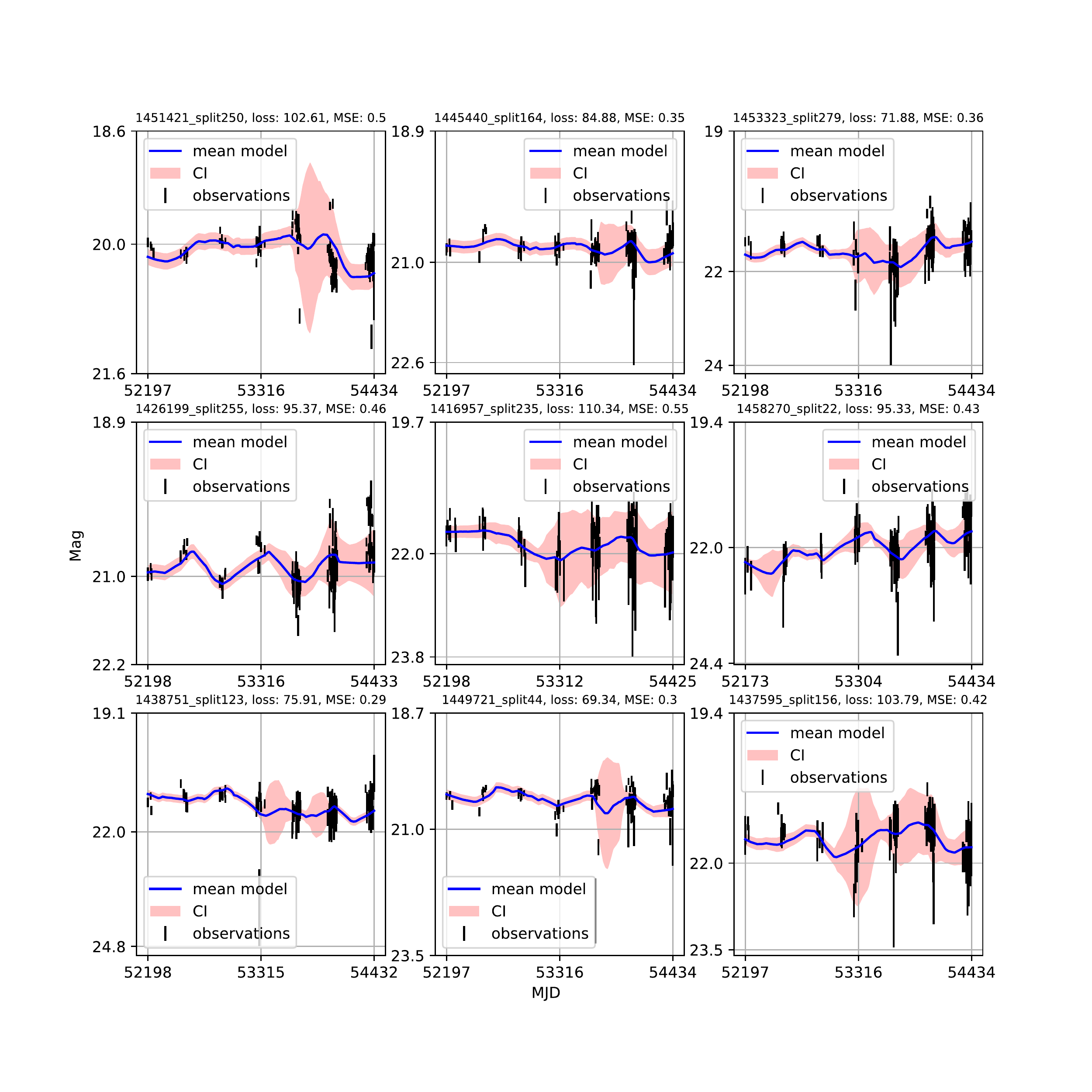}
\end{adjustwidth}
\caption{The same as Figure \ref{fig111}.\label{fig119}}
\end{figure}

\begin{figure}[H]
\begin{adjustwidth}{-\extralength}{0cm}
\centering
\includegraphics[page=2,width=1.3\textwidth]{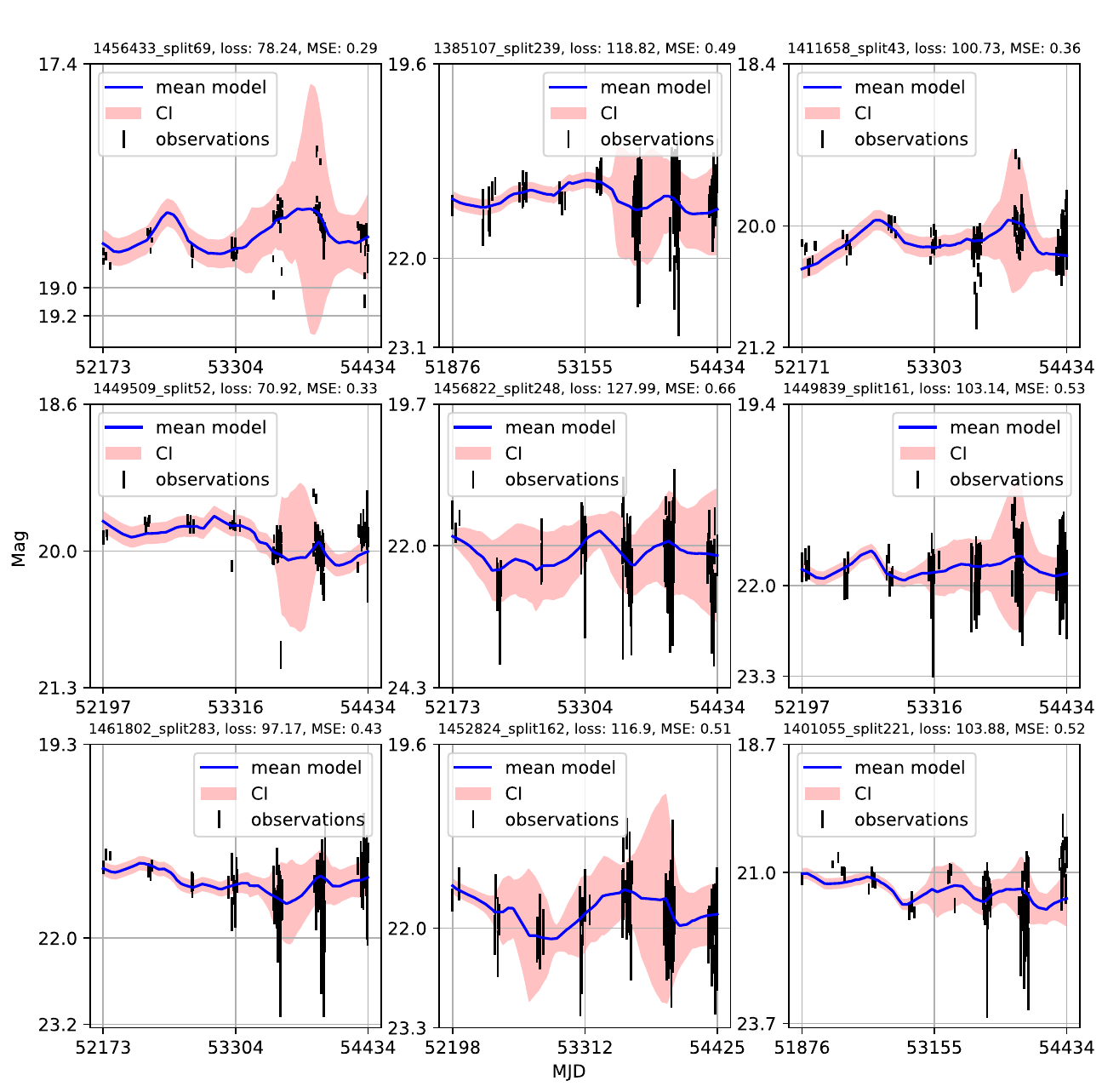}
\end{adjustwidth}
\caption{The same as Figure \ref{fig111}.\label{fig120}}
\end{figure}

\begin{figure}[H]
\begin{adjustwidth}{-\extralength}{0cm}
\centering
\includegraphics[page=3,width=1.3\textwidth]{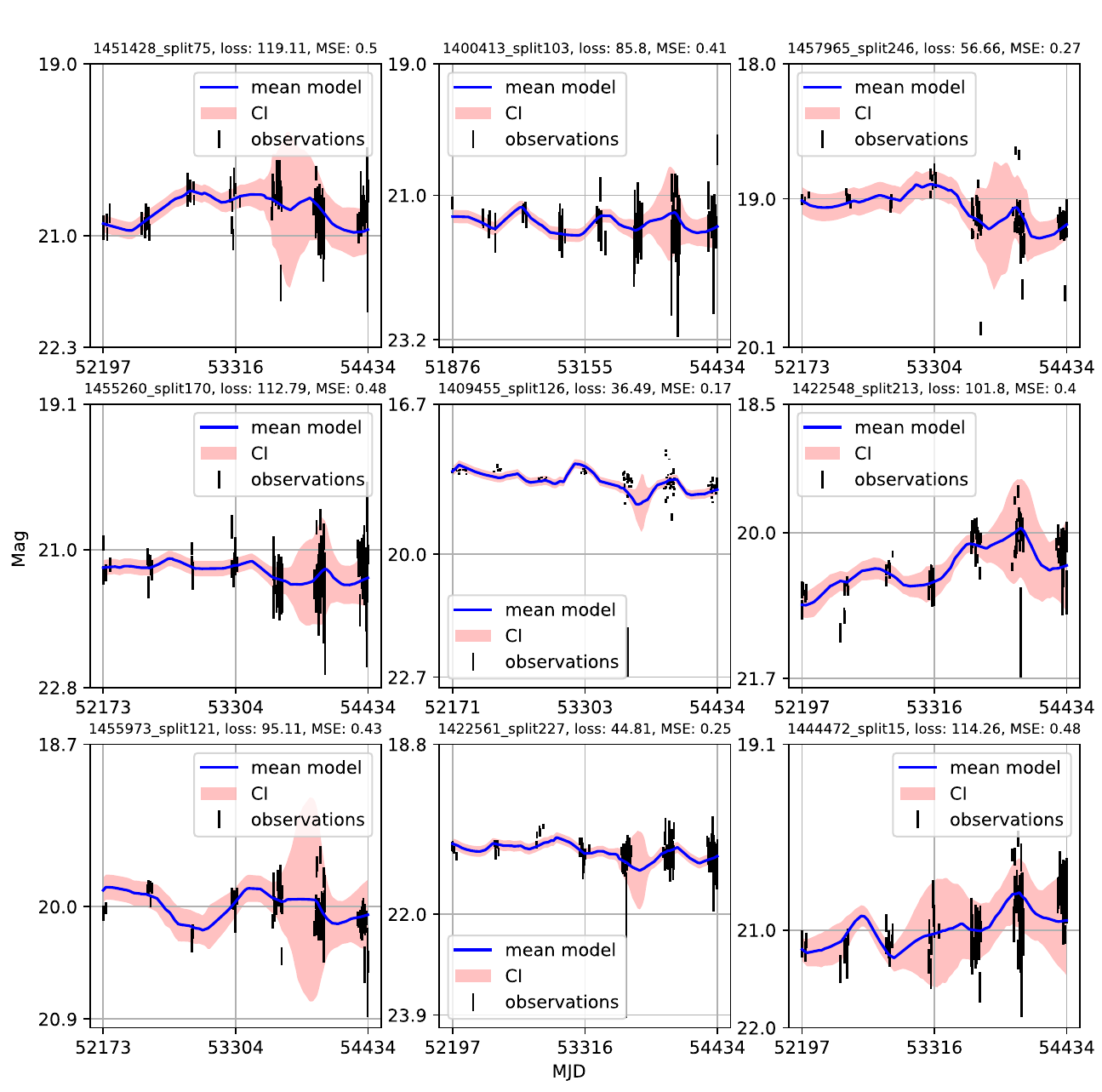}
\end{adjustwidth}
\caption{The same as Figure \ref{fig111}.\label{fig121}}
\end{figure}

\begin{figure}[H]
\begin{adjustwidth}{-\extralength}{0cm}
\centering
\includegraphics[page=4,width=1.3\textwidth]{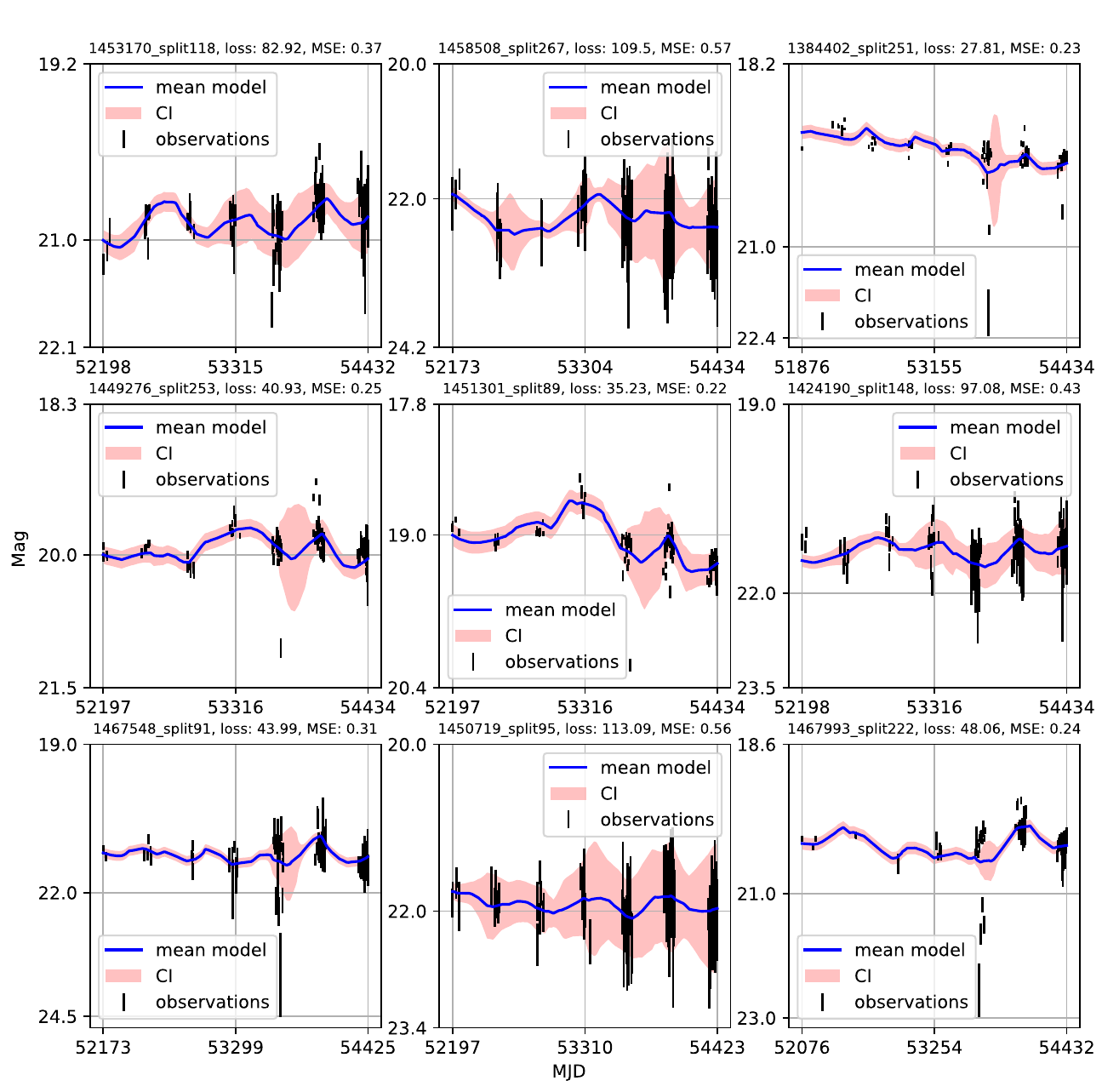}
\end{adjustwidth}
\caption{The same as Figure \ref{fig111}.\label{fig122}}
\end{figure}

\begin{figure}[H]
\begin{adjustwidth}{-\extralength}{0cm}
\centering
\includegraphics[page=5,width=1.3\textwidth]{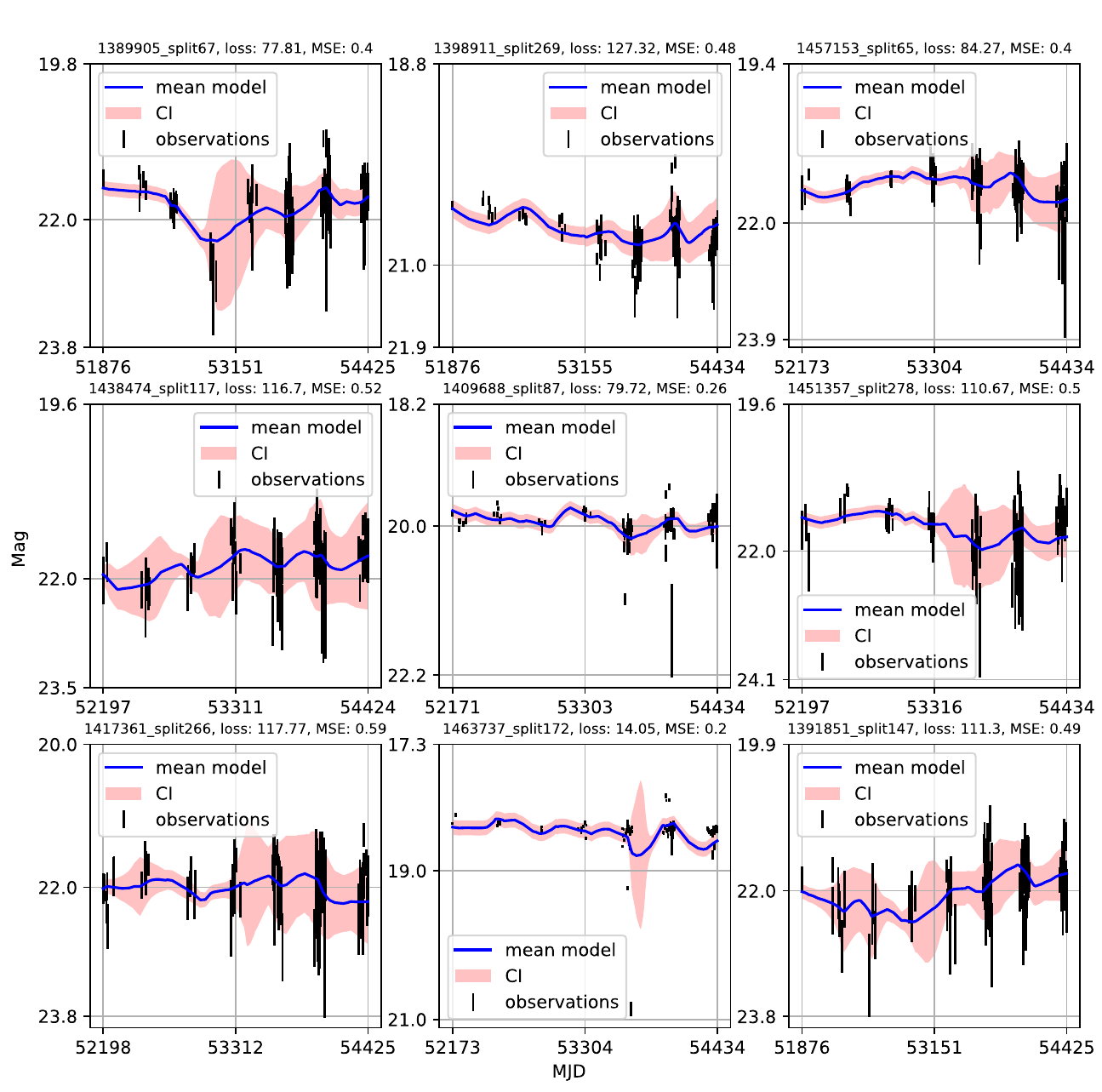}
\end{adjustwidth}
\caption{The same as Figure \ref{fig111}.\label{fig123}}
\end{figure}

\begin{figure}[H]
\begin{adjustwidth}{-\extralength}{0cm}
\centering
\includegraphics[page=6,width=1.3\textwidth]{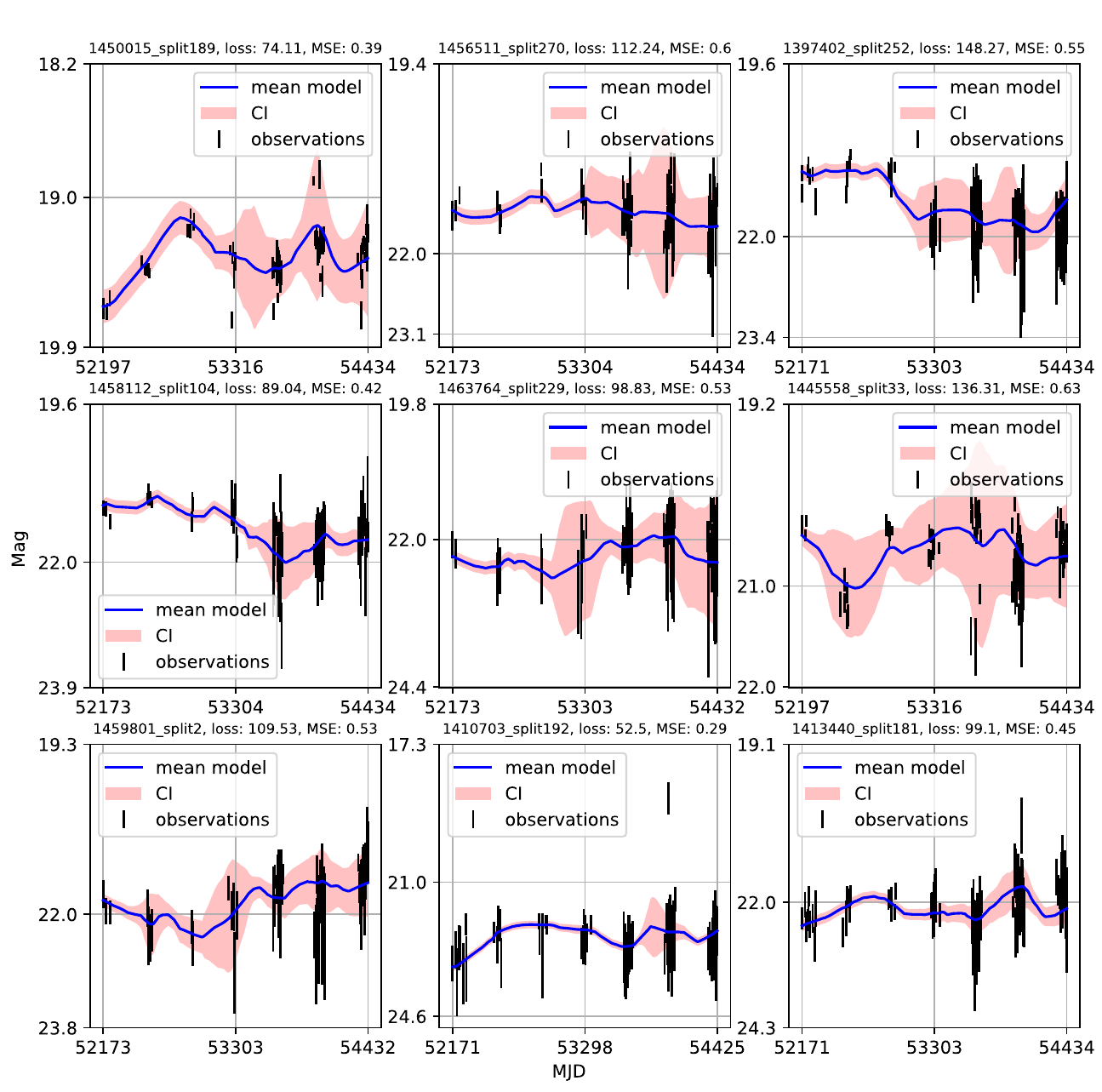}
\end{adjustwidth}
\caption{The same as Figure \ref{fig111}.\label{fig124}}
\end{figure}

\begin{figure}[H]
\begin{adjustwidth}{-\extralength}{0cm}
\centering
\includegraphics[page=7,width=1.3\textwidth]{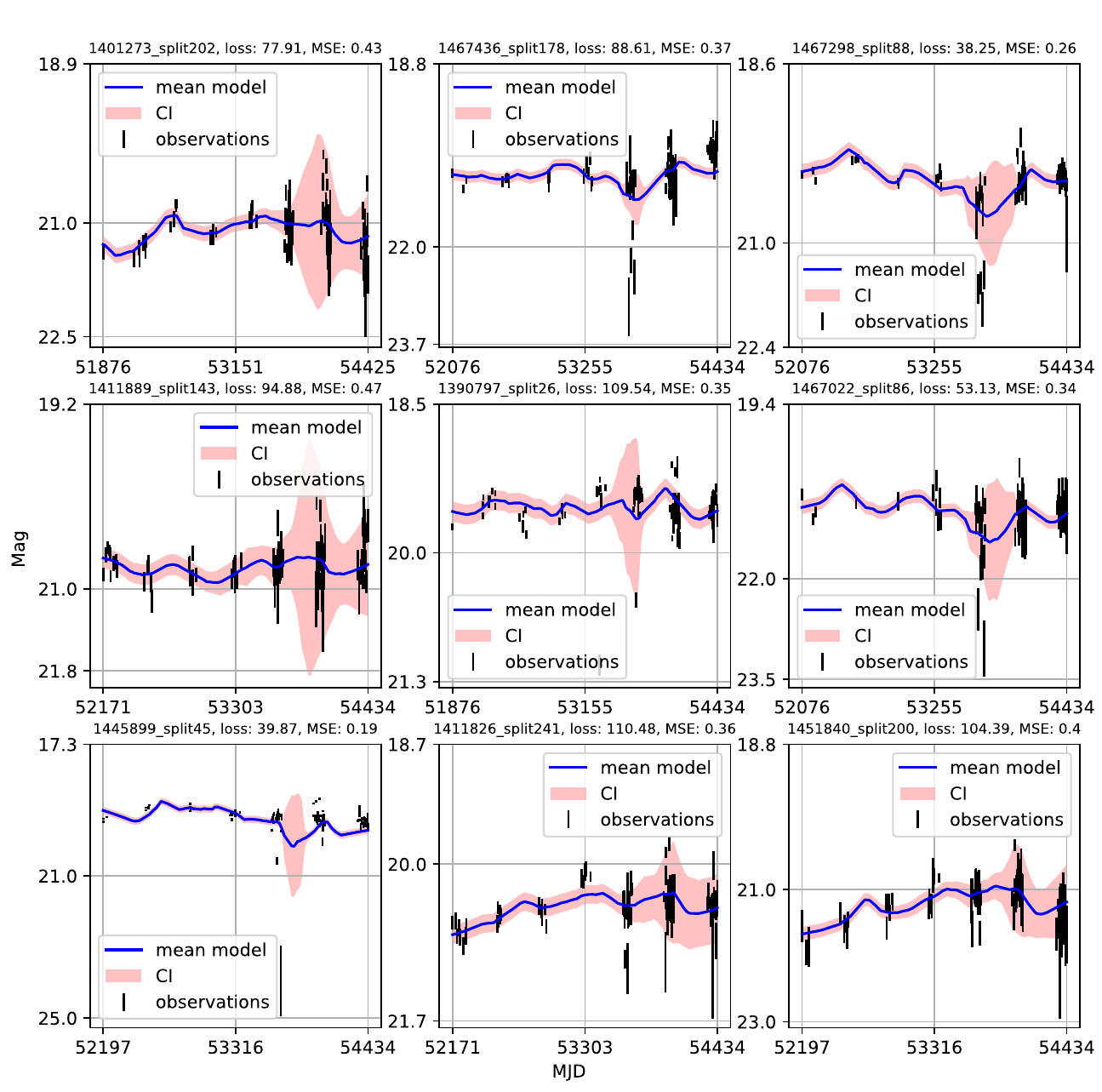}
\end{adjustwidth}
\caption{The same as Figure \ref{fig111}.\label{fig125}}
\end{figure}

\begin{figure}[H]
\begin{adjustwidth}{-\extralength}{0cm}
\centering
\includegraphics[page=8,width=1.3\textwidth]{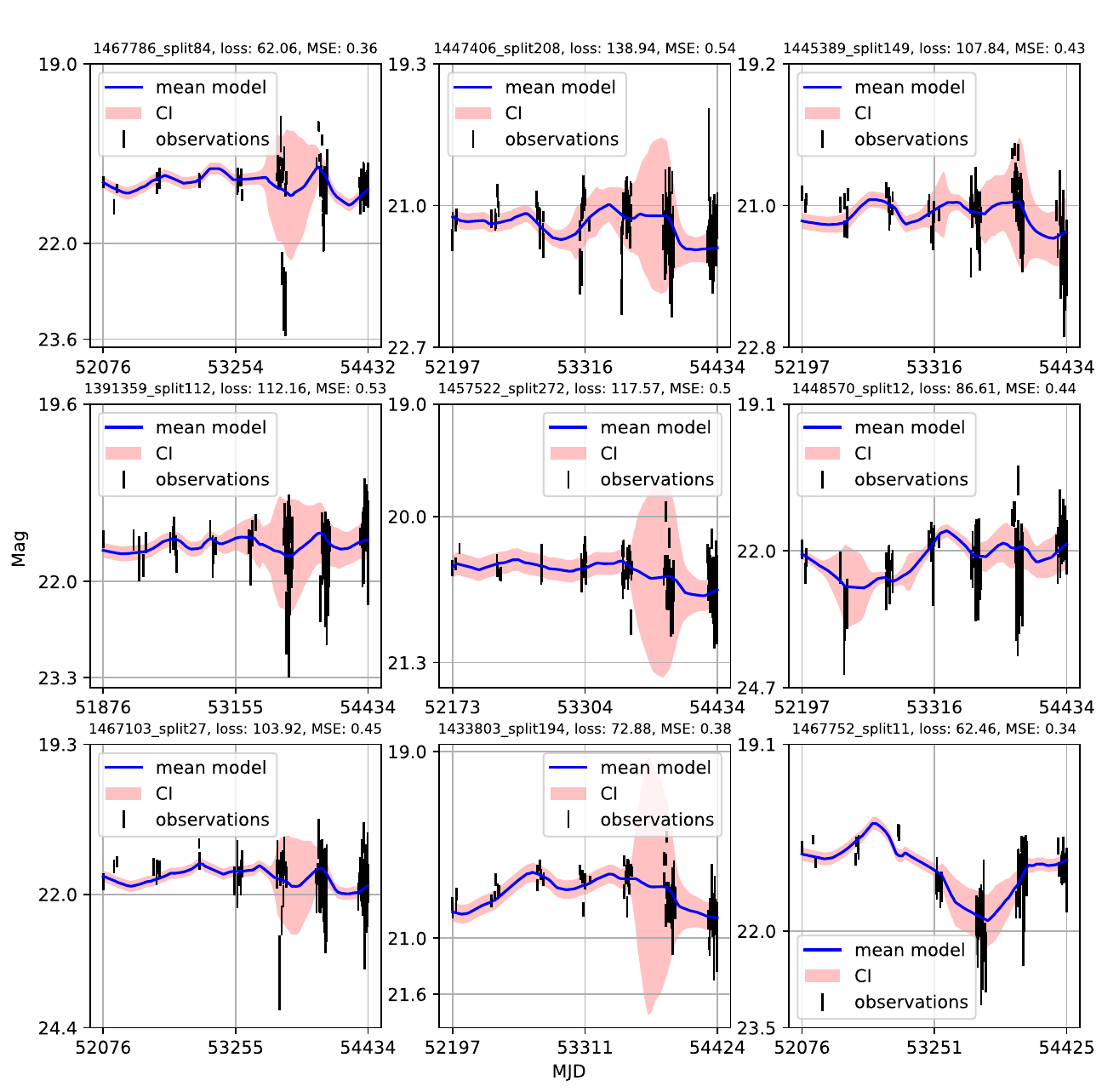}
\end{adjustwidth}
\caption{The same as Figure \ref{fig111}.\label{fig126}}
\end{figure}

\begin{figure}[H]
\begin{adjustwidth}{-\extralength}{0cm}
\centering
\includegraphics[page=9,width=1.3\textwidth]{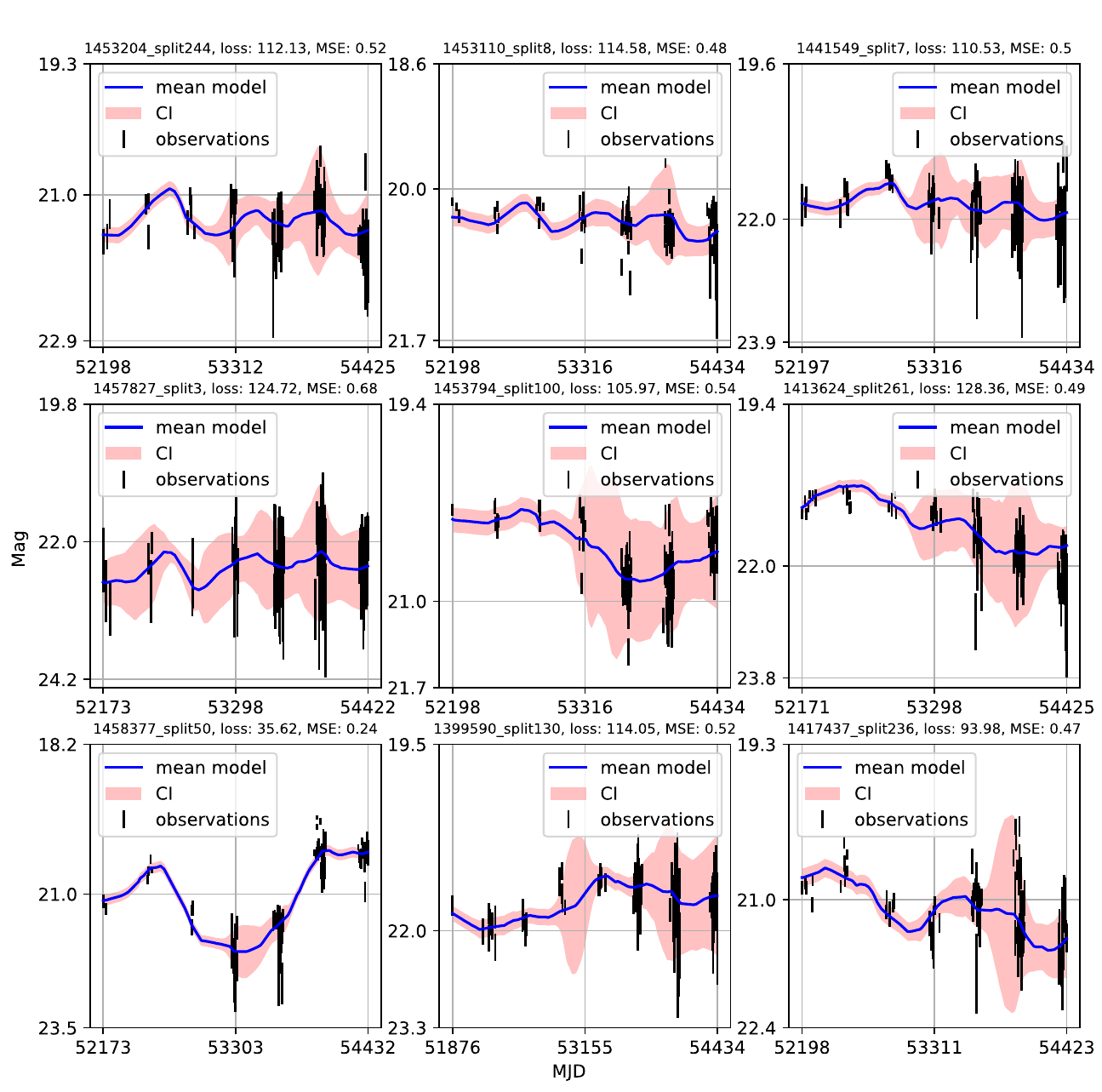}
\end{adjustwidth}
\caption{The same as Figure \ref{fig111}.\label{fig127}}
\end{figure}

\begin{figure}[H]
\begin{adjustwidth}{-\extralength}{0cm}
\centering
\includegraphics[page=10,width=1.3\textwidth]{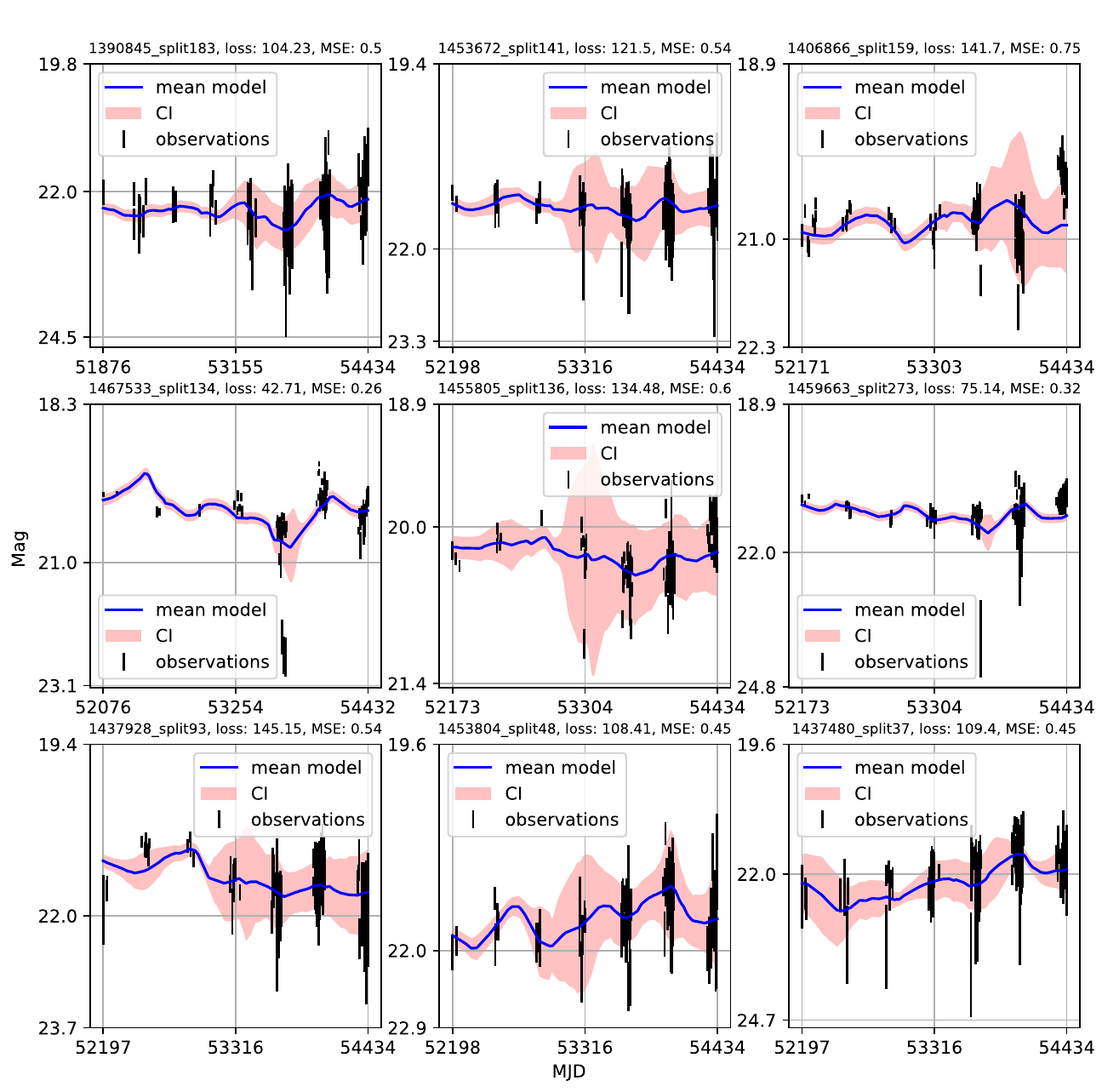}
\end{adjustwidth}
\caption{The same as Figure \ref{fig111}.\label{fig128}}
\end{figure}

\begin{figure}[H]
\begin{adjustwidth}{-\extralength}{0cm}
\centering
\includegraphics[page=11,width=1.3\textwidth]{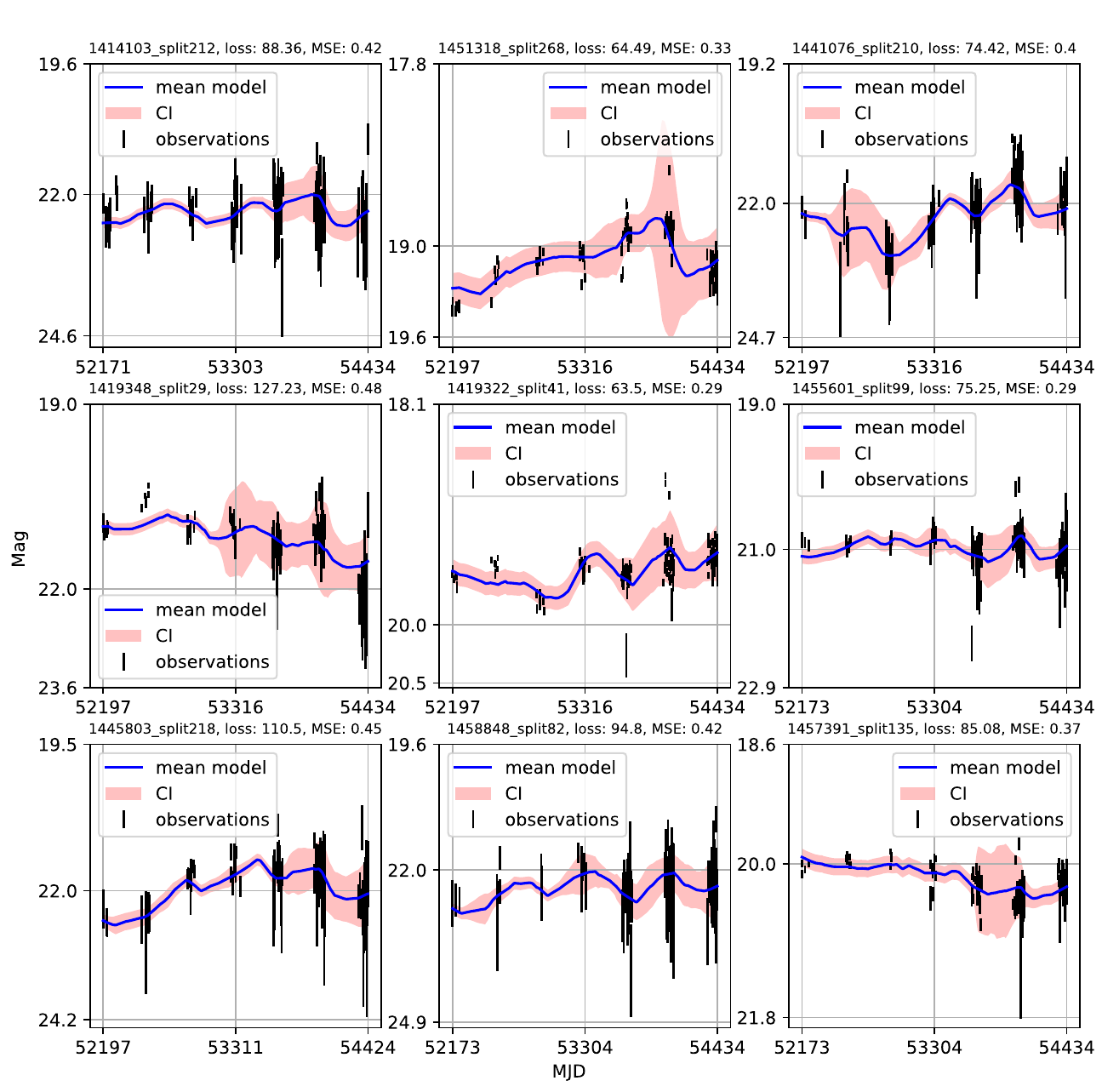}
\end{adjustwidth}
\caption{The same as Figure \ref{fig111}.\label{fig129}}
\end{figure}

\begin{figure}[H]
\begin{adjustwidth}{-\extralength}{0cm}
\centering
\includegraphics[page=12,width=1.3\textwidth]{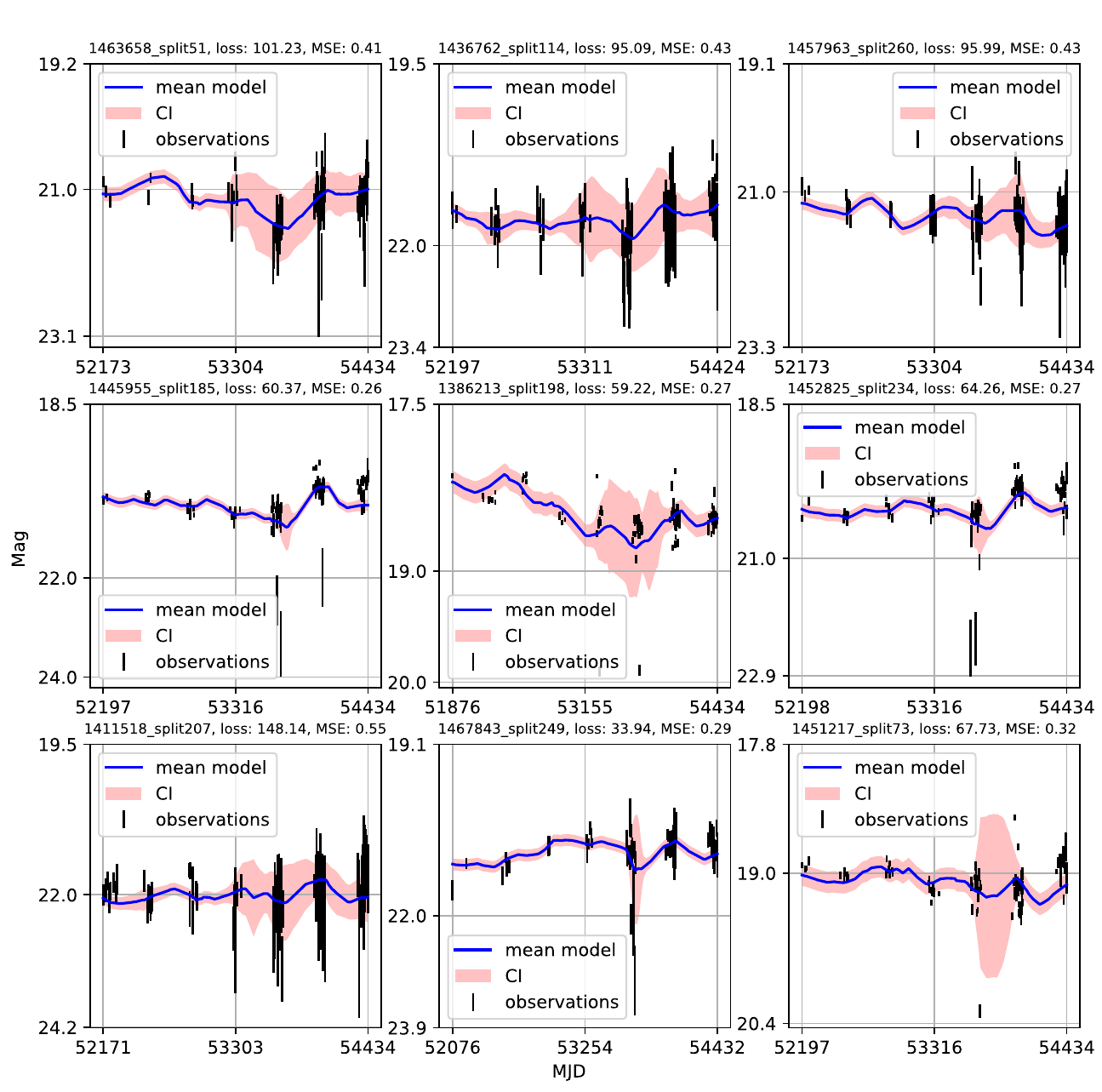}
\end{adjustwidth}
\caption{The same as Figure \ref{fig111}.\label{fig130}}
\end{figure}

\begin{figure}[H]
\begin{adjustwidth}{-\extralength}{0cm}
\centering
\includegraphics[page=13,width=1.3\textwidth]{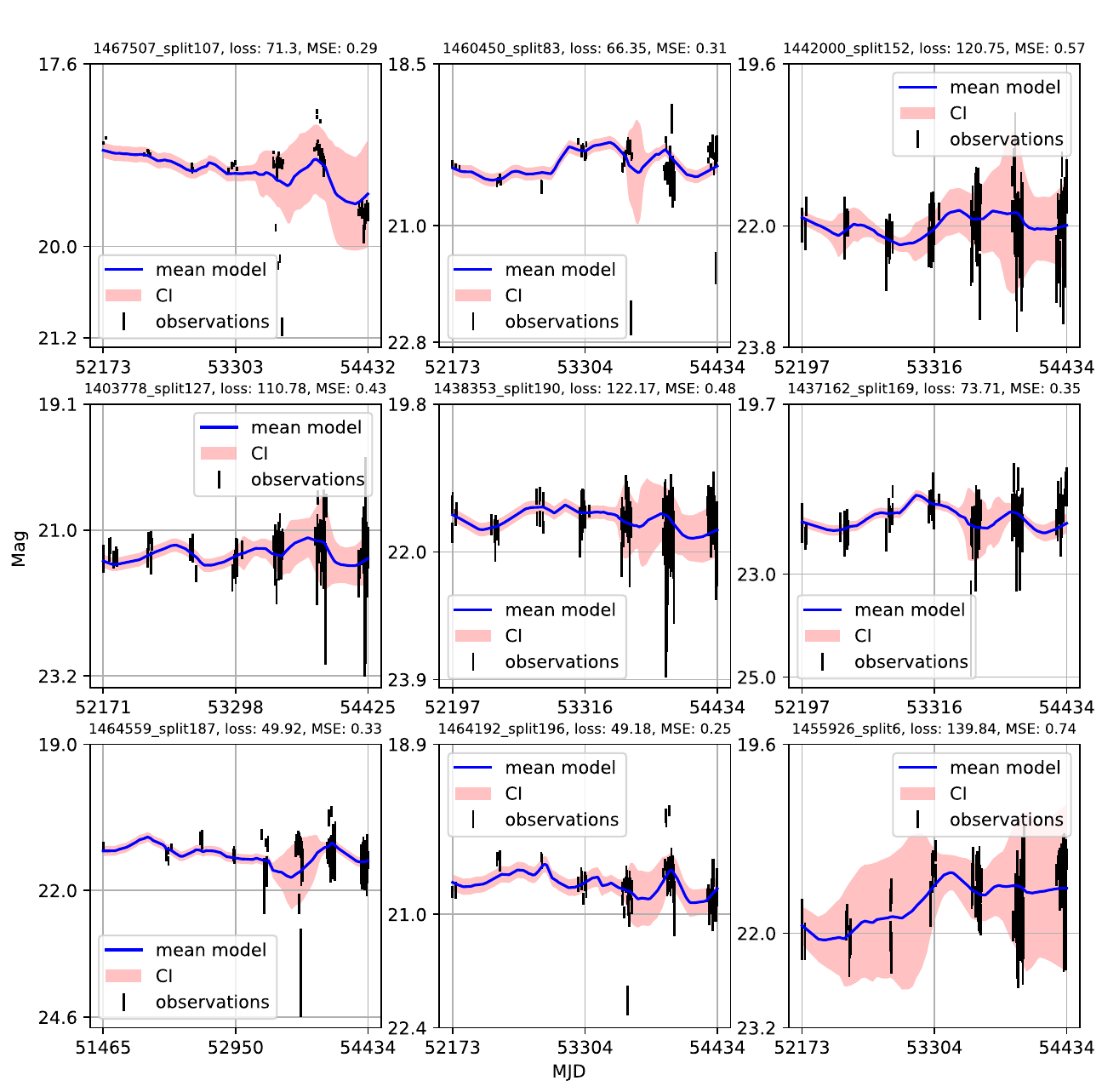}
\end{adjustwidth}
\caption{The same as Figure \ref{fig111}.\label{fig131}}
\end{figure}

\begin{figure}[H]
\begin{adjustwidth}{-\extralength}{0cm}
\centering
\includegraphics[page=14,width=1.3\textwidth]{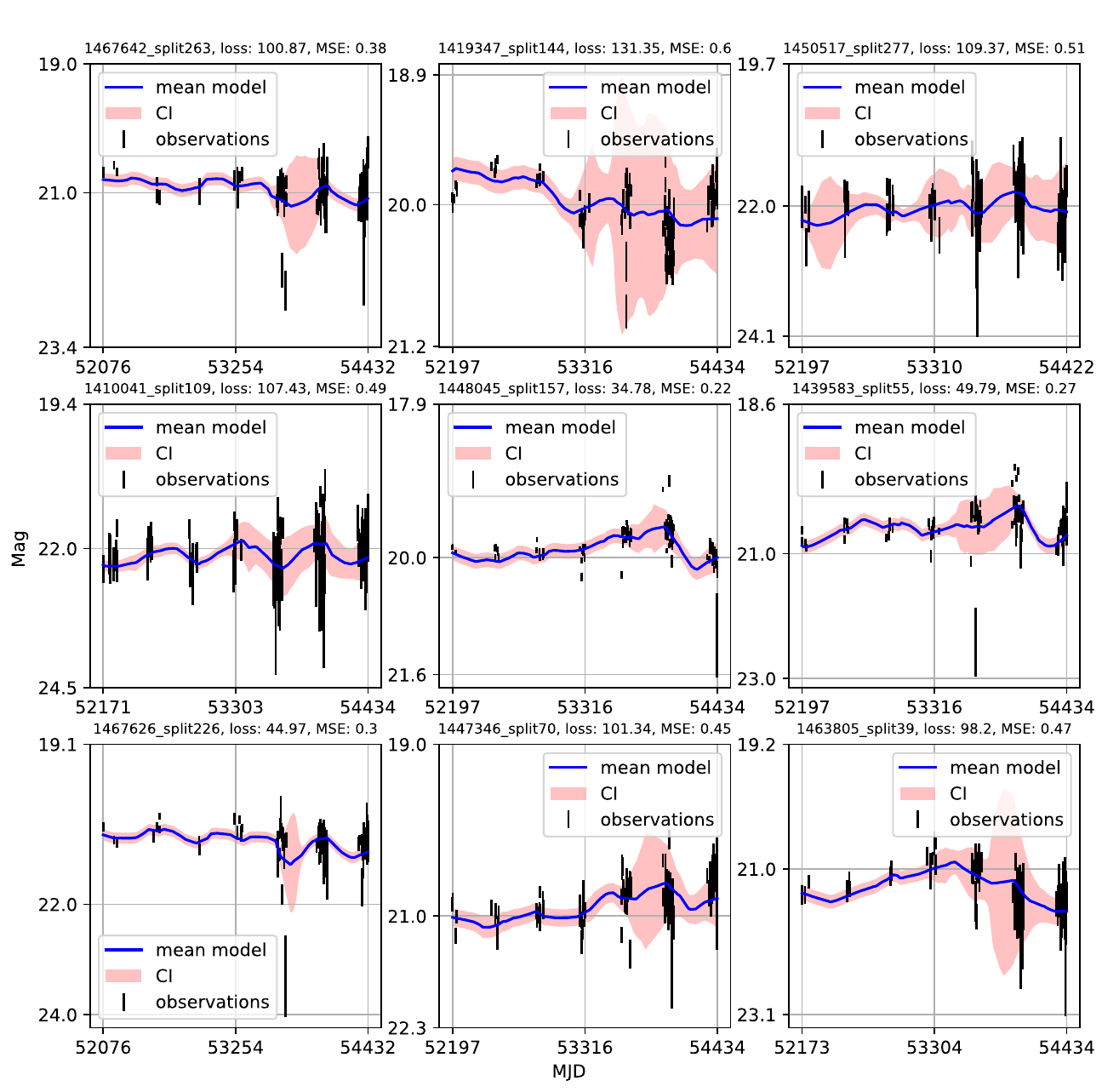}
\end{adjustwidth}
\caption{The same as Figure \ref{fig111}.\label{fig132}}
\end{figure}

\begin{figure}[H]
\begin{adjustwidth}{-\extralength}{0cm}
\centering
\includegraphics[page=15,width=1.3\textwidth]{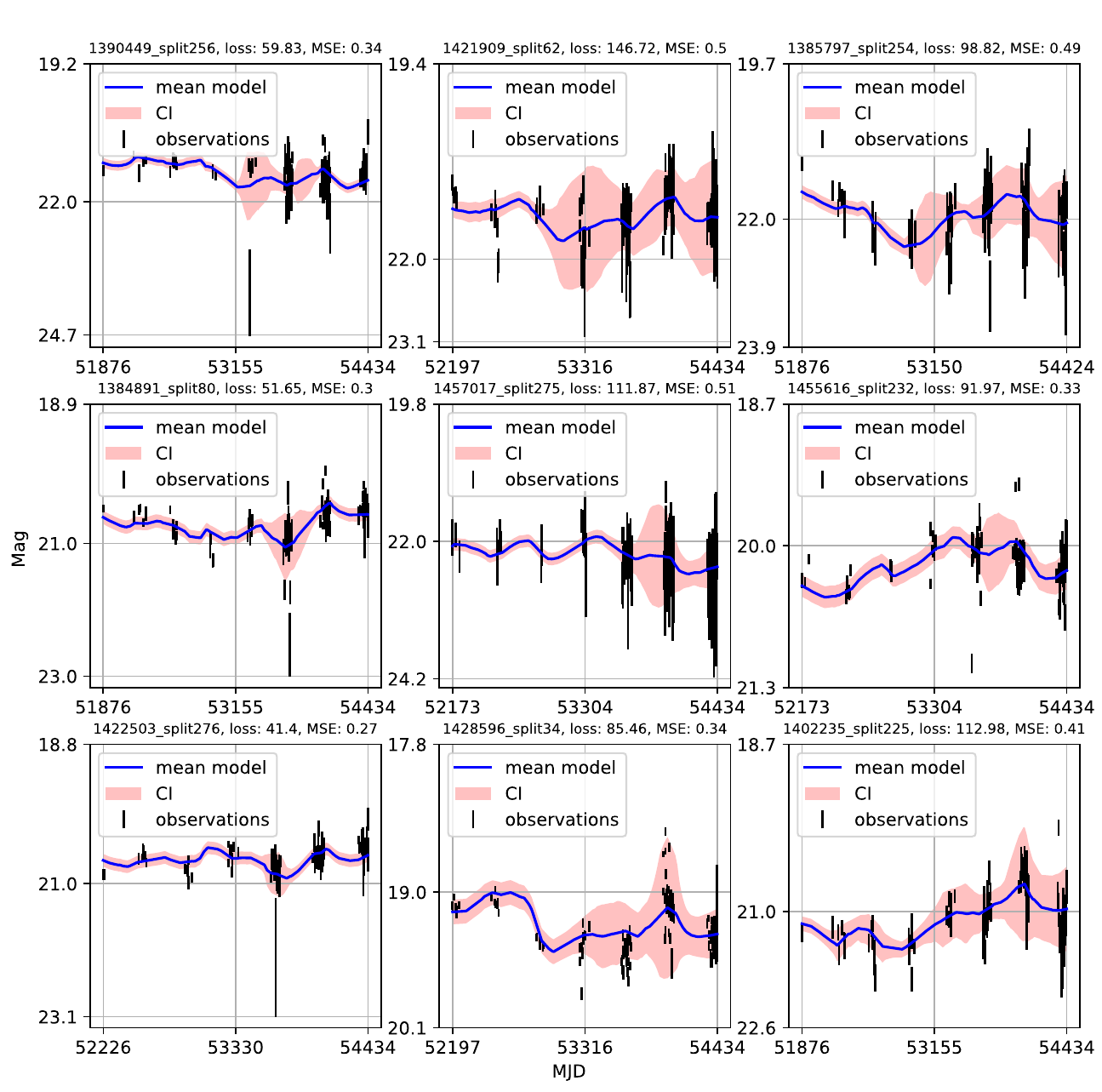}
\end{adjustwidth}
\caption{The same as Figure \ref{fig111}.\label{fig133}}
\end{figure}

\begin{figure}[H]
\begin{adjustwidth}{-\extralength}{0cm}
\centering
\includegraphics[page=16,width=1.3\textwidth]{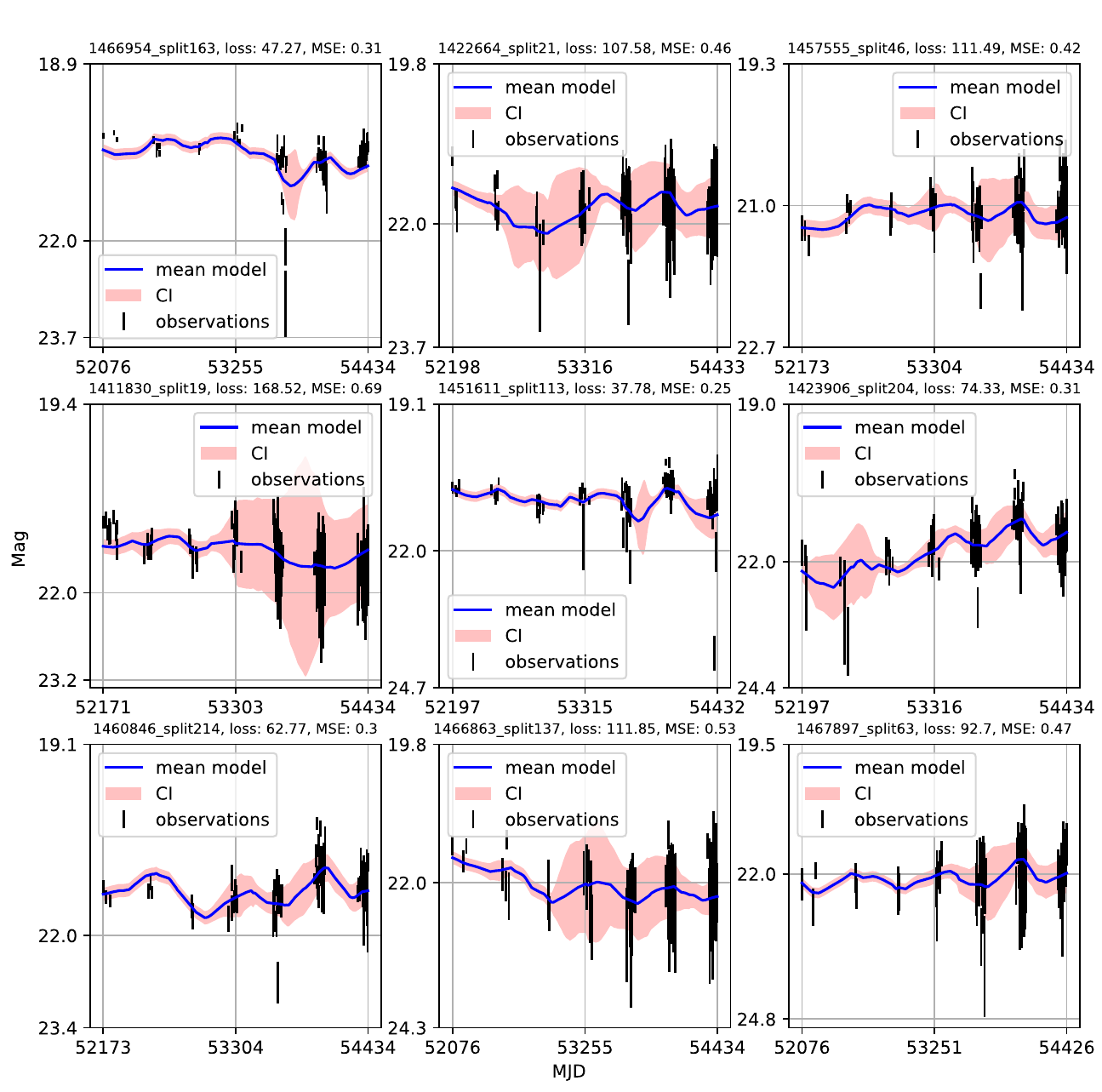}
\end{adjustwidth}
\caption{The same as Figure \ref{fig111}.\label{fig134}}
\end{figure}

\begin{figure}[H]
\begin{adjustwidth}{-\extralength}{0cm}
\centering
\includegraphics[page=17,width=1.3\textwidth]{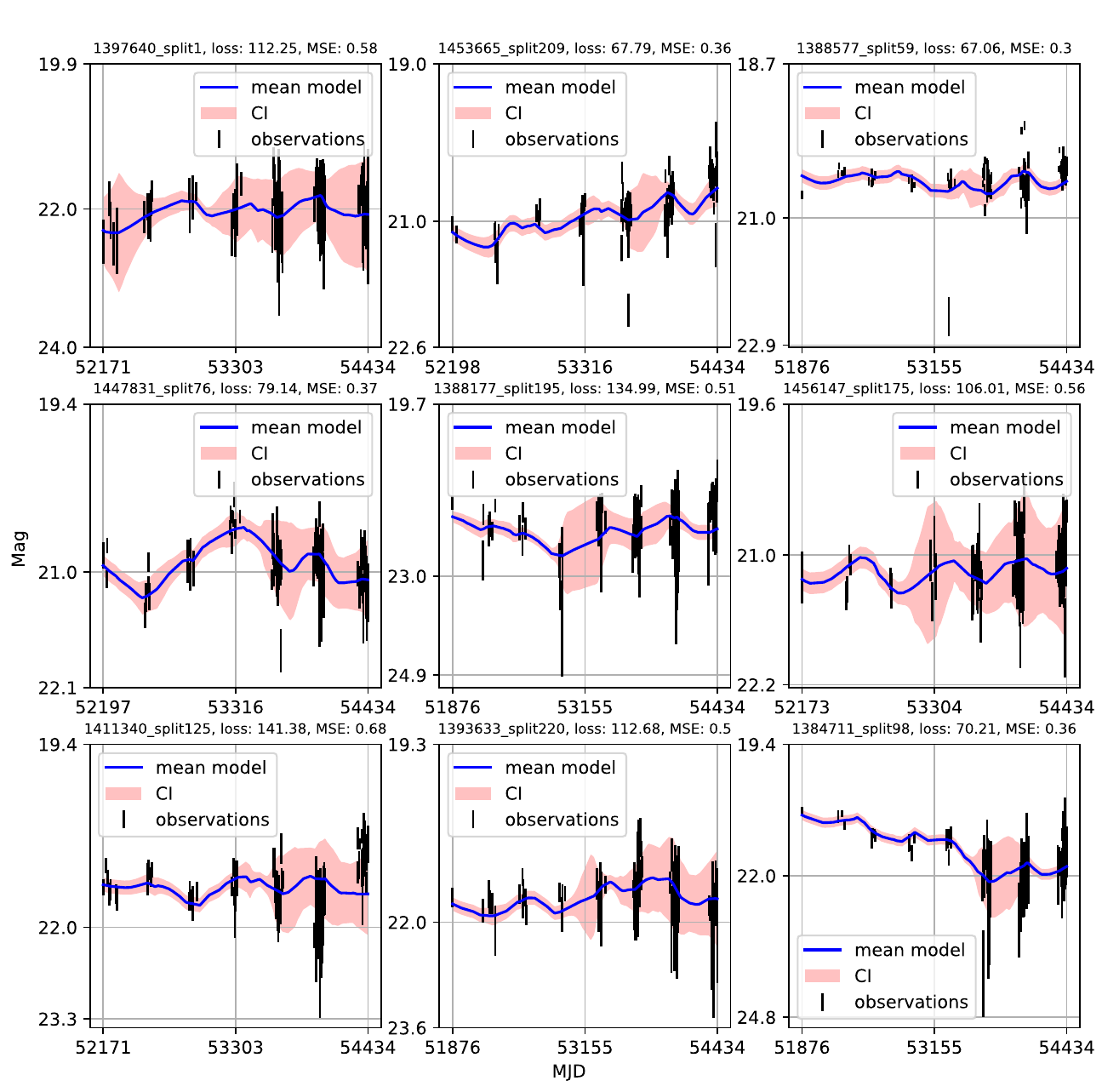}
\end{adjustwidth}
\caption{The same as Figure \ref{fig111}.\label{fig135}}
\end{figure}

\begin{figure}[H]
\begin{adjustwidth}{-\extralength}{0cm}
\centering
\includegraphics[page=18,width=1.3\textwidth]{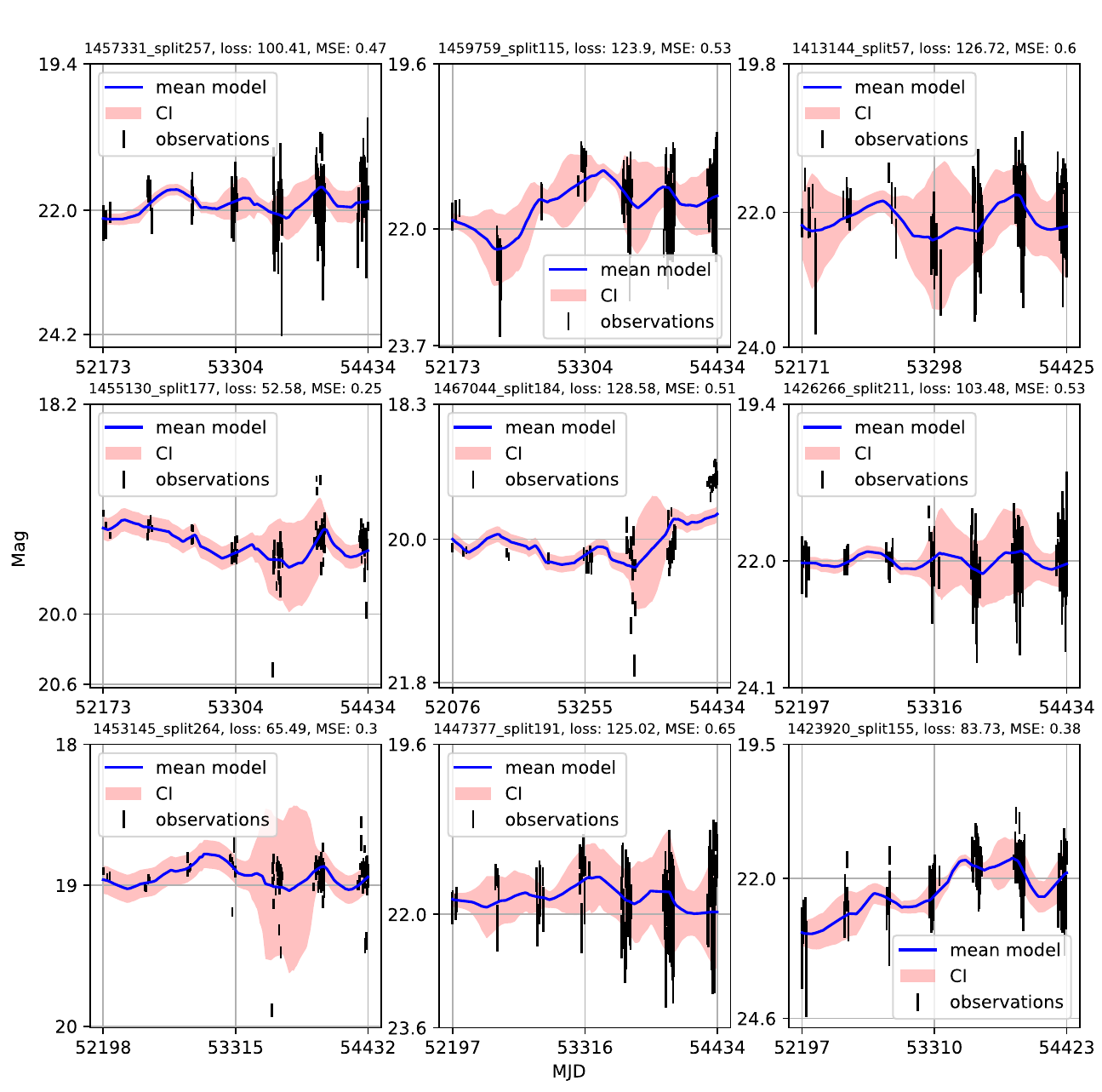}
\end{adjustwidth}
\caption{The same as Figure \ref{fig111}.\label{fig136}}
\end{figure}

\begin{figure}[H]
\begin{adjustwidth}{-\extralength}{0cm}
\centering
\includegraphics[page=19,width=1.3\textwidth]{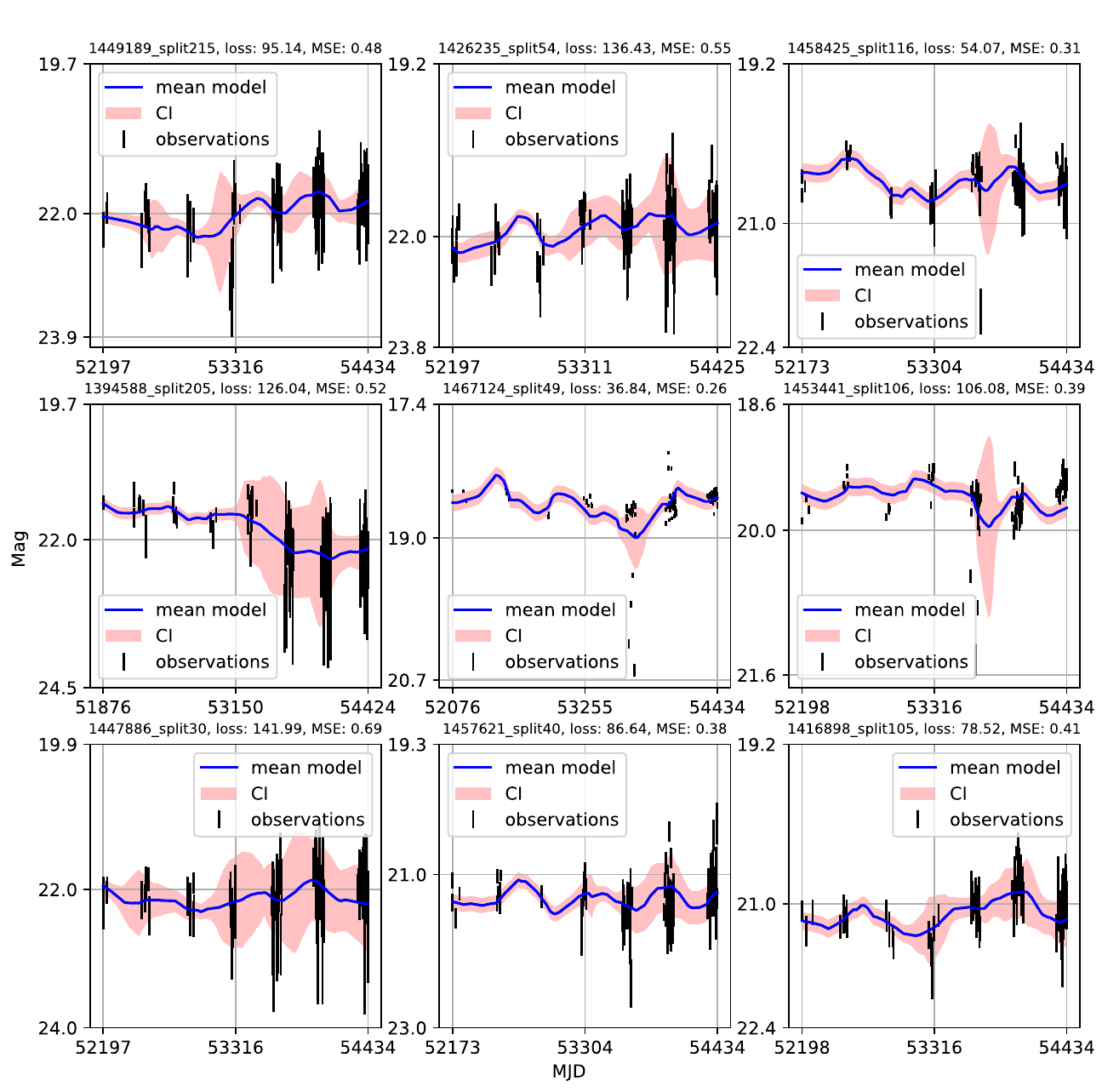}
\end{adjustwidth}
\caption{The same as Figure \ref{fig111}.\label{fig137}}
\end{figure}

\begin{figure}[H]
\begin{adjustwidth}{-\extralength}{0cm}
\centering
\includegraphics[page=20,width=1.3\textwidth]{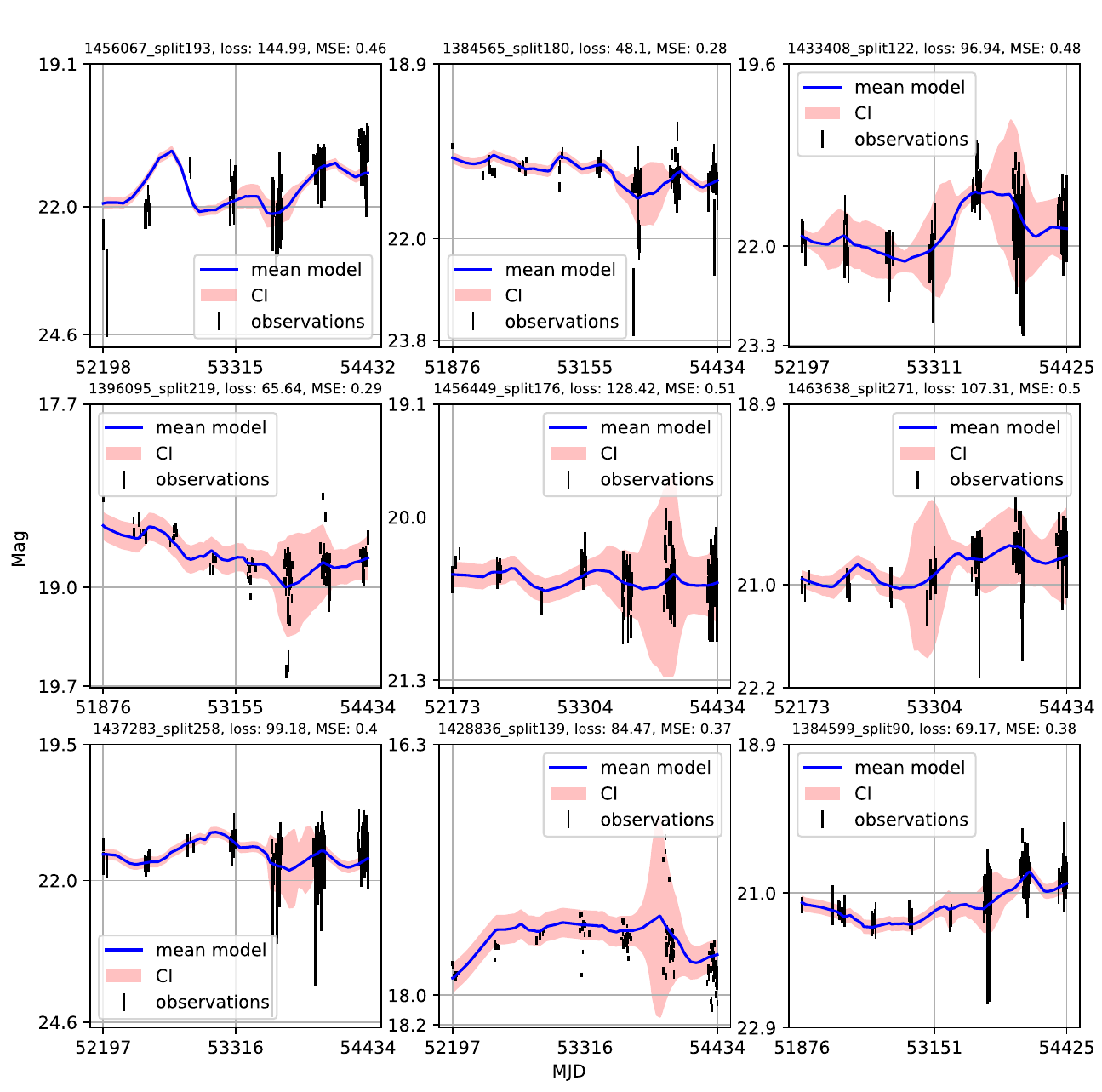}
\end{adjustwidth}
\caption{The same as Figure \ref{fig111}.\label{fig138}}
\end{figure}

\begin{figure}[H]
\begin{adjustwidth}{-\extralength}{0cm}
\centering
\includegraphics[page=21,width=1.3\textwidth]{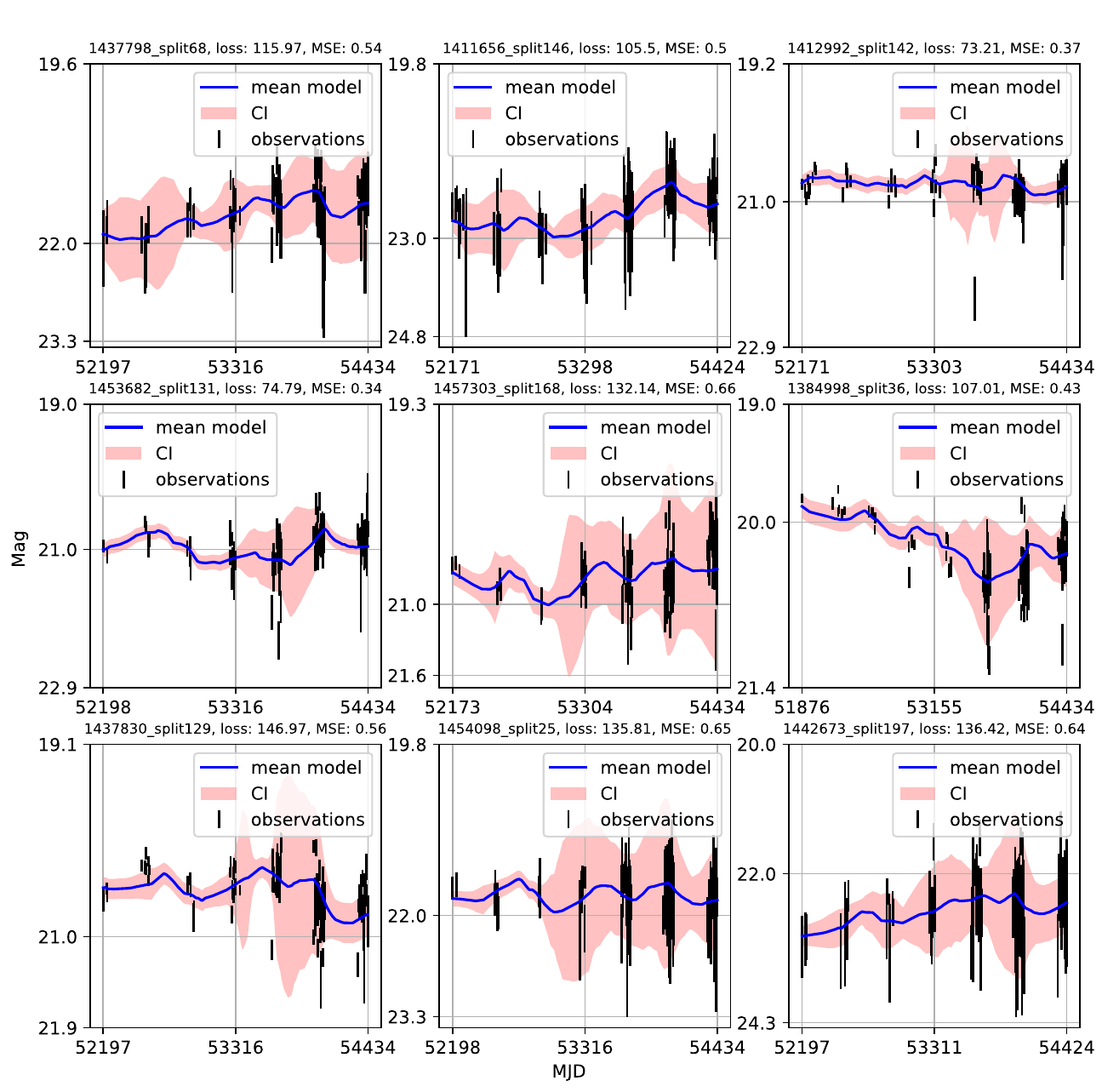}
\end{adjustwidth}
\caption{The same as Figure \ref{fig111}.\label{fig139}}
\end{figure}

\begin{figure}[H]
\begin{adjustwidth}{-\extralength}{0cm}
\centering
\includegraphics[page=22,width=1.3\textwidth]{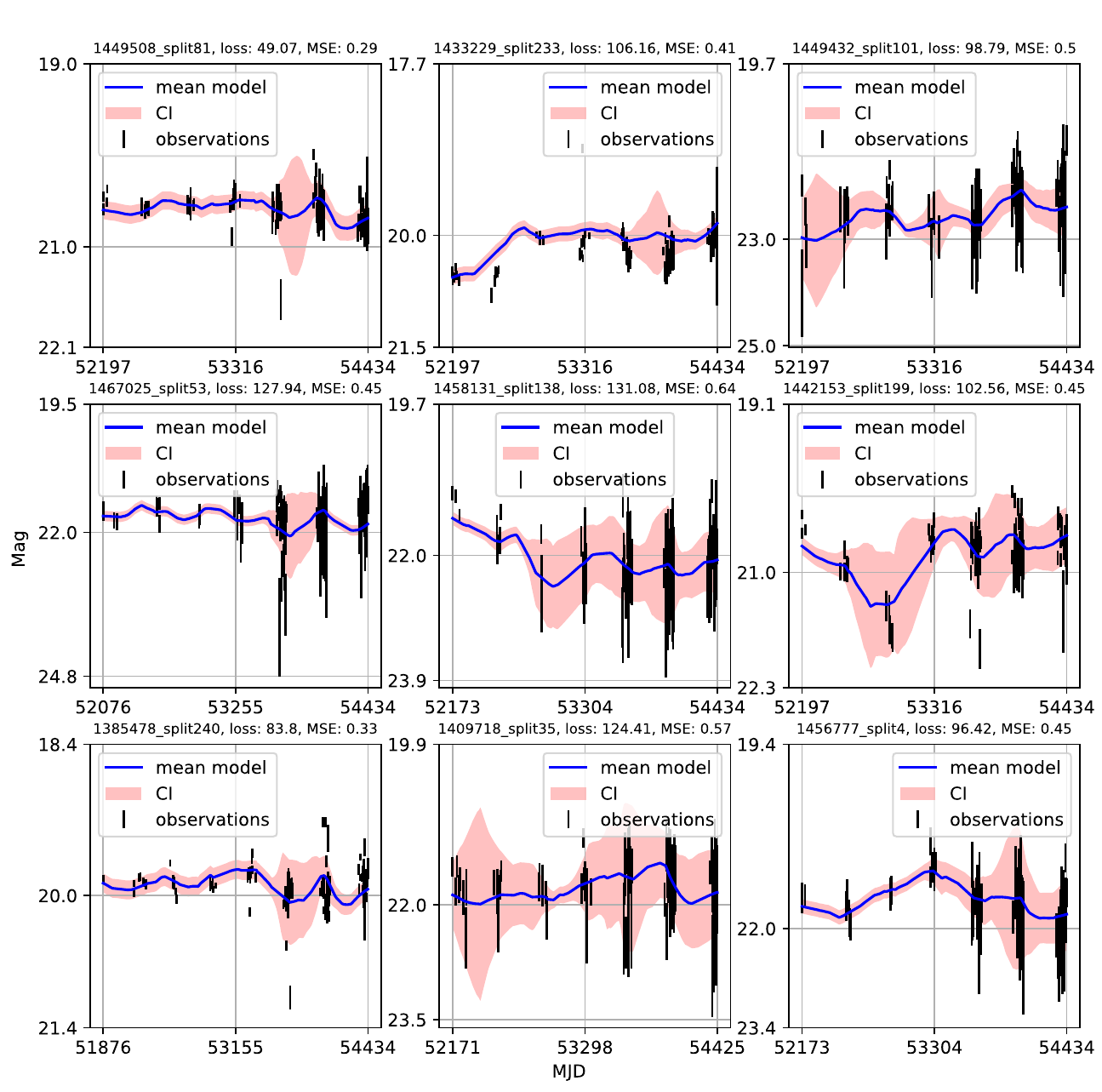}
\end{adjustwidth}
\caption{The same as Figure \ref{fig111}.\label{fig140}}
\end{figure}

\begin{figure}[H]
\begin{adjustwidth}{-\extralength}{0cm}
\centering
\includegraphics[page=23,width=1.3\textwidth]{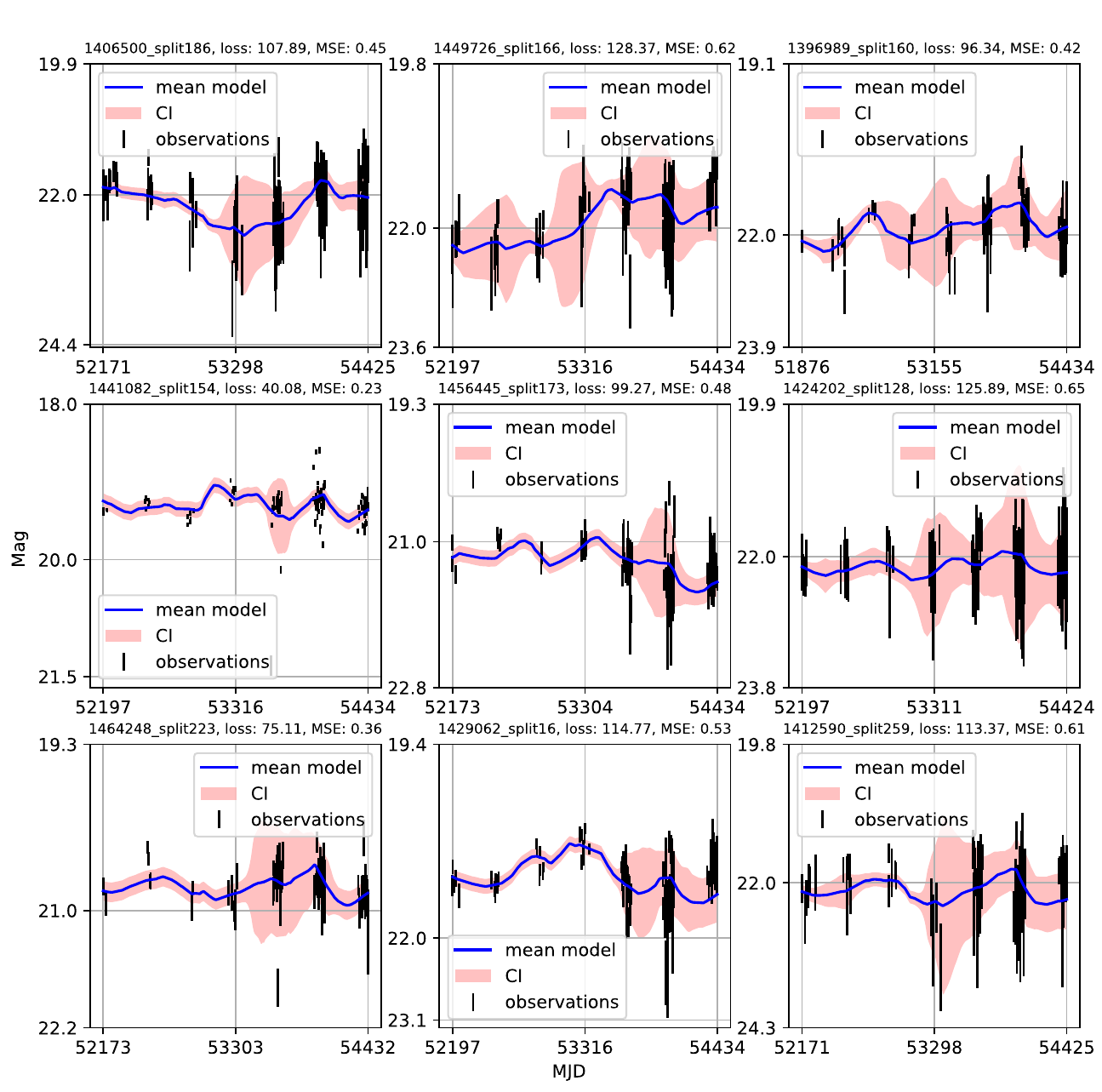}
\end{adjustwidth}
\caption{The same as Figure \ref{fig111}.\label{fig141}}
\end{figure}

\begin{figure}[H]
\begin{adjustwidth}{-\extralength}{0cm}
\centering
\includegraphics[page=24,width=1.3\textwidth]{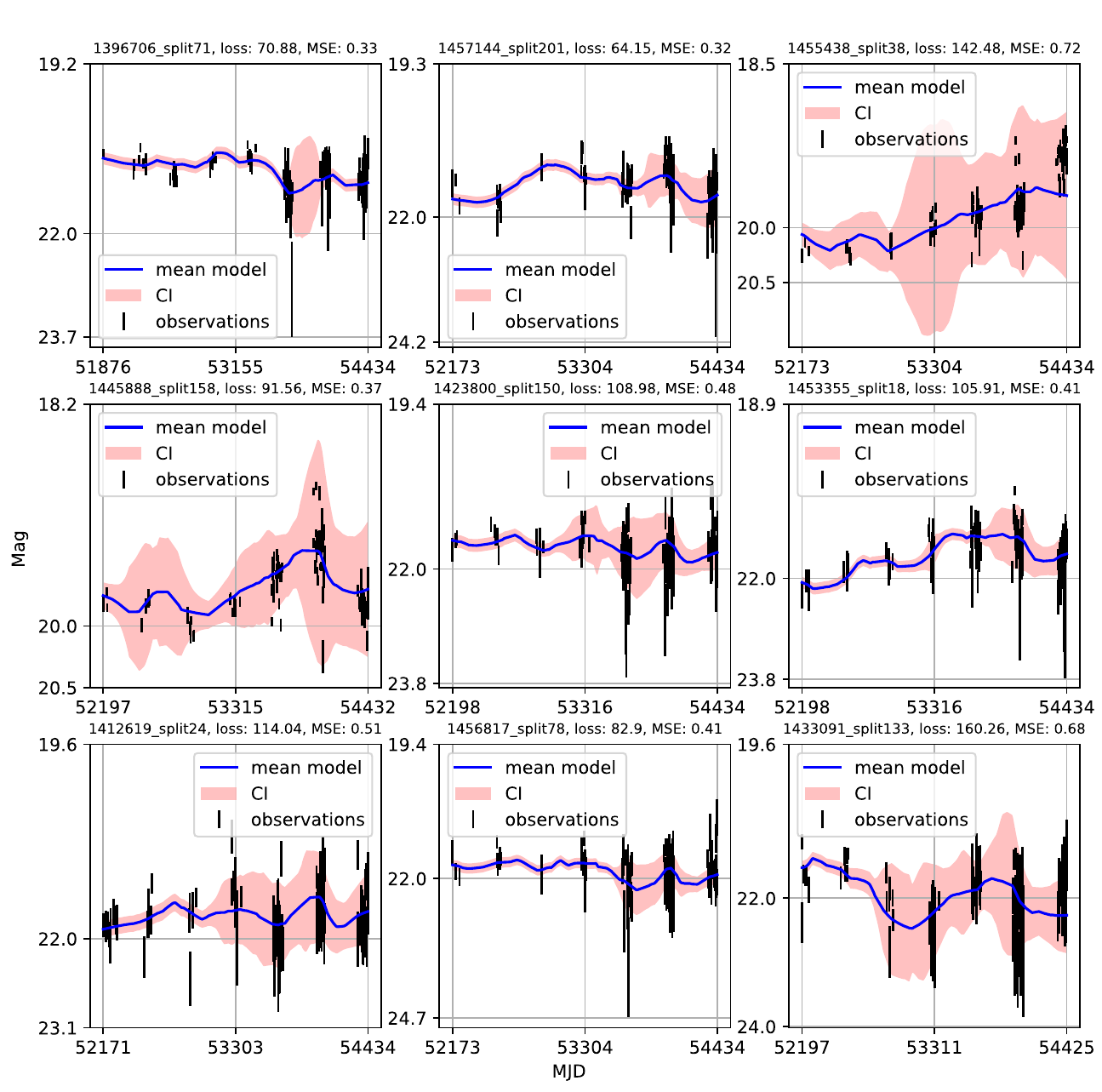}
\end{adjustwidth}
\caption{The same as Figure \ref{fig111}.\label{fig142}}
\end{figure}

Figures \ref{fig111}-\ref{fig142} show a collection of plots of the observed \texttt{u}-band and corresponding predicted light curves for selected 283 objects using a training set for the CNP model.

\begin{adjustwidth}{-\extralength}{0cm}

\reftitle{References}


\bibliography{bibliography}

%


\end{adjustwidth}
\end{document}